\documentstyle[fleqn,aps,rmp,epsf,graphicx,rotating]{revtex}
\newcommand{\sig}{\:\lower0.6ex\hbox{$\stackrel{\textstyle >}{\sim}$}\:}
\newcommand{\sil}{\:\lower0.6ex\hbox{$\stackrel{\textstyle <}{\sim}$}\:}
\newcommand{\sigs}{\:\lower0.4ex\hbox{$\stackrel{\scriptstyle
      >}{\scriptstyle \sim}$}\,}
\newcommand{\sils}{\:\lower0.4ex\hbox{$\stackrel{\scriptstyle
      <}{\scriptstyle \sim}$}\,}
\def\etal{{\em et {al.}}}%
\def\S{{Section}}%

\def\araa{{\em Ann.\ Rev.\ Astron.\ Astrophys.}}

\def\aj{{\em Astron.\ J.}}
\def\anap{{\em Ann.\ Astrophys.}}   
\def\an{{\em Astron. Nach.}}
\def\apj{{\em Astrophys.\ J.}}
\def\apjl{{\em Astrophys.\ J.\ Lett.}}
\def\apjs{{\em Astrophys.\ J.\ Suppl.\ Ser.}}
\def\aap{{\em Astron.\ Astrophys.}}
\def\apss{{\em Astrophys.\ Space Science}}
\def\baas{{\em Bull.\ Amer.\ Astron.\ Soc.}}
\def\bain{{\em Bull.\ Astron.\ Inst.\ Netherlands}}
\def\fcp{{\em Fund.\ Cosm.\ Phys.}}
\def\jcam{{\em J.\ Comput.\ Appl.\ Math.}}

\def\jfm{{\em J.\ Fluid Mech.}}
\def\mnras{{\em Mon.\ Not.\ R.\ Astron.\ Soc.}}
\def\nat{{\em Nature}}
\def\pta{{\em Phil.\ Trans.\ A.}}
\def\ptp{{\em Prog.\ Theo.\ Phys.}}
\def\prd{{\em Phys.\ Rev.\ D}}
\def\pre{{\em Phys.\ Rev.\ E}}
\def\prl{{\em Phys.\ Rev.\ Lett.}}
\def\prsa{{\em Proc.\ R.\ Soc.\ London A}}
\def\pasj{{\em Pub.\ Astron.\ Soc.\ Japan}}
\def\pasp{{\em Pub.\ Astron.\ Soc.\ Pacific}}
\def\pfl{{\em Phys.\ Fluids}}
\def\ppl{{\em Phys.\ Plasmas}}
\def\rpp{{\em Rep.\ Prog.\ Phys.}}
\def\rmp{{\em Rev.\ Mod.\ Phys.}}
\def\zp{{\em Z.\ Phys.}}
\def\za{{\em Z.\ Astrophys.}}

\begin{document}
\title{Control of star formation by supersonic turbulence}
\author{Mordecai-Mark Mac Low}
\address{Department of Astrophysics, American Museum of Natural History, \\
79th Street at Central Park West, New York, NY 10024-5192, USA
\footnote{Electronic address: mordecai@amnh.org}}
\author{Ralf S. Klessen}
\address{Astrophysikalisches Institut Potsdam, An der Sternwarte 16,
D-14482 Potsdam, Germany\footnote{Electronic address: rklessen@aip.de}\\
and UCO/Lick Observatory, University of California, 
Santa Cruz, CA 95064, USA}

\maketitle

\begin{abstract}
Understanding the formation of stars in galaxies is central to much of
modern astrophysics.  However, a quantitative prediction of the star
formation rate and the initial distribution of stellar masses remains
elusive. For several decades it has been thought that the star
formation process is primarily controlled by the interplay between
gravity and magnetostatic support, modulated by neutral-ion drift
(known as ambipolar diffusion in astrophysics).  Recently, however,
both observational and numerical work has begun to suggest that
supersonic turbulent flows rather than static magnetic fields control
star formation.  To some extent, this represents a return to ideas
popular before the importance of magnetic fields to the interstellar
gas was fully appreciated.  This review gives a historical overview of
the successes and problems of both the classical dynamical theory, and
the standard theory of magnetostatic support from both observational
and theoretical perspectives.  The outline of a new theory relying on
control by driven supersonic turbulence is then presented.  Numerical
models demonstrate that although supersonic turbulence can provide
global support, it nevertheless produces density enhancements that
allow local collapse.  Inefficient, isolated star formation is a
hallmark of turbulent support, while efficient, clustered star
formation occurs in its absence.  The consequences of this theory are
then explored for both local star formation and galactic scale star
formation.  It suggests that individual star-forming cores are likely
not quasi-static objects, but dynamically collapsing.  Accretion onto
these objects varies depending on the properties of the surrounding
turbulent flow; numerical models agree with observations showing
decreasing rates.  The initial mass distribution of stars may also be
determined by the turbulent flow.  Molecular clouds appear to be
transient objects forming and dissolving in the larger-scale turbulent
flow, or else quickly collapsing into regions of violent star
formation. We suggest that global star formation in galaxies is
controlled by the same balance between gravity and turbulence as
small-scale star formation, although modulated by cooling and
differential rotation.  The dominant driving mechanism in star-forming
regions of galaxies appears to be supernovae, while elsewhere coupling
of rotation to the gas through magnetic fields or gravity may be
important.\\[0.2cm]
{\em Accepted for publication in Reviews of Modern Physics}
\end{abstract}

\tableofcontents

\section{INTRODUCTION}
\label{sec:introduction}

\subsection{Overview}

Stars are important. They are the dominant source of radiation (with
competition from the cosmic microwave background and from accretion
onto black holes, which themselves probably formed from stars), and of
all chemical elements heavier than the H, He, and Li that made up the
primordial gas.  The Earth itself consists mainly of these heavier
elements, called metals in astronomical terminology.  Metals are
produced by nuclear fusion in the interior of stars, with the heaviest
elements produced during the passage of the final supernova shockwave
through the most massive stars.  To reach the chemical abundances
observed today in our solar system, the material had to go through
many cycles of stellar birth and death.  In a literal sense, we are
star dust.

Stars are also our primary source of astronomical information and,
hence, are essential for our understanding of the universe and the
physical processes that govern its evolution. At optical wavelengths
almost all natural light we observe in the sky originates from stars.
During day this is obvious, but it is also true at night. The Moon,
the second brightest object in the sky, reflects light from our Sun,
as do the planets, while virtually every other extraterrestrial source
of visible light is a star or collection of stars. Throughout the
millenia, these objects have been the observational targets of
traditional astronomy, and define the celestial landscape, the
constellations.  

\begin{figure}[ht]
\begin{center}\unitlength1cm
\begin{picture}( 8.00, 9.00)
\put( 0.00, 0.50){\epsfxsize=7.4cm \epsfbox{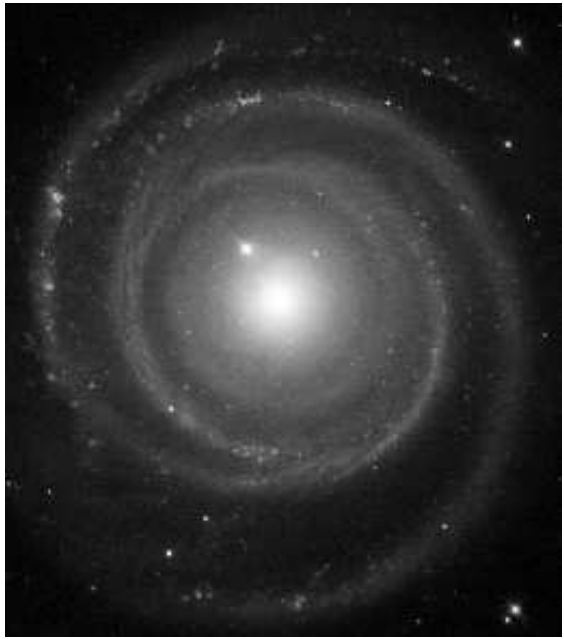}}
\end{picture}
\end{center}%
\caption{\label{fig:NGC4622} Optical image of the
  spiral galaxy NGC$\,$4622 observed with the Hubble Space Telescope.
  (Courtesy of NASA and The Hubble Heritage Team --- STScI/AURA) }
\end{figure}

When we look at the sky on a clear night, we can also note dark patches of
obscuration along the band of the Milky Way.  These are clouds of dust
and gas that block the light from stars further away.  For roughly the
last century we have known that these clouds give birth to stars. The
advent of new observational instruments made it
possible to observe astronomical objects at wavelengths ranging from
$\gamma$-rays to radio frequencies.  Especially useful for
studying the dark clouds are radio, sub-millimeter and far-infrared
wavelengths, at which they are transparent. Observations now show that
{\em all} star formation occurring in the Milky Way is associated with
the dark clouds of molecular hydrogen and dust.

Stars are common. The mass of the Galactic disk plus bulge is about
$6\times 10^{10}\,$M$_{\odot}$ (e.g.\ Dehnen \& Binney 1998), where
$1\,$M$_{\odot} = 1.99 \times10^{33}\,$g is the mass of our Sun. Thus,
there are of order $10^{12}$ stars in the Milky Way, assuming standard
values for the stellar mass distribution (e.g.\ Kroupa 2002). Stars
form continuously. Roughly 10\% of the disk mass of the Milky Way
is in the form of gas, which is forming stars at a rate of about
$1\,$M$_{\odot}\,$yr$^{-1}$.  Although stars dominate the baryonic
mass in the Galaxy, dark matter determines the overall mass
budget: invisible material that reveals its presence only by its
contribution to the gravitational potential. The dark matter halo of
our Galaxy is about ten times more massive than gas and stars together.
At larger scales this imbalance is even more pronouced. Stars are
estimated to make up only 0.4\% of the total mass of the Universe
(Lanzetta, Yahil, \& Fernandez-Soto 1996), and about 17\% of the total
baryonic mass (Walker \etal\ 1991).

Mass is the most important parameter determining the evolution of
individual stars. Massive stars with high pressures at their centers
have strong nuclear fusion there, making them short-lived but very
luminous, while low-mass stars are long-lived but extremely faint.
For example, a star with $5\,$M$_{\odot}$ only lives for $2.5\times
10^7\,$yr, while a star with $0.2\,$M$_{\odot}$ survives for
$1.2\times 10^{13}\,$yr, orders of magnitude longer than the current
age of the universe. For comparison the Sun with an age of $4.5\times
10^9\,$yr has reached approximately half of its life span. The
relationship between mass
and luminosity is quite steep with roughly $L\propto M^{3.2}$
(Kippenhahn \& Weigert 1990). During its short life a
$5\,$M$_{\odot}$ star will shine with a luminosity of $1.5\times
10^4\,$L$_{\odot}$, while the luminosity of an $0.2\,$M$_{\odot}$
star is only $\sim10^{-3}\,$L$_{\odot}$. For reference, the luminosity
of the Sun is $1\,$L$_{\odot} = 3.85\times 10^{33}$erg$\,$s$^{-1}$.

The light from star-forming external galaxies in the visible and blue
wavebands is dominated by young, massive stars. This is the reason why
we observe beautiful spiral patterns in many disk galaxies, like
NGC$\,$4622 shown in Figure \ref{fig:NGC4622}, as spiral density waves
lead to gas compression and subsequent star formation at the wave
locations. Massive stars dominate the optical emission from external
galaxies. In their brief lifetimes, massive stars do not have
sufficient time to disperse in the galactic disk, so they still trace
the characteristics of the instability that triggered their formation.
Hence, understanding the dynamical properties of galaxies requires an
understanding of how, where, and under which conditions stars form.

In a simple approach, galaxies can be seen as gravitational potential
wells containing gas that has been able to radiatively cool in less
than the current age of the universe.  In the absence of any
hindrance, the gas then collapses gravitationally to form stars on a
free-fall time (Jeans 1902)
\begin{equation}
\label{eqn:free-fall-time}
\tau_{\rm ff} = \left(\frac{3\pi}{32 G \rho}\right)^{1/2} = 140 \mbox{ Myr}
\left(\frac{n}{0.1\,{\rm cm}^{-3}}\right)^{-1/2},
\end{equation}
where $n$ is the number density of the gas. Interstellar gas in the
Milky Way consists of one part He for every ten parts H. The mass
density $\rho = \mu n$, where we take the Galactic value for the mean
mass per particle in neutral atomic gas of $\mu = 2.11 \times 10^{-24}\,$g,
and $G$ is the gravitational constant. The free-fall time $\tau_{\rm
ff}$ is very short compared to the age of the Milky Way, about
$10^{10}\,$yr.  However, gas remains in the Galaxy and stars continue
to form from gas that must have already been cooled below its virial
temperature for many billions of years.  What physical processes
regulate the rate at which gas turns into stars?  Another way of
asking the question is, what prevented the Galactic gas from forming
stars at an extremely high rate immediately after it first cooled, and
being completely used up?

Observations of the star formation history of the universe demonstrate
that stars did indeed form more vigorously in the past than today
(e.g.\ Lilly \etal\ 1996, Madau \etal\ 1996, Baldry \etal\ 2002,
Lanzetta \etal\ 2002), with as much as 80\% of star formation in the
Universe being complete by redshift $z=1$, less than half of the
current age of 13 Gyr ago.  What mechanisms allowed rapid star
formation in the past, but reduce its rate today?

The clouds of gas and dust in which stars form are dense enough, and
well enough protected from dissociating UV radiation by self-shielding
and dust scattering in their surface layers, for hydrogen to be mostly
in molecular form in their interior.  The density and velocity
structure of these molecular clouds is extremely complex and follows
hierarchical scaling relations that appear to be determined by
supersonic turbulent motions (e.g.\ Blitz \& Williams 1999).
Molecular clouds are large, and their masses exceed the threshold for
gravitational collapse by far when taking only thermal pressure into
account.  Just like galaxies as a whole, naively speaking, they should
be contracting rapidly and forming stars at very high rate. This is
generally not observed.  
We can define a star formation efficiency of
a region as 
\begin{equation} \label{eqn:sfe}
\epsilon_{SF} = \dot{M_*} \tau/M,
\end{equation}
where $\dot{M_*}$ is the star formation rate, $\tau$ is the lifetime
of the region, and $M$ is the total gas mass in the region (e.g.\
Elmegreen \& Efremov 1997). The star formation efficiency of molecular
clouds in the solar neighborhood is estimated to be of order of a few
percent (Zuckerman \& Evans 1974).

For many years it was thought that support by magnetic pressure
against gravitational collapse offered the best explanation for the
slow rate of star formation.  In this theory, developed by Shu (1977;
and see Shu, Adams, \& Lizano 1987), Mouschovias (1976; and see
Mouschovias 1991b,c), Nakano (1976), and others, interstellar magnetic
field prevents the collapse of gas clumps with insufficient mass to
flux ratio, leaving dense cores in magnetohydrostatic equilibrium.
The magnetic field couples only to electrically charged ions in the
gas, though, so neutral atoms can only be supported by the field if
they collide frequently with ions.  The diffuse interstellar medium
(ISM) with number densities $n \simeq 1$~cm$^{-3}$ (see Ferri\`ere
2001 for a general review of ISM properties) remains highly
ionized, enough that neutral-ion collisional coupling is very efficient
(as we discuss below in \S~\ref{sub:standard}).  In dense cores, where
$n > 10^5$~cm$^{-3}$, ionization fractions drop below parts per ten
million.  Neutral-ion collisions no longer couple the neutrals tightly
to the magnetic field, so the neutrals can diffuse through the field.
This neutral-ion drift allows gravitational collapse to proceed in
the face of magnetostatic support, but on a timescale as much as an
order of magnitude longer than the free-fall time, drawing out the
star formation process.

In this paper we review a body of work that suggests that
magnetohydrostatic support modulated by neutral-ion drift fails to
explain the star formation rate, and indeed appears inconsistent with
observations of star-forming regions.  Instead, we suggest that
control of molecular cloud formation and subsequent support by
supersonic turbulence is both sufficient to explain star formation
rates, and more consistent with observations.  Our review focuses on
how gravitationally collapsing regions form.  The recent comprehensive
review by Larson (2003) goes into more detail on the final stages of
disk accretion and protostellar evolution.

\subsection{Turbulence}
\label{sub:turbulence}
At this point, we need to briefly discuss the concept of turbulence, and the
differences between supersonic, compressible (and magnetized) turbulence, and
the more commonly studied incompressible turbulence.  We mean by turbulence,
in the end, nothing more than the gas flow resulting from random motions at
many scales.  We furthermore will use in our discussion only the very general
properties and scaling relations of turbulent flows, focusing mainly on
effects of compressibility. For a more detailed discussion of the complex
statistical characteristics of turbulence, we refer the reader to the book by
Lesieur (1997). 

Most studies of turbulence treat incompressible turbulence,
characteristic of most terrestrial applications.
Root-mean-square (rms) velocities are subsonic, and the density remains
almost constant.  Dissipation of energy occurs primarily in the 
smallest vortices, where the dynamical scale $\ell$ is shorter than
the length on which viscosity acts $\ell_{\rm visc}$.  Kolmogorov (1941a)
described a heuristic theory based on dimensional analysis that
captures the basic behavior of incompressible turbulence surprisingly
well, although subsequent work has refined the details substantially.
He assumed turbulence driven on a large scale $L$, forming eddies at
that scale.  These eddies interact to form slightly smaller eddies,
transferring some of their energy to the smaller scale.  The smaller
eddies in turn form even smaller ones, until energy has cascaded all
the way down to the dissipation scale $\ell_{\rm visc}$.  

In order to maintain a steady state, equal amounts of energy must be
transferred from each scale in the cascade to the next, and eventually
dissipated, at a rate
\begin{equation}
\dot{E} = \eta v^3/L,
\end{equation}
where $\eta$ is a constant determined empirically. This leads to a
power-law distribution of kinetic energy $E\propto v^2 \propto
k^{-11/3}$, where $k = 2\pi/\ell$ is the wavenumber, and density does
not enter because of the assumption of incompressibility.  Most of the
energy remains near the driving scale, while energy drops off steeply
below $\ell_{\rm visc}$.  Because of the apparently local nature of
the cascade in wavenumber space, the viscosity only determines the
behavior of the energy distribution at the bottom of the cascade below
$\ell_{\rm visc}$, while the driving only determines the behavior near
the top of the cascade at and above $L$.  The region in between is
known as the inertial range, in which energy transfers from one scale
to the next without influence from driving or viscosity.  The behavior
of the flow in the inertial range can be studied regardless of the
actual scale at which $L$ and $\ell_{\rm visc}$ lie, so long as they
are well separated.  One statistical description of incompressible
turbulent flow, the structure functions $S_p(\vec{r}) = \langle
\{v(\vec{x}) - v(\vec{x}+\vec{r})\}^p \rangle$, has been successfully
modeled by assuming that dissipation occurs in filamentary vortex
tubes (She \& Leveque 1994).

Gas flows in the ISM, however, vary from this idealized picture in
three important ways.  First, they are highly compressible, with Mach
numbers ${\cal M}$ ranging from order unity in the warm ($10^4$~K),
diffuse ISM, up to as high as 50 in cold (10~K), dense molecular
clouds.  Second, the equation of state of the gas is very soft due to
radiative cooling, so that pressure $P\propto \rho^{\gamma}$ with the
polytropic index falling in the range $0.4 < \gamma < 1.2$ as a
function of density and temperature (e.g.\ Spaans \& Silk 2000,
Ballesteros-Paredes, V{\'a}zquez-Semadeni, \& Scalo 1999, Scalo \etal\
1998). Third, the driving of the turbulence is not uniform, but rather
comes from blast waves and other inhomogeneous processes.

Supersonic flows in highly compressible gas create strong density
perturbations.  Early attempts to understand turbulence in the ISM
(von Weizs\"acker 1943, 1951, Chandrasekhar 1949) were based on
insights drawn from incompressible turbulence.  An attempt to
analytically derive the density spectrum and resulting gravitational
collapse criterion was first made by Chandrasekhar (1951a,b).  This
work was followed up by several authors, culminating in work by Sasao
(1973) on density fluctuations in self-gravitating media whose
interest has only been appreciated recently.  
Larson (1981) qualitatively applied the basic idea of density
fluctuations driven by supersonic turbulence to the problem of star
formation.
Bonazzola et al.\ (1992) used a renormalization group technique to
examine how the slope of the turbulent velocity spectrum could
influence gravitational collapse.
This approach was combined with low-resolution numerical models to
derive an effective adiabatic index for subsonic compressible
turbulence by Panis \& P\'erault (1998).
Adding to the complexity of the problem, the strong density
inhomogeneities observed in the ISM can be caused not only by
compressible turbulence, but also by thermal phase transitions (Field,
Goldsmith, \& Habing 1969, McKee \& Ostriker 1977, Wolfire \etal\
1995) or gravitational collapse (e.g.\ Kim \& Ostriker 2001).

In supersonic turbulence, shock waves offer additional possibilities
for dissipation.  Shock waves can also transfer energy between widely
separated scales, removing the local nature of the turbulent cascade
typical of incompressible turbulence.  The spectrum may shift only
slightly, however, as the Fourier transform of a step function
representative of a perfect shock wave is $k^{-2}$.  Integrating in
three dimensions over an ensemble of shocks, the differential energy
spectrum $E(k) dk = \rho v^2(k) k^2 dk \propto k^{-2}dk$. This is just the
compressible energy spectrum reported by Porter \& Woodward (1992) and
Porter, Pouquet, \& Woodward (1992, 1994).  They also found that even
in supersonic turbulence, the shock waves do not dissipate all the
energy, as rotational motions continue to contain a substantial
fraction of the kinetic energy, which is then dissipated in small
vortices.  Boldyrev (2002) has proposed a theory of velocity structure
function scaling based on the work of She \& Leveque (1994) using the
assumption that dissipation in supersonic turbulence primarily occurs
in sheet-like shocks, rather than linear filaments at the centers of
vortex tubes.  First comparisons to numerical models show good
agreement with this model (Boldyrev, Nordlund, \& Padoan 2002a), and
it has been extended to the density structure functions by Boldyrev,
Nordlund, \& Padoan (2002b).
Transport properties of supersonic turbulent flows in the
astrophysical context have been discussed by Avillez \& Mac~Low (2002)
and Klessen \& Lin (2003).

The driving of interstellar turbulence is neither uniform nor
homogeneous.  Controversy still reigns over the most important energy
sources at different scales, but we make the argument in
\S~\ref{sub:driving} that isolated and correlated supernovae dominate.
However, it is not yet understood at what scales expanding,
interacting blast waves contribute to turbulence.  Analytic estimates
have been made based on the radii of the blast waves at late times
(Norman \& Ferrara 1996), but never confirmed with numerical models
(much less experiment).  Indeed, the thickness of the blast waves may
be more important than the radii.

Finally, the interstellar gas is magnetized.  Although magnetic field
strengths are difficult to measure, with Zeeman line splitting being
the best quantitative method, it appears that fields within an order
of magnitude of equipartition with thermal pressure and turbulent
motions are pervasive in the diffuse ISM, most likely maintained by a
dynamo driven by the motions of the interstellar gas (e.g.\ Ferri\`ere
1992).  A model for the distribution of energy and the scaling
behavior of strongly magnetized, incompressible turbulence based on
the interaction of shear Alfv\'en waves is given by Goldreich \&
Sridhar (1995, 1997) and Ng \& Bhattacharjee (1996).  They found that
an anisotropic Kolmogorov spectrum $k^{-5/3}$ best describes
the one-dimensional (1D) energy spectrum, rather than the $k^{-3/2}$
spectrum first proposed by Iroshnikov (1963) and Kraichnan (1965).
These results have been confirmed by Verma \etal\ (1996) using
numerical models, and by Verma (1999) using a renormalization group
approach. The scaling properties of the structure functions of such
turbulence was derived from the work of She \& Leveque (1994) by
M\"uller \& Biskamp (2000; also see Biskamp \& M\"uller 2000) by
assuming that dissipation occurs in current sheets.  A theory of
weakly compressible turbulence applicable in particular to small
scales in the ISM has been derived by Lithwick \& Goldreich (2001),
but little progress has been made towards analytic models of strongly
compressible magnetohydrodynamic (MHD) turbulence with ${\cal M} \gg
1$.  See, however, the reviews by Cho, Lazarian, \& Vishniac (2002),
and Cho \& Lazarian (2003).
In particular, an analytic theory of the non-linear density
fluctuations characteristic of such turbulence remains lacking.

\subsection{Outline}

With the above in mind, we suggest that stellar birth is regulated by
interstellar turbulence and its interplay with gravity.  Turbulence,
even if strong enough to counterbalance gravity on global scales, will
usually provoke collapse on smaller scales.  Supersonic turbulence
establishes a complex network of interacting shocks, where converging
flows generate regions of high density. This density enhancement can
be sufficient for gravitational instability. Collapse sets in.
However, the random flow that creates local density enhancements also
may disperse them again.  Hence, the efficiency of star formation
(eq.~\ref{eqn:sfe}) depends strongly on the properties of the
underlying turbulent velocity field, on its driving lengthscale and
strength relative to gravitational attraction.  This principle holds
for star formation throughout all scales considered in this review,
ranging from small star forming regions up to galaxies as a whole.

To lay out this picture of star formation in more detail, we first
outline the observed properties of star-forming interstellar clouds,
and the distribution of stellar masses that form there in
\S~\ref{sec:molecular}.  We then critically discuss the historical
development of star formation theory in \S~\ref{sec:history}.  We
begin the section by describing the classical dynamic theory, and then
move on to the so-called standard theory, where the star formation
process is controlled by magnetic fields.  After describing the
theoretical and observational problems that both approaches have, we
present work in \S~\ref{sec:paradigm} that leads us to the argument
that star formation is controlled by the interplay between gravity and
supersonic turbulence.  The theory is applied to individual star
forming regions in \S~\ref{sec:local}, where we investigate the
implications for stellar clusters, protostellar cores (the direct
progenitors of individual stars), binary stars, protostellar mass
accretion, and the subsequent distribution of stellar masses. In
\S~\ref{sec:galactic}, we discuss the control of star formation by
supersonic turbulence on galactic scales.  We first examine the
formation and destruction of star-forming molecular clouds in light of
models of turbulent flow.  We then ask when is star formation
efficient in galaxies? We review the energetics of the possible
mechanisms that generate and maintain supersonic turbulence in the
interstellar medium, and come to the conclusion that supernova
explosions accompanying the death of massive stars are the most likely
agents.  Then we briefly apply the theory to various types of
galaxies, ranging from low surface brightness galaxies to massive star
bursts.  Finally, in \S~\ref{sec:conclusions} we summarize, and
describe unsolved problems open for future research.

\section{OBSERVATIONS}
\label{sec:molecular}
All present day star formation takes place in molecular clouds (e.g.\
Blitz 1993, Williams, Blitz, \& McKee 2000), so it is vital to
understand the properties, dynamical evolution and fragmentation of
molecular clouds in order to understand star formation. We begin this
section by describing the composition (\S~\ref{sub:mol-clouds}) and
density and velocity structure (\S~\ref{sub:LSS}) of molecular clouds.
We then discuss turbulent support of clouds against gravitational
collapse (\S~\ref{sub:support}), and introduce the observed scaling relations and
their relation to the turbulent flow (\S~\ref{sub:scaling-law}).
Finally, we describe observations of protostellar cores (\S~\ref{sub:cores})
and of the initial mass function of stars (\S~\ref{sub:IMF-observed}).

\subsection{Composition of molecular clouds}
\label{sub:mol-clouds}

Molecular clouds are density enhancements in the interstellar gas dominated by
molecular H$_2$ rather than the atomic H typical of the rest of the
ISM (e.g.\ Ferri\`ere 2001),
mainly because they are opaque to the UV radiation that elsewhere dissociates
the molecules.  In the plane of the Milky Way, interstellar gas has been
extensively reprocessed by stars, so the
metallicity\footnote{Metallicity in astrophysics is
  usually defined as the fraction of heavy elements relative to hydrogen.  It
  averages over local variations in the abundance of the different elements
  caused by varying chemical enrichment histories.}  
is close to the solar value $Z_{\odot}$, while in other galaxies with
lower star formation rates, the metallicity can be as little as
$10^{-3}Z_{\odot}$. 
The refractory elements condense into dust grains,
while others form molecules.  The properties of the dust grains change
as the temperature drops within the cloud, probably due to the
freezing of volatiles such as water and ammonia (e.g. Goodman \etal\ 1995).
This has important consequences for the radiation transport properties
and the optical depth of the clouds.  The presence of heavier elements
such as carbon, nitrogen, and oxygen determines the heating and
cooling processes in molecular clouds (e.g.\ Genzel 1991).  In
addition, 
continuum emission from dust and
emission and absorption lines emitted by molecules formed
from these elements are the main observational tracers of cloud
structure, as cold molecular hydrogen is very difficult to observe.
Radio and sub-millimeter telescopes mostly concentrate on 
thermal continuum from dust and
the rotational transition lines of carbon, oxygen and nitrogen
molecules (e.g.\ CO, NH$_3$, or H$_2$O). By now, several hundred
different molecules have been identified in the interstellar gas. An
overview of the application of different molecules as tracers for
different physical conditions can be found in the reviews by van
Dishoeck \etal\ (1993), Langer \etal\ (2000), van Dishoeck \&
Hogerheijde (2000).

\begin{figure}[htp]
\begin{center}\unitlength1.0cm
\begin{picture}(16,18.5)
\put(1.0,0.0){\epsfysize=18cm \epsfbox{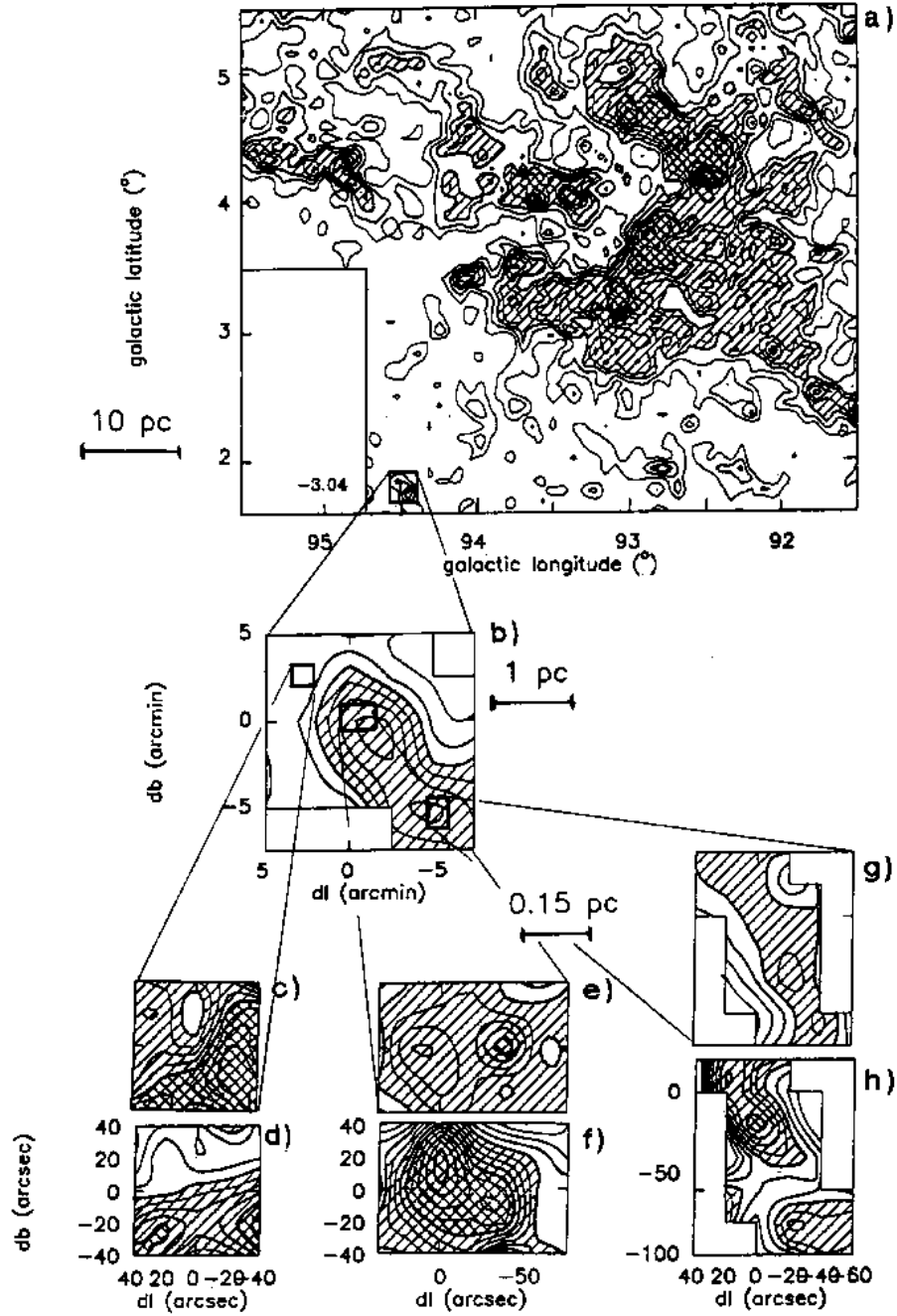}}
\end{picture}
\end{center}\caption{\label{fig:falg92}
Maps of the molecular gas in the Cygnus OB7 complex. (a) Large scale
map of the $^{13}$CO $({\rm J}=1-0)$ emission. The first level and the
contour spacing are $0.25\,$K. (b) Map of the same transition line of a
sub-region with higher resolution (first contour level and spacing are
$0.3\,$K). Both maps were obtained with the Bordeaux telescope. (c)
$^{12}$CO $({\rm J}=1-0)$ and (d) $^{13}$CO $({\rm J}=1-0)$ emission
from the most transparent part of the field. (e) $^{13}$CO $({\rm
J}=1-0)$ and (f) C$^{18}$O $({\rm J}=1-0)$ emission from the most
opaque field. (g) $^{13}$CO $({\rm J}=1-0)$ and (h) C$^{18}$O $({\rm
J}=1-0)$ emission from a filamentary region with medium density. The
indicated linear sizes are given for a distance to Cygnus OB7 of
$750\,$pc. (The figure is from Falgarone \etal\ 1992). 
}
\end{figure}

\subsection{Density and velocity structure of molecular clouds}
\label{sub:LSS}

Emission line observations of molecular clouds reveal clumps and
filaments on all scales accessible by present day telescopes. Typical
parameters of different regions in molecular clouds are listed in
Table \ref{tab:MC-prop}, adopted from Cernicharo (1991). The mass
spectrum of clumps in molecular clouds appears to be well described by
a power law, indicating self-similarity: there is no natural mass or
size scale between the lower and upper limits of the observations.
The largest molecular structures considered to be single objects are
giant molecular clouds (GMCs), which have masses of $10^5$ to
$10^6\,{\rm M}_{\odot}$, and extend over a few tens of parsecs. The
smallest observed structures are protostellar cores with masses of a
few solar masses or less and sizes of $\sil 0.1\,$pc, and less-dense
clumps of similar size.  The volume filling factor $\langle n \rangle
/n$ (where $n$ is the local density, while $\langle n \rangle$ is the
average density of the cloud) of dense clumps, even denser subclumps
and so on, is rather small, ranging from 10\% down to 0.1\% at
densities of $n >10^5$~cm$^{-3}$ (e.g. McKee 1999).  Star formation
always occurs in the densest regions within a cloud, so only a small
fraction of molecular cloud matter is actually involved in building up
stars, while the bulk of the material remains at lower densities.

The density structure of molecular clouds is best inferred from the
column density of dust, which can be observed either via its thermal emission
at millimeter wavelengths in dense regions (e.g.\ Testi \& Sargent
1998, Motte, Andr\'e, \& Neri 1998), or via its extinction of
background stars in the infrared, if a uniform screen of background
stars is present (Lada \etal\ 1994, Alves, Lada, \& Lada 2001).  
Deriving density and mass from thermal emission requires modeling the
temperature profile, which depends on optically thick radiative
transfer through uncertain density distributions.  
Infrared extinction, on the other hand, requires only suitable
background stars.  Reliance on the near-IR color excess to measure column
densities ensures a much greater dynamic range than optical
extinction.  This method has been further developed by Cambr\'esy
\etal\ (2002) who use an adaptive grid to extract maximum information
from non-uniform background star fields. It turns out that the higher
the column density in a region, the higher the variation in extinction
among stars behind that region (Lada \etal\ 1994).  Padoan
\& Nordlund (1999) demonstrated this to be consistent with a
super-Alfv\'enic turbulent flow, while Alves \etal\ (2001) modeled it
with a single cylindrical filament with density $\rho \propto
r^{-2}$. Because turbulence forms many filaments, it is not clear that
these two descriptions are actually contradictory (Padoan, 2001,
private communication), although the identification of a single
filament would then suggest that a minimum scale for the turbulence
has been identified.

A more general technique is emission in optically thin spectral
lines. The best candidates are $^{13}{\rm CO}$ and C$^{18}$O, though
CO freezes out in the very densest regions (with visual extinctions
above $A_V \simeq 10$ magnitudes, see Alves, Lada, \& Lada, 1999).  CO
observations are therefore only sensitive to gas at relatively low
densities $n \sil 10^5$~cm$^{-3}$, and are limited in dynamic range to
at most two decades of column density.  The colder the gas, the lower
the column density at which lines will become optically thick.
Nevertheless, the development of sensitive radio receivers in the
1980's first made it feasible to map an entire molecular cloud region
with high spatial and spectral resolution to obtain quantitative
information about the overall density structure.

The hierarchy of clumps and filaments spans all observable scales (e.g.\ 
Falgarone, Puget, \& Perault 1992, Falgarone \& Phillips 1996, Wiesemeyer
\etal\ 1997) extending down to individual protostars studied with
millimeter-wavelength interferometry (Ward-Thompson \etal\ 1994, Langer \etal\
1995, Gueth \etal\ 1997, Motte \etal\ 1998, Testi \& Sargent 1998,
Ward-Thompson, Motte, \& Andr{\'e} 1999, Bacmann \etal\ 2000, Motte
\etal\ 2001). This is illustrated in Figure \ref{fig:falg92}, which
shows $^{13}$CO, $^{12}$CO and C$^{18}$O maps of a region in the
Cygnus OB7 complex at three levels of successively higher resolution
(from Falgarone \etal\ 1992). At each level, the molecular cloud
appears clumpy and highly structured. When observed with higher
resolution, each clump breaks up into a filamentary network of smaller
clumps. Unresolved features exist even at the highest resolution. The
ensemble of clumps identified in this survey covers a mass range from
about $1\,$M$_{\odot}$ up to a few $100\,$M$_{\odot}$ and densities
$50\,{\rm cm}^{-3} < n({\rm H}_2) < 10^4\,{\rm cm}^{-3}$. These values
are typical for all studies of cloud clump structure, with higher
densities being reached primarily in protostellar cores.

The distribution of clump masses is consistent with a power law of the form
\begin{equation}
\label{eqn:clump-spectrum}
\frac{dN}{dm} \propto m^{\alpha}\:,
\end{equation}
with $-1.3 < \alpha < -1.9$ in molecular line studies (Carr 1987,
Stutzki \& G{\"u}sten 1990, Lada, Bally, \& Stark 1991, Williams, de
Geus, \& Blitz 1994, Onishi \etal\ 1996, Kramer 1998, Heithausen
\etal\ 1998). Dust continuum studies, which pick out the
highest column density regions, find steeper values of $-1.9 <
\alpha < -2.5$ (Testi \& Sargent 1998, Motte \etal\ 1998; also see the
discussion in Ossenkopf, Klessen, \& Heitsch 2001), similar to the stellar mass
spectrum. The power-law mass spectrum is often interpreted as a
manifestation of fractal density structure (e.g.\ Elmegreen \&
Falgarone 1996).  However, the full physical meaning remains
unclear. In most studies molecular cloud clumps are determined either
by a Gaussian decomposition scheme (Stutzki \& G{\"u}sten 1990) or by
the attempt to define (and separate) clumps following density peaks
(Williams \etal\ 1994). There is no one-to-one correspondence between
the identified clumps in either method, however. Furthermore,
molecular clouds are only seen in projection, so one only measures
column density instead of volume density.
It remains unproven that all regions of high density also have high
column density, and vice-versa.  
Even when velocity information is taken into account, the real
3D structure of the cloud remains elusive. In
particular, it can be demonstrated in models of interstellar
turbulence that single clumps identified in simulated observational cubes
(position-position-velocity space) tend to separate into multiple clumps
in real 3D space (Ostriker \etal\ 2001,
Ballesteros-Paredes \& Mac Low 2002).
This effect acts over regions of velocity width similar to the
velocity dispersion, enough to confound clumps even in clouds showing
large-scale velocity gradients.
These projection effects leave clump mass spectra as
poor statistical tools for characterizing molecular cloud structure.

Other means to quantify the structural and dynamical properties of
molecular clouds involve correlations and probability distribution
functions (PDFs) of dynamical variables. Two-point correlation
functions have been studied by many authors, including Scalo (1984),
Kleiner \& Dickman (1987), Kitamura \etal\ (1993), Miesch \& Bally
(1994), LaRosa, Shore \& Magnani (1999), and Ballesteros-Paredes,
V{\'a}zquez-Semadeni, \& Goodman (2002), while other studies have
concentrated on analyzing the PDFs of the column density in
observations, both physical and column density in computational
models, and of dynamical observables such as the centroid velocities
of molecular lines and their differences.  The density PDF has been
used to characterize numerical simulations of the interstellar medium
by V{\'a}zquez-Semadeni (1994), Padoan, Nordlund, \& Jones (1997),
Passot, \& V{\'a}zquez-Semadeni (1998), Scalo
\etal\ (1998), and Klessen (2000). Velocity PDFs for several star-forming
molecular clouds have been determined by Miesch \& Scalo (1995) and
Miesch, Scalo \& Bally (1999). Lis \etal~(1996, 1998) analyzed
snapshots of a numerical simulation of mildly supersonic, decaying
turbulence (without self-gravity) by Porter, Pouquet, \& Woodward
(1994) and applied the method to observations of the $\rho$-Ophiuchus
cloud. The observed PDFs exhibit strong non-Gaussian
features, often being nearly exponential with possible evidence for
power-law tails in the outer parts. Further methods to quantify
molecular cloud structure involve spectral correlation methods
(Rosolowsky \etal\ 1999), principal component analysis (Heyer \&
Schloerb 1997), or pseudometric methods used to describe and rank
cloud complexity (Wiseman \& Adams 1994, Adams \& Wiseman 1994).

A technique especially sensitive to the amount of structure on
different spatial scales is wavelet analysis (e.g.\ Gill \& Henriksen
1990; Langer, Wilson, \& Anderson 1993). In particular, the
$\Delta$-variance, introduced by Stutzki \etal\ (1998), provides a
good separation of noise and observational artifacts from the real
cloud structure. For isotropic systems its slope is directly related
to the spectral index of the corresponding Fourier power spectrum. It
can be applied in an equivalent way both to observational data and
gas dynamic and MHD turbulence simulations, allowing a direct
comparison, as discussed by Mac~Low \& Ossenkopf (2000), Bensch,
Stutzki, \& Ossenkopf (2001), and Ossenkopf \& Mac Low (2002).  They
find that the structure of low-density gas in molecular clouds is
dominated by large-scale modes and, equivalently, the velocity field
by large-scale motions. This means that molecular cloud turbulence is
likely to be driven from the outside, by sources acting external to
the cloud on scales of at least several tens of parsec (Ossenkopf \&
Mac~Low 2002).  

The observational findings are different, however, when focusing on
high-density gas in star forming regions. In this case, the
$\Delta$-variance clearly shows that the density structure is
dominated by individual protostellar cores at the smallest resolved
scales (Ossenkopf \etal\ 2001). This effect is best seen in dust
emission because it is able to trace large density contrasts.
Alternatively, dust extinction maps may also prove to be useful in
this context (see e.g.\ Alves \etal\ 2000 for the Bok globule B68; or
Padoan, Cambr\'esy, \& Langer 2002 for the Taurus molecular cloud).
As CO line emission maps mostly trace the tenuous gas between dense
cores, they miss the small-scale features and pick up the overall
density structure which is dominated by large-scale modes (Ossenkopf
\etal\ 2001).
                                
\subsection{Support of molecular clouds}
\label{sub:support}

Molecular clouds are cold (e.g.\ Cernicharo 1991). The kinetic
temperature inferred from molecular line ratios is typically about
10$\,$K for dark, quiescent clouds and dense cores in GMCs that are
shielded from UV radiation by high column densities of dust, while it
can reach 50--100~K in regions heated by UV radiation from high-mass
stars.  For example, the temperature of gas and dust behind the
Trapezium cluster in Orion is about 50$\,$K.  In cold regions, the
only heat sources are cosmic rays and dissipation of turbulence, while
cooling comes from emission from dust and abundant molecular species.
The thermal structure of the gas is related to its density
distribution and its chemical abundance, so it is remarkable that over
a wide range of gas densities and metallicities the equilibrium
temperature remains almost constant in a small range around $T\approx
10\,$K (Goldsmith \& Langer 1978, Goldsmith 2001).  In the absence of
strong UV irradiation, the approximation of isothermality only breaks
down when the cloud becomes dense enough to be opaque to cooling
radiation, so that heat can no longer be radiated away
efficiently. This occurs at gas density $n({\rm H}_2) >
10^{10}$cm$^{-3}$.  The equation of state then moves from isothermal
with polytropic exponent $\gamma =1$ to adiabatic, with $\gamma = 7/5$
being appropriate for molecular hydrogen (see e.g.\ Tohline 1982 and
references therein).

Despite their low temperatures, the densities in molecular clouds are
so high that their pressures exceed the average interstellar pressure
by an order of magnitude or more.  Typical interstellar pressures lie
around $10^{-13}$ erg cm$^{-3}$ (e.g. Jenkins \& Shaya 1979, Bowyer
\etal\ 1995), while at a temperature of 10~K and a density of
10$^3$~cm$^{-3}$, the pressure in a typical molecular cloud exceeds
$10^{-12}$~erg~cm$^{-3}$.  Gravitational confinement was traditionally
cited to explain the high pressures observed in GMCs (Kutner \etal\
1977; Elmegreen, Lada, \& Dickinson 1979; Blitz 1993; Williams \etal\
2000).  Their masses certainly exceed by orders of magnitude the
critical mass for gravitational stability $M_J$ defined by Eq.\
(\ref{eqn:jeans-mass}), computed from their average density and
temperature.  However, if only thermal pressure opposed gravitational
attraction, they should be collapsing and very efficiently forming
stars on a free-fall timescale
(Eq.\ \ref{eqn:free-fall-time}). That is not the
case. Within molecular clouds, low-mass gas clumps appear highly
transient and pressure confined rather than being bound by
self-gravity. Self-gravity appears to dominate only in the most
massive individual cores, where star formation actually is observed
(Williams, Blitz, \& Stark 1995; Yonekura \etal\ 1997; Kawamura \etal\
1998; Simon \etal\ 2001).

In the short lifetimes of molecular clouds (\S~\ref{sub:clouds}) they
likely never reach a state of dynamical equilibrium
(Ballesteros-Paredes \etal\ 1999a; Elmegreen 2000).  This is in
contrast to the classical picture that sees molecular clouds as
long-lived equilibrium structures (Blitz \& Shu 1980). The overall
star formation efficiency (eq.~\ref{eqn:sfe}) on scales of molecular
clouds as a whole is low in our Galaxy, of order of 10\% or smaller
(Zuckerman \& Evans 1974).  Only a small fraction of molecular cloud
material associated with the highest-density regions is actually
forming stars. The bulk of observed molecular cloud material is
inactive, in a more tenous state between individual star forming
regions.

Except on the scales of isolated protostellar cores, the observed line
widths are always wider than implied by the excitation temperature of
the molecules. This is interpreted as the result of bulk motion
associated with turbulence. We will argue in this review that it is
this interstellar turbulence that determines the lifetime and fate of
molecular clouds, and so their ability to collapse and form stars.

Magnetic fields have long been discussed as a stabilizing agent in
molecular clouds. However, magnetic fields with average field strength
of 10$\,\mu$G (Verschuur 1995a,b; Troland \etal\ 1996, Crutcher 1999)
cannot stabilize molecular clouds as a whole. This is particularly
true on scales of individual protostars, where magnetic fields appear
too weak to impede gravitational collapse in essentially all cases
observed (see \S~\ref{sub:standardprobs}). Furthermore, magnetic
fields cannot prevent turbulent velocity fields from decaying quickly
(see the discussion in \S~\ref{sub:motions}).

Molecular clouds appear to be transient features of the turbulent flow
of the interstellar medium (Ballesteros-Paredes \etal\ 1999a).  Just
as Lyman-$\alpha$ clouds in the intergalactic medium were shown to be
transient objects formed in the larger scale cosmological flow (Cen et
al.\ 1994, Zhang, Anninos, \& Norman 1995) rather than stable objects
in gravitational equilibrium (Rees 1986, Ikeuchi 1986), molecular
clouds may never reach an equilibrium configuration.  The high
pressures seen in molecular clouds can be produced by ram pressure
from converging supersonic flows in the ISM (see \S~\ref{sub:clouds}).
So long as the flow persists, it confines the cloud, and supplies
turbulent energy.  When the flow ends, the cloud begins to expand at
its sound speed, eventually dissipating into the ISM
(V\'azquez-Semadeni, Shadmehri, \& Ballesteros-Paredes 2002).  Further
shocks may help this process along.

\subsection{Scaling relations for molecular clouds}
\label{sub:scaling-law}

Observations of molecular clouds exhibit correlations between various
properties, such as clump size, velocity dispersion, density and
mass. Larson (1981) first noted, using data from several different
molecular cloud surveys, that the density $\rho$ and the velocity
dispersion $\sigma$ appear to scale with the cloud size $R$ as
\begin{eqnarray}
\rho &\propto& R^{\alpha}\;\label{eqn:larson-a}\\
\sigma &\propto& R^{\beta}\;,\label{eqn:larson-b}
\end{eqnarray}
with $\alpha$ and $\beta$ being constant scaling exponents.  Many
studies have been done of the scaling properties of molecular
clouds. The most commonly quoted values of the exponents are $\alpha
\approx -1.15 \pm 0.15$ and $\beta \approx 0.4 \pm 0.1$ (e.g.\ Dame
\etal\ 1986, Myers \& Goodman 1988, Falgarone \etal\ 1992, Fuller \&
Myers 1992, Wood, Myers, \& Daugherty 1994, Caselli \& Myers 1995).
However, the validity of these scaling relations is the subject of
strong controversy and significantly discrepant values have been
reported by Carr (1987) and Loren (1989), for example.

The above standard values are often interpreted in terms of the virial
theorem (Larson 1981, Caselli \& Myers 1995).  If one assumes virial
equilibrium, Larson's relations (Eq.'s \ref{eqn:larson-a} and
\ref{eqn:larson-b}) are not independent. For $\alpha = -1$, which
implies constant column density, a value of $\beta = 0.5$ suggests
equipartition between self-gravity and the turbulent velocity
dispersion, such that the ratio between kinetic and potential
energy is constant with $E_{\rm kin}/|E_{\rm pot}| =
\sigma^2R/(2GM)\approx 1/2$.
Note, that for any arbitrarily chosen value of the
density scaling exponent $\alpha$, a corresponding value of $\beta$
obeying equipartition can always be found (V{\'a}zquez-Semadeni \&
Gazol 1995).  Equipartition is usually interpreted as indicating
virial equilibrium in a static object.  However, Ballesteros-Paredes
\etal\ (1999b) pointed out that in a dynamic, turbulent environment,
the other terms of the virial equation (McKee \& Zweibel 1992) can
have values as large as or larger than the internal kinetic and
potential energy.  In particular, the changing shape of the cloud will
change its moment of inertia, and turbulent flows will produce large
fluxes of kinetic energy through the surface of the cloud.  As a
result, 
equipartition between internal kinetic and potential
energy does not necessarily imply virial equilibrium.

Kegel (1989) and Scalo (1990) proposed that the density-size relation
may be a mere artifact of the limited dynamic range in the
observations, rather than reflecting a real property of interstellar
clouds.  In particular, in the case of molecular line data, the
observations are restricted to column densities large enough for the
tracer molecule to be shielded against photodissociating UV radiation,
but small enough for the lines to remain optically thin.  With limited
integration times, most CO surveys tend to select objects in an even
smaller range of column densities, giving roughly constant column
density, which automatically implies $\rho \propto R^{-1}$.  Surveys
with longer integration times, and therefore larger dynamic ranges,
seem to exhibit an increasingly large scatter in density-size plots,
as seen, for example, in the data of Falgarone \etal\ (1992).  Results
from numerical simulations, which are free from observational bias,
indicate the same trend (V{\'a}zquez-Semadeni, Ballesteros-Paredes, \&
Rodriguez 1997).  Three-dimensional simulations of supersonic
turbulence (Mac Low 1999) were used by Ballesteros-Paredes \& Mac Low
(2002) to perform a comparison of clumps measured in physical space to
clumps observed in position-position-velocity space.  They found no
relation between density and size in physical space, but a clear trend
of $\rho \propto R^{-1}$ in the simulated observations, caused simply
by the tendency of clump-finding algorithms to pick out clumps with
column densities close to the local peak values.  Also, for clumps
within molecular clouds, the structures identified in CO often do not
correspond to those derived from higher-density tracers (see e.g.\
Langer \etal\ 1995, Bergin \etal\ 1997, Motte \etal\ 1998 for
observational discussion, and Ballesteros-Paredes \& Mac Low 2002 for
theoretical discussion).  In summary, the existence of a physical
density-size relation appears doubtful.

The velocity-size relation appears less prone to observational
artifacts.  Although some measurements of molecular clouds do not seem
to exhibit this correlation (e.g.\ Loren 1989, or Plume \etal\ 1997),
it does appear to be a real property of the cloud.  It is often
explained using the standard (though incomplete) argument of virial
equilibrium.  In supersonic turbulent flows, however, the scaling
relation is a natural consequence of the characteristic energy
spectrum $E(k) \propto k^{-2}$ in an ensemble of shocks, even in the
complete absence of self-gravity (Ossenkopf \& Mac Low 2002,
Ballesteros-Paredes \& Mac~Low 2002, Boldyrev, Nordlund, \& Padoan
2002).  Larger scales carry more energy, leading to a relation between
velocity dispersion and size that empirically reproduces the observed
relation.  Thus, although the velocity-size relation probably does
exist, its presence does not argue for virial equilibrium, or even
energy equipartition, but rather for the presence of a supersonic
turbulent cascade.

\subsection{Protostellar cores}
\label{sub:cores}

\subsubsection{From cores to stars}
\label{subsub:4-phases}
Protostellar cores are the direct precursors of stars.  The
transformation of cloud cores into stars can be conveniently
subdivided into four observationally motivated phases (e.g.\ Shu
\etal\ 1987, Andr{\'e} \etal\ 2000).

(a) The {\em prestellar phase} describes the isothermal gravitational
contraction of molecular cloud cores before the formation of the
central protostar. Prestellar cores are cold and are best observed in
molecular lines or dust emission.  The isothermal collapse phase ends
when the inner parts reach densities of $n({\rm H}_2) \approx
10^{10}\,$cm$^{-3}$. Then the gas and dust become optically thick, so
the heat generated by the collapse can no longer freely radiate away
(e.g.\ Tohline 1982). The central region begins to heat up, and
contraction pauses. As the temperature increases to
$T\approx 2000\,$K molecular hydrogen begins to dissociate, absorbing
energy. The core becomes unstable again and collapse sets in anew.
Most of the released gravitational energy goes into the dissociation
of ${\rm H}_2$ so that the temperature rises only slowly.
This situation is similar to the first isothermal collapse phase. When
all molecules in the core are dissociated, the temperature rises
sharply and pressure gradients again halt the collapse.  This second
hydrostatic object is the true protostar. 

(b) The cloud core then enters the {\em class 0 phase} of evolution,
in which the central protostar grows in mass by the accretion of
infalling material from the outer parts of the original cloud core.
Higher angular momentum material first falls onto a disk and then gets
transported inwards by viscous processes. In this phase star and disk
are deeply embedded in an envelope of gas and dust. The mass of the
envelope $M_{\rm env}$ greatly exceeds the total mass $M_{\star}$ of
star and disk together.  The main contribution to the total luminosity
is accretion, and the system is best observed at sub-millimeter and
infrared wavelengths.

(c) At later times, powerful protostellar outflows develop that clear
out the envelope along the rotational axis.  This is the {\em class I
  phase} during which the system is observable in infrared and optical
wavebands, and for which $M_{\rm env}\ll M_{\star}$. In optical
light the central protostar is only visible when looking along
the outflow direction.

(d) In the {\em class II phase}, the envelope has disappeared, because
all available gas has either been accreted or dispersed by the
outflow. The protostar no longer accretes, and it enters the classical
pre-main sequence contraction phase.  It still is surrounded by a
tenous disk of gas and dust with a mass of order 10$^{-3}$ that of the
star.  The disk adds an infrared excess to the spectral energy
distribution of the system, which is dominated by the stellar Planck
spectrum at visible wavelengths (e.g.\ Beckwith 1999).  This is the
stage during which planets are believed to form (e.g.\ Lissauer 1993,
Ruden 1999). Protostellar systems in this stage are commonly called T
Tauri stars (Bertout 1989). As time evolves further the disk becomes
more and more depleted until only a tenuous dusty debris disk remains
that is long-lived and can last (i.e.\ continuously reform from
collisions of planetesimals) into and throughout the stellar
main-sequence phase (Zuckerman 2001).

Detailed calculations of all phases of dynamical collapse 
assuming spherical symmetry 
are presented by Masunaga, Miyama, \&
Inutsuka (1998), Masunaga \& Inutsuka (2000a,b), Wuchterl \& Klessen
(2001), and Wuchterl \& Tscharnuter (2003).

\begin{figure}[ht]
\begin{center}
\unitlength1.0cm
\begin{turn}{180}
\begin{picture}(16,11.7)
\put( 0.0,0.5){\epsfysize=10.5cm \epsfbox{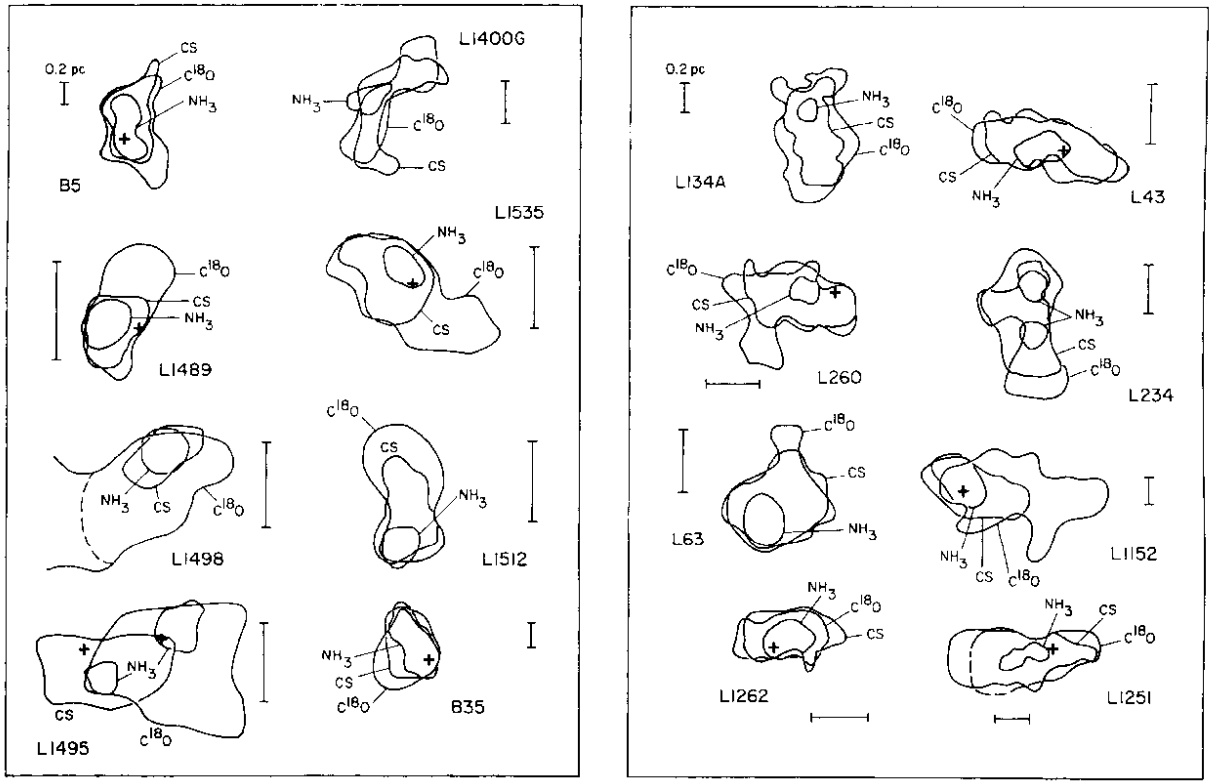}}
\end{picture}
\end{turn}
\end{center}
\caption{\label{fig:core-shapes}Intensity contours at half maximum of
    16 dense cores in dark clouds, in the $1.3\,$cm $(J,k) = (1,1)$
    lines of NH$_3$, in the $3.0\,$mm $J=2\rightarrow 1$ line of CS,
    and the $2.7\,$mm $J=1\rightarrow 0$ line of C$^{18}$O. A linear
    scale of $0.2\,$pc is indicated in each individual map and
    associated protostars are specified by a cross. The figure is from
    Myers \etal\ (1991).}
\end{figure}

\subsubsection{Properties of protostellar cores}
\label{subsub:core-obs}

A number of small, dense molecular cores have been identified by low
angular resolution, molecular line surveys of nearby dark clouds
(e.g.\ Benson \& Myers 1989, Myers \etal\ 1991), as illustrated in
Figure \ref{fig:core-shapes}.  About half of them are associated with
protostars, i.e.\ they are in the class 0 or I phase of evolution as
inferred from the presence of low-luminosity IRAS sources and CO
outflows, while the other half are observed to still be in their
prestellar phase (e.g.\ Beichman \etal\ 1986, Andr{\'e} \etal\ 2000).
One of the most notable properties of the sampled cores are their very
narrow line widths. These are very close to the line widths expected
for thermal broadening alone and, as a result, many of the cores
appear approximately gravitationally virialized (e.g.\ Myers 1983).
They are thought either to be in the very early stage of gravitational
collapse or to have subsonic turbulence supporting the clump. A
comparison of the line widths of cores with embedded protostellar
objects (i.e.~with associated IRAS sources) and the starless cores
reveals a substantial difference.  Typically, cores with infrared
sources exhibit broader lines, which suggests the presence of a
considerable turbulent component not present in starless cores. This
may be caused by the central protostar feeding back energy and
momentum into its surrounding envelope.  Molecular outflows associated
with many of the sources may be a direct indication of this process.

\begin{figure}
\begin{center}\unitlength1cm
\begin{picture}( 8.00, 7.50)
\put( 0.00, 0.50){\epsfxsize=9cm \epsfbox{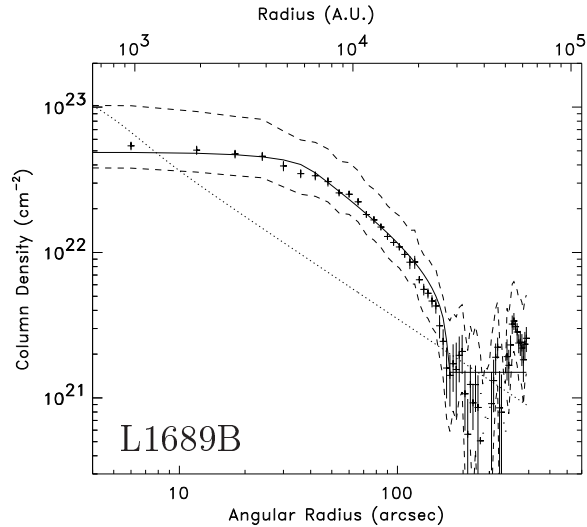}}
\end{picture}
\end{center}%
\caption{\label{fig:L1689B} Radial column density profile of the
prestellar core L1689B derived from combined infrared absorption and
$1.3\,$mm continuum emission maps. Crosses show the observed values
with the corresponding statistical errors, while the total
uncertainties in the method are indicated by the dashed lines. For
comparison, the solid line denotes the best-fitting Bonnor-Ebert
sphere and the dotted line the column density profile of a singular
isothermal sphere. The observed profile is well reproduced by an
unstable Bonnor-Ebert sphere with a density contrast of $\sim 50$, see
Bacmann \etal\ 2000 for a further details.}
\end{figure}
The advent of a new generation of infrared detectors and powerful
receivers in the radio and sub-millimeter wavebands in the late 1990's
made it possible to determine the radial column density profiles of
prestellar cores with high sensitivity and resolution (e.g.\
Ward-Thompson \etal\ 1994, Andr{\'e} \etal\ 1996, Motte, Andr{\'e},
and Neri 1998, Ward-Thompson \etal\ 1999, Bacmann \etal\ 2000, Motte
\& Andr{\'e} 2001). These studies show that starless cores typically
have flat inner density profiles out to radii of a few hundredths of a
parsec, followed by a radial decline of roughly $\rho\propto 1/r^2$
and possibly a sharp outer edge at radii 0.05--0.3~pc (e.g.\ Andr{\'e}
\etal\ 2000). This is illustrated in Figure \ref{fig:L1689B} which
shows the observed column density of the starless core L1689B derived
from combining mid-infrared absorption maps with 1.3 mm dust continuum
emission maps (from Bacmann \etal\ 2000). Similar profiles have been
derived independently from dust extinction studies (Lada \etal\ 1994,
Alves \etal\ 2001).  Protostellar cores often are elongated or
cometary shaped and appear to be parts of filamentary structures that
connect several objects.

The various theoretical approaches to explain the observed core
properties are discussed and compared in Section \ref{sub:core-models}.

\subsection{The observed IMF} 
\label{sub:IMF-observed}
Hydrogen-burning stars can only exist in a finite mass range
\begin{equation}
\label{eqn:mass-range}
0.08 \sil m \sil 100\:,
\end{equation}
where the dimensionless mass $m\equiv M/(1\mbox{ M}_{\odot})$.
Objects with $m \sil 0.08$ do not have central temperatures and
pressures high enough for hydrogen fusion to occur. If they are larger
than about ten times the mass of Jupiter, $m > 0.01$, they are called
brown dwarfs, or more generally substellar objects (e.g.\ Burrows
\etal\ 1993, Laughlin \& Bodenheimer 1993; or for a review Burrows
\etal\ 2001). Stars with $m > 100$, on the other hand, blow themselves
apart by radiation pressure (e.g.\ Phillips 1994).

It is complicated and laborious to estimate the IMF in our Galaxy empirically.
The first such determination from the solar neighborhood (Salpeter 1955)
showed that the number $\xi(m)dm$ of stars with masses in the range $m$ to $m +
dm$ can be approximated by a power-law relation
\begin{equation}
  \label{eqn:salpeter}
  \xi(m)dm \propto m^{-\alpha}dm\;,
\end{equation}
with index $\alpha \approx 2.35$ for stars in the mass range $0.4
\le m \le 10$. However, approximation of the IMF with a single
power-law is too simple.  Miller \& Scalo (1979) introduced a
log-normal functional form, again to describe the IMF for Galactic
field stars in the vicinity of the Sun,
\begin{equation}
  \label{eqn:miller-scalo}
  \log_{10}\xi(\log_{10}\,\!m) = A -
  \frac{1}{2(\log_{10}\sigma)^2} \left[ \log_{10}\left(\frac{m}{m_0}\right)\right]^2\;.
\end{equation}
This analysis has been repeated and improved by Kroupa, Tout, \&
Gilmore (1990), who derive values 
\begin{eqnarray}
  \label{eqn:IMF-kroupa}
   m_0 &=& 0.23, \nonumber \\ 
   \sigma &=& 0.42,\\
   A&=&0.1.\nonumber  
\end{eqnarray}

\begin{figure}[htp]
\begin{center}
\unitlength1.0cm
\begin{picture}(8,10)
\put( 0.00,0.0){\includegraphics[width=0.4\textwidth]{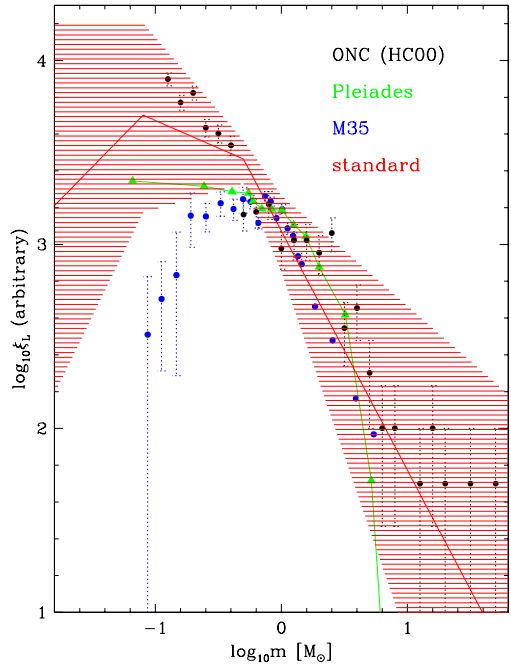}}
\end{picture}
\end{center}
\caption{\label{fig:Kroupa-2002-4a} The measured stellar mass
  function $\xi$ as function of logarithmic mass $\log_{10} m$ in the
  Orion nebular cluster 
  (upper circles), the Pleiades 
  (triangles connected by line), and the cluster M35 
  (lower circles). None of the mass functions is corrected for unresolved
  multiple stellar systems. The average initial stellar mass function
  derived from Galactic field stars in the solar neighborhood is shown
  as 
  a line with the associated uncertainty range indicated by the
  hatched area. (From Kroupa 2002.)}
\end{figure}
The IMF can also be estimated, probably more directly, by studying
individual young star clusters. Typical examples are given in
Figure~\ref{fig:Kroupa-2002-4a} (taken from Kroupa 2002), which
plots the mass function derived from star counts in the Trapezium
Cluster in Orion (Hillenbrand \& Carpenter 2000), in the Pleiades
(Hambly \etal\ 1999) and in the cluster M35 (Barrado~y Navascu\'es
\etal\ 2001). 

The most popular approach to approximating the IMF empirically is to
use a multiple-component power-law of the form 
equation~(\ref{eqn:salpeter})
with the following parameters (Scalo 1998, Kroupa 2002):
\begin{equation}
\label{eqn:3power-law}
\xi(m) =  \left \{
\begin{array}{ll} 0.26\,m^{-0.3}& \;\;\;\;{\rm for } \;\; 0.01\le m<0.08 \;,\nonumber\\
0.035\,m^{-1.3}& \;\;\;\;{\rm for } \;\; 0.08\le m<0.5\;, \\
0.019\,m^{-2.3}& \;\;\;\;{\rm for } \;\; 0.5\le m<\infty\;.\nonumber\\
\end{array}\right.
\end{equation} 

This representation of the IMF is statistically corrected for binary
and multiple stellar systems too close to be resolved, but too far
apart to detect spectroscopically.  Neglecting these systems
overestimates the masses of stars, as well as reducing inferred
stellar densities.  These mass overestimates influence the derived
stellar mass distribution, underestimating the number of low-mass
stars.  The IMF may steepen further towards high stellar masses and a
fourth component could be defined with $\xi(m)=0.019\,m^{-2.7}$ for
$m>1.0$ thus arriving at the IMF proposed by Kroupa, Tout, \& Gilmore
(1993).  In equation~(\ref{eqn:3power-law}), the exponents for masses
$m < 0.5$ are very uncertain due to the difficulty of detecting and
determining the masses of very young low-mass stars.  The exponent for
$0.08\le m<0.5$ could vary between $-0.7$ and $-1.8$, and the value in
the substellar regime is even less certain.

\begin{figure}[ht]
\begin{center}
 \unitlength1.0cm
 \begin{picture}(8,7.3)
 \put( 0.00,-2.0){\includegraphics[width=0.4\textwidth]{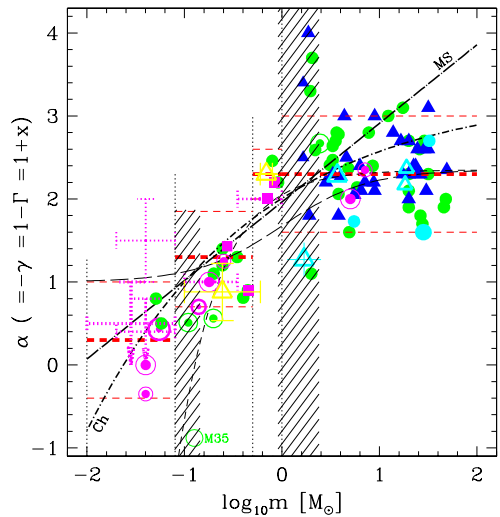}}
 \end{picture}
\end{center}
 \caption{\label{fig:Kroupa-2002-5a} A plot of power-law exponents
   determined for various stellar clusters in the mass range $-2 <
   \log_{10} m < 2$, to illustrate the observed scatter.  The solid
   dots and triangles are from measurements of 
   OB associations and clusters in the Milky Way and the Large
   Magellanic Cloud,
   respectively. Globular cluster data are indicated by open 
   triangles. None of these measurements is corrected for unresolved
   binaries.  The mean values of the exponent $\alpha$ derived in the
   solar neighborhood, equation \ref{eqn:3power-law}, and the
   associated uncertainties are indicated by horizontal 
   dashed lines. Note that for low stellar masses the values of
   $\alpha$ determined from observations in young stellar clusters lie
   systematically lower due to the inability to resolve close binaries
   and multiple stellar systems. 
   Other
   lines indicate alternative functional forms for the IMF: $MS$
   gives the Miller-Scalo (1979) IMF; and $Ch$ the one sugestied by
   Chabrier (2001, 2002).  For a more detailed discussion see Kroupa
   (2002).  } 
\end{figure}

There are some indications that the slope of the mass spectrum
obtained from field stars may be slightly shallower than the one
obtained from observing stellar clusters (Scalo 1998).  The reason for
this difference is unknown.  It is somehow surprising given the fact
most field stars appear to come from dissolved clusters (Adams \& Myers
2001). It is possible that the field star IMF is inaccurate because of
incorrect assumptions about past star formation rates and age
dependences for the stellar scale height. Both issues are either known
or irrelevant for the IMF derived from cluster surveys. On the other
hand, the cluster surveys could have failed to include low-mass stars
due to extinction or crowding.
It has also been claimed that the IMF may vary between different
stellar clusters (Scalo 1998), as the measured exponent $\alpha$ in
each mass interval exhibits considerable scatter when comparing
different star forming regions. This is illustrated in Figure
\ref{fig:Kroupa-2002-5a}, which is again taken from Kroupa
(2002). This scatter, however, may be entirely due to effects related to the
dynamical evolution of stellar clusters (Kroupa 2001). 

Despite these differences in detail, all IMF determinations share the
same basic features, and it appears reasonable to say that the basic
shape of the IMF is a universal property common to all star forming
regions in the present-day Galaxy, perhaps with some intrinsic
scatter. There still may be some dependency on the metallicity of the
star forming gas, but changes in the IMF do not seem to be gross even
in that case. There is no compelling evidence for qualitatively
different behavior such as truncation at the low or high-mass end.

\section{HISTORICAL DEVELOPMENT}
\label{sec:history}
 
Stars form from gravitational contraction of gas and dust in molecular
clouds.  A first estimate of the stability of such a system against
gravitational collapse can be made by simply considering its energy
balance. For instability to occur, gravitational attraction must
overcome the combined action of all dispersive or resistive forces. In
the simplest case, the absolute value of the potential energy of a
system in virial equilibrium is exactly twice the total kinetic
energy, $E_{\rm pot} + 2\,E_{\rm kin} = 0$. If $E_{\rm pot} +
2\,E_{\rm kin}<0$ the system collapses, while for $E_{\rm pot} +
2\,E_{\rm kin}>0$ it expands. This estimate can easily be extended by
including the surface terms and additional physical forces (see further
discussion in \S~\ref{sub:scaling-law}). In particular taking magnetic
fields into account may become important for describing
interstellar clouds (Chandrasekhar, 1953; see also McKee
\etal, 1993, for a more recent discussion). In the presence of
turbulence, the total kinetic energy includes not only the internal
energy but also the contribution from turbulent gas motions. General
energy considerations can provide qualitative insight into the
dynamical behavior of a system (Bonazzola \etal\ 1987,
Ballesteros-Paredes, \etal\ 1999b).

A thorough investigation, however, requires a linear stability
analysis.  For the case of a non-magnetic, isothermal, infinite,
homogeneous, self-gravitating medium at rest (i.e.\ without turbulent
motions) Jeans (1902) derived a relation between the oscillation
frequency $\omega$ and the wavenumber $k$ of small perturbations,
\begin{equation}
\omega^2 - c_{\rm s}^2 k^2 + 4\pi G\,\rho_0 = 0\;,
\label{eqn:jeans-dispersion-rel}
\end{equation}
where $c_{\rm s}$ is the isothermal sound speed, $G$ the gravitational
constant, and $\rho_0$ the initial mass density. The derivation
neglects viscous effects and assumes that the
linearized version of the Poisson equation describes only the relation
between the perturbed potential and the perturbed density (neglecting
the potential of the homogeneous solution, the so-called `Jeans
swindle', see e.g.\ Binney and Tremaine, 1997).  The third term
in Eq.\ (\ref{eqn:jeans-dispersion-rel}) is responsible for the
existence of decaying and growing modes, as pure sound waves stem from
the dispersion relation $\omega^2 - c_{\rm s}^2 k^2 =0$. Perturbations
are unstable against gravitational contraction if their wavenumber is
below a critical value, the Jeans wavenumber $k_{\rm J}$, i.e.~if
\begin{equation}
k^2 < k_{\rm J}^2 \equiv \frac{4 \pi G \rho_0}{c_{\rm s}^2}\;, 
\label{eqn:jeans-wave-number}
\end{equation}
or equivalently, if the wavelength of the perturbation exceeds a
critical size given by $\lambda_{\rm J} \equiv 2 \pi k_{\rm J}^{-1}$.
Assuming the perturbation is spherical with diameter  $\lambda_{\rm
J}$, this directly translates into a mass limit
\begin{equation}
M_{\rm J} \equiv  \frac{4\pi}{3}\rho_0 \left(\frac{\lambda_{\rm J}}{2}\right)^3 =  \frac{\pi}{6}\left( \frac{\pi}{G}
\right)^{3/2} \rho_0^{-1/2} {c_{\rm s}^3}.
\label{eqn:jeans-mass}
\end{equation}
All perturbations exceeding the Jeans mass $M_{\rm J}$ will collapse
under their own weight.  For isothermal gas $c_{\rm s}^2 \propto T$,
so $M_{\rm J}\propto \rho_0^{-1/2} T^{3/2}$. The critical mass $M_{\rm
J}$ decreases when the density $\rho_0$ grows or when the temperature
$T$ sinks.

The Jeans instability has a simple physical interpretation in terms of
the energy budget. The energy density of a sound wave is
positive. However, its gravitational energy is negative, because the
enhanced attraction in the compressed regions outweighs the reduced
attraction in the dilated regions. The instability sets in at the
wavelength $\lambda_{\rm J}$ where the net energy density becomes
negative. The perturbation grows, allowing the energy to decrease
further. For a fundamental derivation of this instability from the
canonical ensemble in statistical physics, see Semelin, S\'anchez, \&
de~Vega (2001), and de~Vega \& S{\'a}nchez (2002a,b). In isothermal
gas, there is no mechanism that prevents complete collapse. In
reality, however, during the collapse of molecular gas clumps, the
opacity increases and at densities of $n({\rm H}_2) \approx
10^{10}\:$cm$^{-3}$ the equation of state becomes adiabatic rather
than isothermal. Then collapse proceeds slower. Finally at very high
central densities ($\rho \approx 1\:$g$\,$cm$^{-3}$) fusion sets
in. This energy source leads to a new equilibrium (e.g.\ Tohline
1982): a new star is born.

Attempts to include the effect of turbulent motions into the star
formation process were already being made in the middle of the
twentieth century by von Weizs\"acker (1943, 1951) based on
Heisenberg's (1948a,b) concept of turbulence. He also considered the
production of interstellar clouds from the shocks and density
fluctuations in compressible turbulence. A more quantitative theory
was proposed by Chandrasekhar (1951a,b), who
investigated the effect of microturbulence in the subsonic regime. In
this approach the scales of interest, e.g.\ for gravitational collapse,
greatly exceed the outer scale of the turbulence.  If turbulence is
isotropic (and more or less incompressible), it contributes to
the pressure on large scales, and Chandrasekhar derived a dispersion
relation similar to Eq.\ (\ref{eqn:jeans-dispersion-rel}) except for the
introduction of an effective sound speed 
\begin{equation}
c_{{\rm s},e\!\!\!\;f\!\!f}^2 = c_{\rm s}^2 + 1/3 \, \langle v^2 \rangle\,,
\label{eqn:eff-sound-speed}
\end{equation}
where $\langle v^2 \rangle$ is the rms velocity dispersion due to
turbulent motions.  

Sasao (1973) noted that Chandrasekhar's derivation neglected the
effect of the inertia of the turbulent flow in forming density
enhancements while focussing only on the effective turbulent pressure.
The developments through the mid-eighties are reviewed by Scalo
(1985). 
Both Sasao (1973) and Chandrasekhar (1951a,b) made the
microturbulent assumption that the outer scale of the turbulence is
smaller than that of the turbulent clouds.  However, 
the outer scales of molecular cloud turbulence typically exceed or are
at least comparable to the size of the system (e.g.\ Ossenkopf and
Mac~Low, 2001), so the assumption of microturbulence is invalid.
Bonazzola \etal\ (1987) therefore suggested a wavelength-dependent
effective sound speed $c_{{\rm s},e\!\!\!\;f\!\!f}^2(k) = c_{\rm s}^2
+ 1/3 \, v^2(k)$ for Eq.\ (\ref{eqn:jeans-dispersion-rel}).  In this
description, the stability of the system depends not only on the total
amount of energy, but also on the wavelength distribution of the
energy, since $v^2(k)$ depends on the turbulent power spectrum.  A
similar approach was also adopted by V{\'a}zquez-Semadeni and Gazol
(1995), who added Larson's (1981) empirical scaling relations to the
analysis.

An elaborate investigation of the stability of turbulent,
self-gravitating gas was undertaken by Bonazzola \etal\ (1992), who used
renormalization group theory to derive a dispersion relation with a
generalized, wavenumber-dependent, effective sound speed and an
effective kinetic viscosity that together account for turbulence at
all wavelengths shorter than the one in question.  According to their
analysis, turbulence with a power spectrum steeper than $P(k)\propto
1/k^3$ can support a region against collapse at large scales, and
below the thermal Jeans scale, but not in between.  On the other hand,
they claim that turbulence with a shallower slope, as is expected for
incompressible turbulence (Kolmogorov 1941a,b), Burgers turbulence
(Lesieur 1997, p. 238), or shock dominated flows (Passot, Pouquet
\& Woodward 1988), cannot support clouds against collapse at scales
larger than the thermal Jeans wavelength.
It may even be possible to describe the equilibrium state of
self-gravitating gas as an inherently inhomogeneous thermodynamic
critical point (de~Vega, S{\'a}nchez and Combes, 1996a,b; de~Vega and
S{\'a}nchez, 2000).  This may render all applications of incompressible
turbulence to the theory of star formation meaningless. In fact, it is
the main goal of this review to introduce and stress the importance of
compressional effects in supersonic turbulence for determining the
outcome of star formation.

In order to do that, we need to recapitulate the development of our
understanding of the star formation process over the last few decades.
We begin with the classical dynamical theory (\S~\ref{sub:classical})
and describe the problems that it encounters in its original form
(\S~\ref{sub:classicalprobs}). In particular the timescale problem lead
astrophysicists to think about the influence of magnetic fields. This
line of reasoning resulted in the construction of the paradigm of
magnetically mediated star formation, which we discuss in
\S~\ref{sub:standard}.  However, it became clear that this so-called
``standard theory'' has a variety of serious shortcomings
(\S~\ref{sub:standardprobs}). These lead to the rejuvenation of the
earlier dynamical concepts of star formation and their reconsideration in
the modern framework of compressible supersonic turbulence which we
discuss in \S~\ref{sec:paradigm}.

\subsection{Classical dynamical theory}
\label{sub:classical}

The classical dynamical theory focuses on the
interplay between self-gravity on the one side and pressure gradients
on the other. Turbulence is taken into account, but only on
microscopic scales significantly smaller than the collapse scales. In
this microturbulent regime, random gas motions yield an isotropic
pressure that can be absorbed into the equations of
motion by defining an effective sound speed as in Eq.\
(\ref{eqn:eff-sound-speed}). The dynamical behavior of the system
remains unchanged, and we do not distinguish between the
effective and thermal sound speed $c_{\rm s}$ in this and the
following two sections.

Because of the importance of gravitational instability for stellar
birth, Jeans' (1902) pioneering work has triggered numerous attempts
to derive solutions to the collapse problem, both analytically and
numerically.  Particularly noteworthy are the studies by Bonnor (1956)
and Ebert (1957), who independently derived analytical solutions for
the equilibrium structure of spherical density perturbations in
self-gravitating, isothermal, ideal gases, as well as a criterium for
gravitational collapse. See Lombardi and Bertin (2001) for a recent
analysis, and studies by Schmitz (1983, 1984, 1986, 1988) and Schmitz
\& Ebert (1986, 1987) for the treatment of rotation and generalized,
polytropic equations of state. It has been argued recently that this
may be a good description for the density distribution in quiescent
molecular cloud cores just before they begin to collapse and form
stars (Bacmann \etal\ 2000, Alves, Lada, and Lada 2001). The first
numerical calculations of protostellar collapse became possible in the
late 1960's (e.g.\ Bodenheimer \& Sweigart, 1968; Larson, 1969;
Penston, 1969a,b). They showed that gravitational contraction proceeds
in a highly nonhomologous manner, contrary to what had previously been
assumed (Hayashi 1966). 

\begin{figure}
\unitlength1cm
\begin{picture}( 16.00, 9.00)
\put( 0.00, 0.50){\epsfxsize=16cm \epsfbox{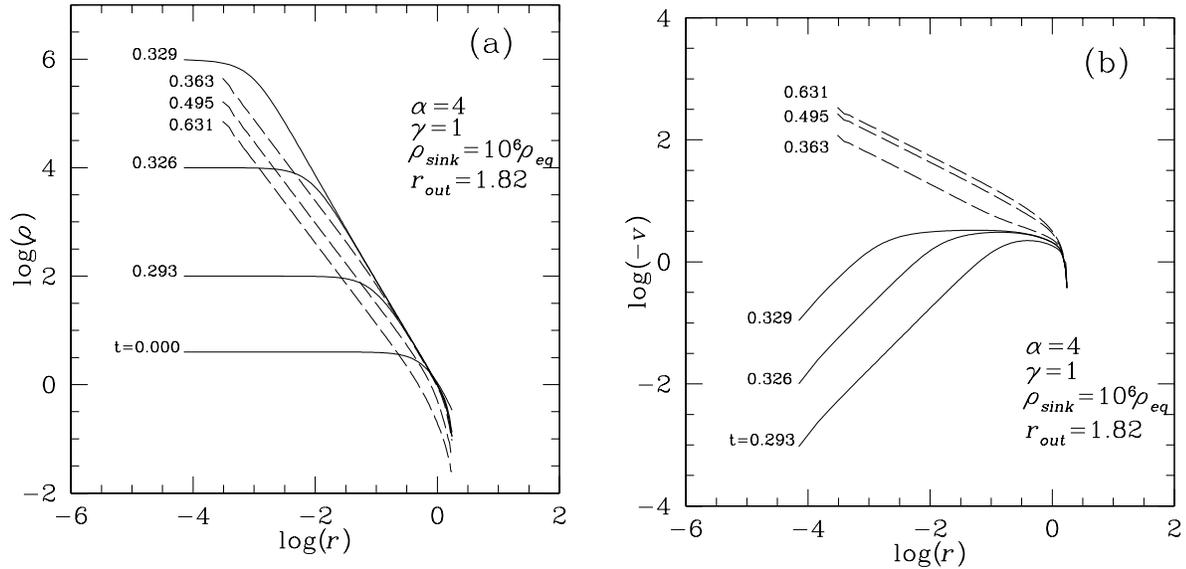}}
\end{picture}
\caption{\label{fig:larson-density-evolution} Radial density profile
(a) and infall velocity profile (b)  at various stages of
dynamical collapse. All quantities are given in normalized units. The
initial configuration at $t=0$ corresponds to a critical isothermal
($\gamma=1$) Bonnor-Ebert sphere with outer radius $r_{\rm out} =
1.82$. It has $\alpha=4$ times more mass than allowed by
hydrostatic equilibrium, and therefore begins to contract. The numbers
on the left denote the evolutionary time and illustrate the `runaway'
nature of collapse. Since the relevant collapse timescale, the
free-fall time $\tau_{\rm ff}$, scales with density as $\tau_{\rm
ff}\propto \rho^{-1/2}$ central collapse speeds up as $\rho$
increases. When density contrast reaches a value of $10^6$ a ``sink''
cell is created in the center, which subsequently accretes all
incoming matter.  This time roughly corresponds to the formation of
the central protostar, and allows for following its subsequent
accretion behavior. The profiles before the formation of the central
point mass indicated by solid lines, and for later times by dashed
lines. The figure is from Ogino \etal\ (1999).  }
\end{figure}
This is illustrated in Figure \ref{fig:larson-density-evolution},
which shows the radial density distribution of a protostellar core at
various stages of the isothermal collapse phase. The gas sphere
initially follows a Bonnor-Ebert critical density profile but has four
times more mass than allowed in an equilibrium state. Therefore it is
gravitationally unstable and begins to collapse. As the inner part has
no pressure support, it falls freely. As matter moves inwards, the
density in the interior grows, while the density decreases in the
outer parts. This builds up pressure gradients in the outer parts,
where contraction is retarded from free fall. In the interior,
however, the collapse remains in approximate free fall. Thus, it
actually speeds up, because the free-fall timescale
scales with density as $\tau_{\rm ff}\propto \rho^{-1/2}$. Changes in
the density structure occur in a smaller and smaller region near the
center and on shorter and shorter timescales, while practically nothing
happens in the outer parts.  As a result the overall matter
distribution becomes strongly centrally peaked,
approaching $\rho \propto r^{-2}$. This is the density profile
of an isothermal sphere. The establishment of a central singularity
corresponds to the formation of a protostar that grows in mass by
accreting the remaining envelope until the reservoir of gas is
exhausted.

It was Larson (1969) who realized that the dynamical evolution in the
initial isothermal collapse phase can be described by an analytical
similarity solution. This was independently discovered also by Penston
(1969b), and later extended by Hunter (1977) into the regime after the
protostar has formed. This so called Larson-Penston solution describes
the isothermal collapse of homogeneous ideal gas spheres initially at
rest. Its properties are summarized in Table \ref{tab:larson-penston}.  This solution makes
two important predictions. The first is the occurrence of supersonic
infall velocities that extend over the entire collapsing core. Before
the formation of the central protostar, the infall velocity tends
towards $-3.3 c_{\rm s}$, while afterwards it approaches free fall
collapse in the center with $v\propto r^{-1/2}$ ,while still
maintaining $v\approx -3.3 c_{\rm s}$ in the outer envelope (Hunter
1977). Second, the Larson-Penston solution predicts constant
protostellar accretion rates $\dot{M} \approx 30 c^3_{\rm s}/G$. 

\begin{figure}
\begin{center}\unitlength1cm
\begin{picture}( 8.00, 9.00)
\put( 0.00, 0.10){\epsfxsize=9.0cm \epsfbox{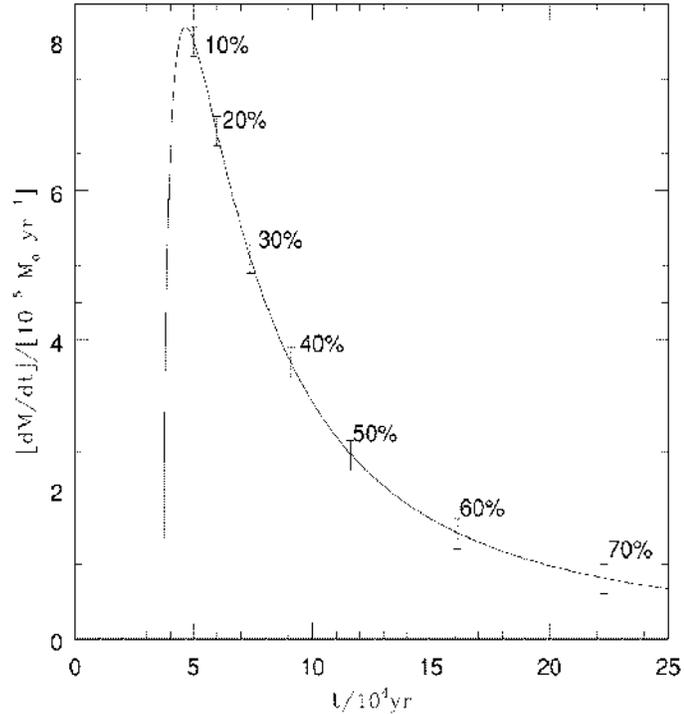}}
\end{picture}
\end{center}%
\caption{\label{fig:larson-accretion-rate}
  Time evolution of the protostellar mass accretion rate for the collapse
  of a gas clump with Plummer-type density distribution similar to
  observed protostellar cores. For details see Whitworth \&
  Ward-Thompson (2001) 
}
\end{figure}
In general, the dynamical models conceptually allow for time-varying
protostellar mass accretion rates, if the gradient of the density
profile of a collapsing cloud core varies with radius. In particular,
if the core has a density profile with a flat inner region and then a
decrease outwards, as is observed in low-mass cores (see
\S~\ref{sub:standardprobs}), then $\dot{M}$ has a high initial peak,
while the flat core is accreted, and later declines as the
lower-density outer-envelope material is falling in (e.g.\ Ogino
\etal\ 1999). 
The time evolution of $\dot{M}$ for the collapse of a sphere with a
generalized Plummer (1911) profile, with parameters fit to the
protostellar core L1544, is illustrated in Figure
\ref{fig:larson-accretion-rate} (see Whitworth \& Ward-Thompson 2001).
Plummer-type spheres have flat inner density profiles with radius
$R_0$ and density $\rho_0$ followed by an outer power-law decline,
\begin{equation}
\rho(r) = \rho_0 \left[\frac{R_0}{(R_0^2 + r^2)^{1/2}} \right]^{\eta},
\end{equation}
where $\eta =5$ is the classical Plummer sphere, while $\eta = 4$ is
adopted by Whitworth \& Ward-Thompson (2001) to reproduce observed
cloud cores.  Such a profile has the basic properties of a
Larson-Penston sphere in mid-collapse.

The dynamical properties of the Larson-Penston solution set it
clearly apart from the inside-out collapse model (Shu 1977) derived
for magnetically mediated star formation
(\S~\ref{sub:standard}). One-dimensional numerical simulations of the
dynamical collapse of homogeneous, isothermal spheres typically
demonstrate global convergence to the Larson-Penston solution, but
also show that certain deviations occur, e.g.\ in the time evolution
$\dot{M}$, due to pressure effects (Bodenheimer \& Sweigart 1968;
Larson 1969, Hunter 1977; Foster \& Chevalier 1993; Tomisaka 1996b;
Basu 1997; Hanawa \& Nakayama, 1997; Ogino \etal\ 1999).

The formation of clusters of stars (as opposed to binary or small multiple
stellar systems) is accounted for in the classical dynamical theory by
simply considering larger and more massive molecular cloud regions. The
proto-cluster cloud will fragment and build up a cluster of stars if it has
highly inhomogeneous density structure similar to the observed clouds (Keto,
Lattanzio, \& Monaghan 1991; Inutsuka \& Miyama 1997, Klessen \& Burkert 2000,
2001) or, equivalently, if it is subject to strong external perturbations,
e.g.\ from cloud-cloud collisions (Whitworth \etal\ 1995; Turner \etal\ 
1995), or is highly turbulent (see \S~\ref{sub:new}).

\subsection{Problems with classical theory}
\label{sub:classicalprobs}

The classical theory of gravitational collapse balanced by pressure
and microturbulence did not take into account the conservation of
angular momentum and magnetic flux during collapse.  It became clear
from observations of polarized starlight (Hiltner 1949, 1951) that
substantial magnetic fields thread the interstellar medium
(Chandrasekhar \& Fermi 1953a).  This forced the magnetic flux problem to
be addressed, but also raised the possibility that the solution to
the angular momentum problem might be found in the action of magnetic
fields.  The typical strength of the magnetic field in the diffuse ISM
was not known to an order of magnitude, though, with estimates ranging
as high as 30 $\mu$G from polarization (Chandrasekhar \& Fermi 1953a)
and synchrotron emission (e.g.\ Davies \& Shuter 1963).  Lower values
from Zeeman measurements of H{\sc i} (Troland \& Heiles 1986) and from
measurements of pulsar rotation and dispersion measures (Rand \&
Kulkarni 1989, Rand \& Lyne 1994) comparable to the modern value of
around 3~$\mu$G only gradually became accepted over the next two
decades.  Even now, measurements of synchrotron emission leave open
the possibility that there is a stronger disordered field in the Milky
Way, although their interpretation depends critically on the
assumption of equipartition between magnetic field energy and other
forms (Beck 2001).

The presence of a field, especially one much stronger than 3~$\mu$G,
formed a major problem for the classical theory of star formation.  To
see why, let us consider the behavior of a field in a region of
isothermal, gravitational collapse (Mestel \& Spitzer 1956, Spitzer
1968).  If we neglect all surface terms except thermal pressure $P_0$
(a questionable assumption as shown by Ballesteros-Paredes \etal\
1999b, but the usual one at the time), and assume that the field,
$\vec{B}$ is uniform, and passes through a spherical region of average
density $\rho$ and radius $R$, we can write the virial equation as
(Spitzer 1968)
\begin{equation} \label{virial}
4 \pi R^3 P_0 = 3 \frac{Mk_BT}{\mu} - \frac{1}{R} 
   \left(\frac35 GM^2 - \frac13 R^4 B^2\right),
\end{equation}
where $M = (4/3) \pi R^3 \rho$ is the mass of the region, $k_B$ is
Boltzmann's constant, $T$ is the temperature of the region, and $\mu$
is the mean mass per particle.  So long as the ionization is
sufficiently high for the field to be frozen to the matter, the flux
through the cloud $\Phi = \pi R^2 B$ must remain constant.  Therefore,
the opposition to collapse due to magnetic energy given by the last
term on the right hand side of Eq.~(\ref{virial}) will remain
constant during collapse.  If it cannot prevent collapse
at the beginning, it remains unable to do so as the field is compressed.

If we write the radius $R$ in terms of the mass and density of the
region, we can rewrite the two terms in parentheses on the right hand
side of Eq.~(\ref{virial}) to show that gravitational attraction
can only overwhelm magnetic repulsion if
\begin{equation} \label{eqn:crit-rho}
M > M_{\rm cr} \equiv \frac{5^{3/2}}{48\pi^2} \frac{B^3}{G^{3/2} \rho^2} = 
(4 \times 10^6 \mbox{M}_{\odot})\left(\frac{n}{1\,{\rm
cm}^{-3}}\right)^{-2} \left(\frac{B}{\mbox{3 $\mu$G}}\right)^3,
\end{equation}
where the numerical constant is correct for a uniform sphere, and the
number density $n$ is computed with mean mass per particle $\mu = 2.11
\times 10^{-24}$~g~cm$^{-3}$.  Mouschovias \& Spitzer (1976) noted
that the critical mass can also be written in terms of a critical
mass-to-flux ratio
\begin{equation} \label{eqn:crit-phi}
\left(\frac{M}{\Phi}\right)_{\rm cr} = \frac{\zeta}{3\pi}
\left(\frac{5}{G}\right)^{1/2} = 490 \:\mbox{g}\,{\rm G}^{-1}\,\mbox{cm}^{-2},
\end{equation}
where the constant $\zeta = 0.53$ for uniform spheres (or flattened
systems, as shown by Strittmatter 1966) is used in the final
equality. Assuming a constant mass-to-flux ratio in a region results
in $\zeta = 0.3$ (Nakano \& Nakamura 1978).  For a typical
interstellar field of 3~$\mu$G, the critical surface density for
collapse is $7 \mbox{ M}_{\odot}$~pc$^{-2}$, corresponding to a number
density of 230~cm$^{-2}$ in a layer of thickness 1~pc $\approx 3.09
\times 10^{18}\,\mbox{cm}$.  A cloud is termed {\em subcritical} if it
is magnetostatically stable and {\em supercritical} if it is not.

The very large value for the magnetic critical mass in the diffuse ISM
given by Eq.~(\ref{eqn:crit-rho}) forms a crucial objection to the
classical theory of star formation.  Even if such a large mass could
be assembled, how could it fragment into objects with stellar masses
of 0.01--100~M$_{\odot}$, when the critical mass should remain invariant
under uniform spherical gravitational collapse?

Two further objections to the classical theory were also prominent.
First was the embarrassingly high rate of star formation predicted by
a model governed by gravitational instability, in which objects should
collapse on roughly the free-fall timescale, Eq.\ (\ref{eqn:free-fall-time}), 
orders of magnitude shorter than the ages of typical galaxies.  

Second was the gap between the angular momentum contained in a parcel
of gas participating in rotation in a galactic disk and the much
smaller angular momentum contained in stars 
(Spitzer 1968, Bodenheimer 1995).  The disk of the Milky Way rotates
with angular velocity $\Omega \simeq 10^{-15}$~s$^{-1}$.  A uniformly
collapsing cloud with initial radius $R_0$ formed from material with
density $\rho_0 = 2 \times 10^{-24}$~g~cm$^{-3}$ rotating with the
disk will find its angular velocity increasing as $(R_0/R)^2$, or as
$(\rho/\rho_0)^{2/3}$.  By the time it reaches a typical stellar
density of $\rho = 1$~g~cm$^{-3}$, its angular velocity has increased
by a factor of $6 \times 10^{15}$, giving a rotation period of well
under a second.  The centrifugal force $\Omega^2 R$ exceeds the
gravitational force by eight orders of magnitude for solar
parameters. 
This is unphysical, and indeed typical solar-type stars
have rotational periods of several tens of days instead. A detailed
discussion including a demonstration that 
binary formation does not solve this problem can be found in
Mouschovias (1991b).

The observational discovery of bipolar outflows from young stars
(Snell, Loren \& Plambeck 1980) was a surprise that was unanticipated
by the classical model of star formation.  It has become clear that
the driving of these outflows is one part of the solution of the
angular momentum problem, and that magnetic fields transfer the
angular momentum from infalling to outflowing gas (e.g. K{\"o}nigl \&
Pudritz 2001).

Finally, millimeter-wave observations of emission lines from dense
molecular gas revealed a further puzzle: extremely superthermal
linewidths indicating that the gas was moving randomly at hypersonic
velocities (Zuckerman \& Palmer 1974). Such motions generate shocks
that would dissipate the energy of the motions within a crossing time
because of shock formation (e.g. Field 1978).  Attempts were made
using clump models of turbulence to show that the decay time might be
longer (Scalo \& Pumphrey 1982, Elmegreen 1985).  In hindsight,
moving spherical gas clumps turn out not to be a good model for
turbulence, however, so these models failed to accurately predict its
behavior (Mac Low \etal\ 1998).

\subsection{Standard theory of isolated star formation}
\label{sub:standard}

The problems outlined in the preceeding subsection were addressed in
what we call the standard theory of star formation, which has formed
the base of most work in the field for the past two decades.  Mestel
\& Spitzer (1956) first noted that the problem of magnetic support
against fragmentation could be resolved if mass could move across
field lines, and proposed that this could occur in mostly neutral gas
through the process of ion-neutral drift, usually known as ambipolar
diffusion in the astrophysical community.\footnote{In plasma physics,
the term ambipolar diffusion is applied to ions and electrons held
together electrostatically rather than magnetically while drifting
together out of neutral gas.}  The other problems outlined then
appeared solvable by the presence of strong magnetic fields, as we now
describe.

Ambipolar diffusion can solve the question of how magnetically
supported gas can fragment if it allows neutral gas to gravitationally
condense across field lines.  The local density can then increase
without also increasing the magnetic field, thus lowering the critical
mass for gravitational collapse $M_c$ given by
Eq.\ (\ref{eqn:crit-rho}).  This can also be interpreted as increasing
the local mass-to-flux ratio towards the critical value given by
Eq.\ (\ref{eqn:crit-phi}).  

The timescale $\tau_{\rm AD}$ on which this occurs can be derived by
considering the relative drift velocity of neutrals and ions
$\vec{v}_{\rm D} =\vec{v}_i - \vec{v}_n$ under the influence of the
magnetic field $\vec{B}$ (Spitzer 1968).  So long as the ionization
fraction is small and we do not care about instabilities such as that
found by Wardle (1990), the inertia and pressure of the ions may be
neglected.  The ion momentum equation then reduces in the steady-state
case to a balance between Lorentz forces and ion-neutral drag,
\begin{equation} \label{eqn:drift}
\frac{1}{4\pi} (\nabla \times \vec{B}) \times \vec{B} = \alpha \rho_i
\rho_n (\vec{v}_i -  \vec{v}_n),
\end{equation}
where the coupling coefficient 
\begin{equation} \label{eqn:couple}
\alpha = \langle \sigma v \rangle/(m_i + m_n) \approx  9.2 \times
10^{13}\,\mbox{cm}^3\,\mbox{s}^{-1}\,\mbox{g}^{-1}
\end{equation}
(e.g.\ Smith \& Mac Low 1997), with $m_i$ and $m_n$ the mean mass per
particle for the ions and neutrals, and $\rho_i$ and $\rho_n$ the ion
and neutral densities.  Typical values in molecular clouds are $m_i =
10 \,m_{\rm H}$ and $m_n = (7/3) m_{\rm H}$.  The value of $\alpha$ is
roughly independent of the mean velocity, as the ion-neutral
cross-section
$\sigma$ scales inversely with velocity in the regime of interest
(Osterbrock 1961, Draine 1980).  To estimate the typical timescale,
consider drift occurring across a cylindrical region of radius $R$,
with a typical bend in the field also of order $R$ so the Lorentz
force can be estimated as roughly $B^2/4\pi R$.  Then the ambipolar
diffusion timescale can be derived by solving for $v_{\rm D}$ in Eq.\
(\ref{eqn:drift}) to be
\begin{equation}
\label{eqn:AD}
\tau_{\rm AD}  =  \left.\frac{R}{v_{\rm D}}\right. = \left.\frac{4\pi \alpha \rho_i \rho_n R}{(\nabla \times \vec{B}) \times
\vec{B}}\right. \approx \frac{4\pi \alpha
\rho_i \rho_n R^2} {B^2}
           = (25\,{\rm{Myr}}) 
                \left(\frac{B}  {3\,\mu\mbox{G}      }\right)^{-2}
                \left(\frac{n_n}{10^2\,\mbox{cm}^{-3}}\right)^{2}
                \left(\frac{R}  {1\,\mbox{pc}        }\right)^{2}
                \left(\frac{x}  {10^{-6}             }\right)^{},
\end{equation}

For ambipolar diffusion to solve the magnetic flux problem on an
astrophysically relevant timescale, the ionization fraction $x$ must be extremely
small.  With the direct observation of dense molecular gas (Palmer \&
Zuckerman 1967, Zuckerman \& Palmer 1974) more than a decade after the
original proposal by Mestel \& Spitzer (1956), such low ionization
fractions came to seem plausible.  Nakano (1976, 1979) and Elmegreen
(1979) computed the detailed ionization balance of molecular clouds
for reasonable cosmic ray ionization rates, showing that at densities
greater than $10^4\,$cm$^{-3}$, 
the ionization fraction was roughly
\begin{equation}
x \approx (5 \times 10^{-8}) \left(\frac{n}{10^5 \mbox{ cm}^{-3}}\right)^{-1/2}
\end{equation}
(Elmegreen 1979), becoming constant at densities higher than
$10^7\,$cm$^{-3}$ or so.  Below densities of $10^4\,$cm$^{-3}$, the
ionization increases because of the external UV radiation field, and
the gas is tightly coupled to the magnetic field.

With typical molecular cloud parameters $\tau_{\rm AD}$ is of order
$10^7\,$yr (eq.~\ref{eqn:AD}). 
The ambipolar diffusion timescale $\tau_{\rm AD}$ is thus about
10--20 times longer than the corresponding dynamical timescale
$\tau_{\rm ff}$ of the system (e.g.\ McKee \etal\ 1993).  The delay
induced by waiting for ambipolar diffusion to occur 
was then taken as a way to explain the low star formation rates
observed in normal galaxies, as well as the
long lifetimes of molecular clouds, which at that time
were thought to be about 30--100$\,$Myr (Solomon
\etal\ 1987, Blitz \& Shu 1980). See \S~\ref{sub:clouds} for arguments
that they are under 10 Myr, however.

These considerations lead to the investigation of star formation
models based on ambipolar diffusion as a dominant physical process
rather than relying solely on gas dynamical collapse. In particular
Shu (1977) proposed the self-similar collapse of initially
quasi-static singular isothermal spheres as the most likely
description of the star formation process.  He assumed that ambipolar
diffusion in a magnetically subcritical, isothermal cloud core would
lead to the build-up of a quasi-static $1/r^2$-density structure that
contracts on timescales of order of $\tau_{\rm AD}$.  This
evolutionary phase is denoted quasi-static because $\tau_{\rm
AD}\gg\tau_{\rm ff}$. Ambipolar diffusion is supposed to eventually
lead to the formation of a singularity in central density, at which
point the system becomes unstable and undergoes inside-out collapse.
During collapse this model assumes that magnetic fields are no longer
dynamically important and they are subsequently ignored in the
original formulation of the theory. A rarefaction wave moves outward
with the speed of sound, with the cloud material behind the wave
falling freely onto the core and matter ahead still being at rest.  

The Shu (1977) model predicts constant mass accretion onto the central
protostar at a rate $\dot{M}=0.975\,c_{\rm s}^3/G$.  This is
significantly below the values derived for Larson-Penston collapse. In
the latter case the entire system is collapsing dynamically and
delivers mass to the center very efficiently, while in the former case
inward mass transport is comparatively inefficient as the cloud
envelope remains at rest until reached by the rarefaction wave.  The
density structure of the inside-out collapse, however, is essentially
indistinguishable from the predictions of dynamical collapse. To
observationally differentiate between the two models one needs to
obtain kinematical data and determine the magnitude and spatial extent
of infall motions with high accuracy. The basic predictions of
inside-out collapse are summarized in Table \ref{tab:shu}.  
As singular isothermal spheres by definition have infinite mass, the
growth of the central protostar is taken to come to a halt when
feedback processes (like bipolar outflows, stellar winds, etc.) become
important and terminate further infall.

Largely within the framework of the standard theory, numerous
analytical extensions to the original inside-out collapse
model have been proposed. The stability of isothermal gas clouds with
rotation, for example, has been investigated by Schmitz (1983, 1984,
1986), Tereby, Shu, \& Cassen (1984), Schmitz \& Ebert (1986, 1987),
Inutsuka \& Miyama (1992), Nakamura, Hanawa, \& Nakano (1995),
and Tsuribe \& Inutsuka (1999b).

The effects of magnetic fields on the equilibrium structure of clouds
and later during the collapse phase (where they have been neglected in
the original inside-out scenario) are considered by Schmitz (1987),
Baureis, Ebert, \& Schmitz (1989), Tomisaka, Ikeuchi, \& Nakamura
(1988a,b, 1989a,b, 1990), Tomisaka (1991, 1995, 1996a,b), Galli \& Shu
(1993a,b), Li \& Shu (1996, 1997), Galli \etal\ (1999, 2001), and Shu
\etal\ (2000). The proposed picture is that ambipolar diffusion of
initially subcritical cores that are threaded by uniform magnetic
fields will lead to the build-up of disk-like structures with constant
mass-to-flux ratio. These disks are called {\em isopedic}. The
mass-to-flux ratio increases steadily with time. As it exceeds the
maximum value consistent with magnetostatic equilibrium, the entire
core becomes supercritical and begins to collapse from the inside out
with the mass-to-flux ratio assumed to remain approximately constant.
It can be shown (Shu \& Li 1997), that for isopedic disks the forces
due to magnetic tension are just a scaled version of the disk's
self-gravity with oposite sign (i.e.\ obstructing gravitational
collapse), and that the magnetic pressure scales as the gas pressure
(although the proportionality factor in general is spatially varying
except in special cases). These findings allow the application of many
results derived for unmagnetized disks to the magnetized regime with
only slight modifications to the equations. One application of this
result is that for isopedic disks the derived mass accretion rate is
just a scaled version of the original Shu (1977) rate, i.e.\
$\dot{M}\approx (1+{H}_0)\,c_{\rm s}^3/G$, with the dimensionless
parameter $H_0$ depending on the effective mass-to-flux ratio.

However, the basic assumption of constant mass-to-flux ratio during
the collapse phase appears inconsistent with detailed numerical
calculations of ambipolar diffusion processes (see
\S~\ref{subsub:SIS}). In these computations the mass-to-flux ratio in
the central region increases more rapidly than in the outer parts of
the cloud. This leads to a separation into a dynamically collapsing
inner core with $(M/\Phi)_{\rm n} > 1$, and an outer envelope with
$(M/\Phi)_{\rm n} < 1$ that is still held up by the magnetic
field. The parameter $(M/\Phi)_{\rm n}$ is the dimensionless
mass-to-flux ratio normalized to the critical value given by Eq.\
(\ref{eqn:crit-phi}). The isopedic description may therefore only be
valid in the central region with $(M/\Phi)_{\rm n} > 1$.

The presence of strong magnetic fields was
suggested as a way to explain the universally observed (Zuckerman \&
Palmer 1974) presence of hypersonic random motions in molecular clouds
by Arons \& Max (1975). They noted that linear Alfv\'en waves have no
dissipation associated with them, as they are purely transverse. In a
cloud with Alfv\'en speed $v_{\rm A} = B/(4\pi\rho)^{1/2}$ much greater than
the sound speed $c_{\rm s}$, such Alfv\'en waves could produce the observed
motions without necessarily forming strong shocks.  This was
generally, though incorrectly, interpreted to mean that these waves
could therefore survive from the formation of the cloud, explaining
the observations without reference to further energy input into the
cloud.  The actual work acknowledged that ambipolar diffusion would
still dissipate these waves (Kulsrud \& Pearce 1969, Zweibel \&
Josafatsson 1983) at a rate substantial enough to require energy input
from a driving source to maintain the observed motions.

Strong magnetic fields furthermore provided a mechanism to reduce the angular
momentum in collapsing molecular clouds through magnetic braking.
Initially this was treated assuming that clouds were rigid rotating
spheres (Ebert, von Hoerner, \& Temesv\'ary 1960), but was accurately
calculated by Mouschovias \& Paleologou (1979, 1980) for both
perpendicular and parallel cases.  They showed that the criterion for
braking to be effective was essentially that the outgoing helical
Alfv\'en waves from the rotating cloud had to couple to a mass of gas
equal to the mass in the cloud.  Mouschovias \& Paleologou (1980)  show
that this leads to a characteristic deceleration time for a parallel
rotator of density $\rho$ and thickness $H$ embedded in a medium of
density $\rho_0$ and Alfv\'en velocity $v_{\rm A} = B/(4\pi\rho_0)^{1/2}$ of
\begin{equation}
\tau_{\parallel} = (\rho/\rho_0) (H/2v_{\rm A}),
\end{equation}
and a characteristic time for a perpendicular rotator with radius $R$,
\begin{equation}
\tau_{\perp} = \frac12 \left[\left(1 +
\frac{\rho}{\rho_0}\right)^{1/2} - 1\right] \frac{R}{v_{\rm A}}.
\end{equation}
For typical molecular cloud parameters, these times can be less than
the free-fall time, leading to efficient transfer of angular momentum
away from collapsing cores. This may help to resolve the
angular momentum problem in star formation described in
\S~\ref{sub:classicalprobs}.

\subsection{Problems with standard theory} 
\label{sub:standardprobs}

During the 1980's the theory of magnetically mediated star formation
discussed in the previous section was widely advocated and generally
accepted as the standard theory of low-mass star formation, almost
completely replacing the earlier dynamical models. However, despite
its success and intellectual beauty, the picture of magnetically
mediated star formation 
occurring on timescales an order of magnitude longer than the
free-fall timescale suffers from a series of severe
observational and theoretical
shortcomings. 
This became obvious
in the 1990's with improved numerical simulations and the advent of
powerful new observational techniques, especially at sub-millimeter and
infrared wavelengths. Critical summaries are given by Whitworth
\etal\ (1996) and Nakano (1998). 

The standard theory introduces an artificial dichotomy to the star
formation process.  It suggests that low-mass stars form from
low-mass, magnetically subcritical cores, whereas high-mass stars and
stellar clusters form from magnetically supercritical cloud cores (Shu
\etal\ 1987, Lizano \& Shu 1989). This distinction became necessary when
it was understood that the formation of very massive stars or stellar
clusters cannot be regulated by magnetic fields and ambipolar
diffusion processes (see \S~\ref{subsub:obs-stars}). We will argue in
\S~\ref{sec:paradigm} that this is true for low-mass stars
also, and therefore that star formation is {\em not} mediated by
magnetic fields on {\em any} scale, but instead is controlled by
interstellar turbulence (\S~\ref{sub:new}). The new theory gives a
unified description of both low-mass {\em and} high-mass star
formation, thus removing the undesired artificial dichotomy introduced
by the standard theory.

Before we introduce the new theory of star formation based
on interstellar turbulence, we need to analyze in detail the
properties and shortcomings of the theory we seek to replace. 
We begin with the theoretical considerations that make the inside-out
collapse of quasistatic, singular, isothermal spheres unlikely to be
an accurate description of stellar birth. We then discuss the
disagreement of the theory with observations.

\subsubsection{Singular isothermal spheres}
\label{subsub:SIS}
The collapse of singular isothermal spheres is the astrophysically
most unlikely and unstable member of a large family of self-similar
solutions to the 1D collapse problem. Ever since the
studies by Bonnor (1956) and Ebert (1957), and by Larson (1969) and
Penston (1969a) much attention in the star formation community has
been focused on finding astrophysically relevant, analytic, asymptotic
solutions to this problem (see \S~\ref{sub:classical}). The standard
solution derived by Shu (1977) considers evolution of initially
singular, isothermal spheres as they leave equilibrium. His findings
subsequently were extended by Hunter (1977, 1986).  Whitworth \&
Summers (1985) demonstrated that all solutions to the isothermal
collapse problem are members of a two-parameter family with the
Larson-Penston-type solutions (collapse of spheres with uniform
central density) and the Shu-type solutions (expansion-wave collapse
of singular spheres) populating extreme ends of parameter space. The
solution set has been extended to include a polytropic equation of
state (Suto \& Silk 1988), shocks (Tsai and Hsu 1995), and cylinder
and disk-like geometries (Inutsuka and Miyama 1992; Nakamura, Hanawa,
\& Nakano 1995). In addition, mathematical generalization using a
Lagrangian formulation has been proposed by Henriksen (1989, see also
Henriksen, Andr{\'e}, and Bontemps 1997).

Of all proposed initial configurations for protostellar collapse, 
quasi-static, singular, isothermal spheres seem to be the most difficult
to realize in nature. Stable equilibria for self-gravitating, spherical,
isothermal gas clouds embedded in an external medium of given pressure
are only possible up to a density contrast of $\rho_{\rm c}/\rho_{\rm s}\approx
14$ between the cloud center and surface.  More
centrally concentrated clouds can only reach unstable
equilibrium states. Hence, all evolutionary paths that could yield a
central singularity lead through instability, so collapse will 
set in long before a $1/r^2$ density profile is established at
small radii $r$ (Whitworth \etal\ 1996; also Silk \& Suto 1988, and
Hanawa \& Nakayama 1997). External perturbations also tend to break
spherical symmetry in the innermost region and flatten the overall
density profile at small radii. 
The resulting behavior in the central region then more closely
resembles the Larson-Penston description of collapse.  Similar
behavior is found if outward propagating shocks are considered (Tsai
and Hsu 1995).  As a consequence, the existence of physical processes
that are able to produce singular, isothermal, equilibrium spheres in
nature is highly questionable. 

The original proposal of
ambipolar diffusion processes in magnetostatically supported gas does
not yield the desired result either.
Ambipolar diffusion in magnetically supported gas clouds
results in a dynamical Larson-Penston-type collapse of the central
region where magnetic support is lost, while the outer part is still
hold up primarily by the field (and develops a $1/r^2$ density
profile). Mass is fed to the center not by an outward moving
expansion wave, but by ambipolar diffusion in the outer envelope.
The proposal that singular isothermal spheres may form through
ambipolar diffusion processes in magnetically subcritical cores has
been extensively studied by Mouschovias and collaborators in a series
of numerical simulations with ever increasing accuracy and
astrophysical detail (Mouschovias 1991, Mouschovias \& Morton 1991,
1992a,b, Fiedler \& Mouschovias 1992, 1993, Morton \etal\ 1994,
Ciolek \& Mouschovias 1993, 1994, 1995, 1996, 1998, Basu and
Mouschovias 1994, 1995a,b, Desch \& Mouschovias 2001; see however
also Nakano 1979, 1982, 1983, Lizano \& Shu 1989, or Safier \etal\
1997). The numerical results indicate that the decoupling between
matter and magnetic fields occurs over several orders of magnitude in
density, becoming important at $n({\rm H}_2) >
10^{10}\,$cm$^{-3}$.  There is no single critical density below
which matter is fully coupled to the field and above which it is not,
although ambipolar diffusion is indeed the dominant physical
decoupling process (e.g.\ Desch \& Mouschovias 2001). As a
consequence of ambipolar diffusion, initially subcritical
gas clumps separate into a central nucleus that becomes both
thermally and magnetically supercritical, and an extended envelope
that is still held up magnetostatically. The central region
goes into rapid collapse sweeping up much of its residual magnetic
flux with it
(Basu 1997).  

Star formation from singular isothermal spheres is also biased against
binary formation. The collapse of rotating singular isothermal spheres
very likely will result in the formation of single stars, as the
central protostellar object forms very early and rapidly increases in
mass with respect to a simultaneously forming and growing rotationally
supported protostellar disk (e.g.\ Tsuribe \& Inutsuka 1999a,b).  By
contrast, the collapse of cloud cores with flat inner density profiles
will deliver a much smaller fraction of mass directly into the central
protostar within a free-fall time. More matter will go first into a
rotationally supported disk-like structure. These disks tend to be
more massive with respect to the central protostar in a
Larson-Penston-type collapse compared to collapsing singular
isothermal spheres, they are more likely become unstable to
subfragmentation, resulting in the formation of binary or higher-order
stellar systems (see the review by Bodenheimer \etal\ 2000).  Since
the majority of stars seems to form as part of binary or higher-order
system (e.g.\ Mathieu \etal\ 2000), star formation in nature appears
incompatible with collapse from strongly centrally peaked initial
conditions (Whitworth \etal\ 1996).

\subsubsection{Observations of clouds and cores}
\label{subsub:obs-clouds}
Before we consider the observational evidence against the standard
theory of magnetically mediated star formation, let us recapitulate
its basic predictions as introduced in Section \ref{sub:standard}. The
theory predicts (a) constant accretion rates and (b) infall motions
that are confined to regions that have been passed by a rarefaction
wave that moves outwards with the speed of sound, while the parts of a
core that lie further out remain static.  The theory furthermore (c)
relies on the presence of magnetic field strong enough to hold up the
gas in molecular cloud cores sufficiently long, so it predicts that
cores should be magnetically subcritical during most of their
lifetimes. In the following we demonstrate {\em all} these predictions
appear to be contradicted by observations.

\begin{figure}[t]
\begin{center}
\unitlength1cm
\begin{picture}( 8.00, 11.00)
\put( 0.00, 0.50){\epsfxsize=8cm \epsfbox{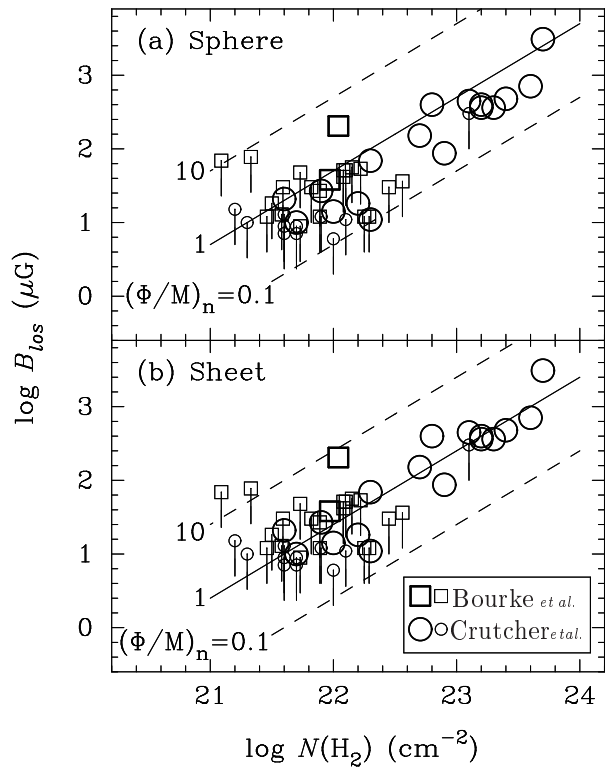}}
\end{picture}
\end{center}
\caption{\label{fig:B-vs-N} Line-of-sight magnetic field strength
$B_{\rm los}$ versus column density $N({\rm H_s}$ for various
molecular cloud cores. Squares are observations of Bourke \etal\
(2001) and circles are observations summarized by Crutcher (1990),
Sarma \etal\ (2000), and Crutcher \& Troland (2000).  Large symbols
represent clear detections of the Zeeman effect, whereas small ones
are $3\sigma$ upper limits to the field strength. The lines drawn for
the upper limits connect $3\sigma$ to $1\sigma$ limits. To guide the
eye, lines of constant flux-to-mass ratio $(\Phi/M)_{\rm n}$ are
given, normalized to the critical value, i.e.\ to the inverse of
equation (\ref{eqn:crit-phi}). The observed line-of-sight component
$B_{\rm los}$ of the field is statistically deprojected to obtain the
absolute value of the field $B$. The upper panel (a) assumes spherical
core geometry, while the lower panel (b) assumes a sheetlike
geometry. A value of $(\Phi/M)_{\rm n}<1$ corresponds to a
magnetically supercritical core with magnetic field strengths too weak
to support against gravitational contraction, while $(\Phi/M)_{\rm
n}>1$ allows magnetic support as required by the standard theory. Note
that almost all observed cores are magnetically supercritical. This is
evident when assuming spherical symmetry, but even in the case of
sheetlike protostellar cores the average ratio is $\langle
(\Phi/M)_{\rm n} \rangle \approx 0.4$ when considering the $1\sigma$
upper limits. This is significantly lower than the critical value. The
one clear exception is RCW57, but it has two velocity components,
leaving its Zeeman value in some doubt. For further discussion see
Bourke \etal\ (2001), from which this figure was drawn.}
\end{figure}

\paragraph{Magnetic Support}
In his critical review of the standard theory of star formation,
Nakano (1998) pointed out that {\em no} convincing magnetically
subcritical core had been found up to that time. Similar conclusions
still hold today. {\em All} magnetic field measurements are consistent
with cores being magnetically supercritical or at most marginally
critical.

When the theory was formulated in the late 1970's and 1980's,
accurate measurements of magnetic field strength in molecular clouds
and cloud cores did not exist or were highly uncertain. Consequently,
magnetic fields in molecular cloud were essentially {\em assumed} to
have the properties necessary for the theory to work and to circumvent
the observational problems associated with the classical dynamical
theory (Section \ref{sub:classicalprobs}). In particular, the field
was thought to have a strong fluctuating component associated with
magnetohydrodynamic waves that give rise to the superthermal
linewidths ubiquitiously observed in molecular cloud material, as well
as offering non-thermal support agains self-gravity.

Even today, accurate determinations of magnetic field strengths in
molecular cloud cores are rare. Most field estimates rely on measuring
the Zeeman splitting in molecular lines, typically OH, which is
observationally challenging (e.g.\ Crutcher \etal\ 1993). The
Zeeman effect has only been detected above the $3\sigma$ significance level
in a few dozen clouds, while the number of non-detections or
upper limits is considerably larger. For a compilation of field
strengths in low-mass cores see Crutcher (1999), or more recent work
by Bourke \etal\ (2001). Their basic results are summarized in Figure
\ref{fig:B-vs-N}, which plots the observed line-of-sight magnetic
field $B_{\rm los}$ against the column density $N({\rm H_2})$
determined from CO measurements. 

More detailed interferometric measurements may shift some of the data
points closer to the critical value or even slightly into the
subcritical regime (Crutcher, private communication), but Nakano's
(1998) objection still remains valid.  There are no observations of
strongly subcritical cores, and the inferred potential energies
typically exceed or at most are in approximate equipartition with the
magnetic energies. This means that even in the nominally subcritical
cases dynamical collapse will set in quickly as it only requires very
little ambipolar diffusion to reach a critical mass to flux ratio
(Ciolek \& Basu 2001). If both non-detections and upper limits are
taken into account, magnetic fields are on average too weak to prevent
or significantly retard the gravitational collapse of cores.  The
basic assumption of the standard theory of magnetically mediated star
formation therefore seems at odds with the observational facts.

This contradiction could, in principle, be weakened by making the
extreme geometrical assumption that all cores are highly
flattened, essentially sheetlike objects (Shu \etal\ 1999).
Flux-to-mass ratios can then be derived that come closer to the
critical value of equilibrium between magnetic pressure and gravity.
But even for sheetlike `cores', the average flux-to-mass ratio lies
somewhat below the critical value when taking all measurements into
account, including the upper limits at the $1\sigma$ level (Bourke
\etal\ 2001). In addition, highly flattened morphologies appear
inconsistent with the observed density structure of 
cores. They typically appear as `roundish' objects (like the dark
globule B68 described by
Alves \etal\ 2001) and more likely are moderately prolate (with
axis ratios of about 2:1) than highly oblate (with axis ratios $\sim
6$:1) when statistically deprojected (Myers \etal\ 1991, Ryden 1996;
however, some authors do prefer the oblate interpretation, see Li \& Shu
1996, Jones, Basu, \& Dubinski 2001). 

Bertoldi \& McKee (1992) already argued that the very massive clumps
that form stellar clusters need to be magnetically supercritical.
This conclusion was extended to low-mass cloud cores by Nakano (1998).
He noted that clumps and cores in molecular clouds are generally
observed as regions of significantly larger column density compared to
the cloud as a whole (e.g.\ Benson and Myers 1989, Tatematsu \etal\
1993). If a core were strongly magnetically subcritical
it would need to be confined by the mean cloud pressure or mean
magnetic field in the cloud.  Otherwise it would quickly expand and
disappear. Calculations of the collapse of strongly subcritical cores
such as those by Ciolek \& Mouschovias (1994) fix the magnetic field
at the outer boundaries, artificially confining the cloud. Under the
assumption of virial equilibrium, typical values for the mean pressure
and mean magnetic field in molecular clouds demand column densities in
magnetostatic cores that are comparable to those in the ambient
molecular cloud material. This contradicts the observed large column
density contrast between cloud cores and the rest of the cloud, giving
additional evidence that most low-mass cores are also magnetically
supercritical and collapsing.

\paragraph{Infall Motions}
Protostellar infall motions are observed on scales larger than and with
velocities greater than predicted by the standard theory. One of the
basic assumptions of the standard theory
is the existence of a long-lasting quasi-static phase in protostellar
evolution while ambipolar diffusion acts. 
Once ambipolar diffusion establishes the central singularity, a
rarefaction wave expands transsonically. Gas
beyond the rarefaction wave remains at rest. Therefore the theory
predicts that prestellar cores (cloud cores without central
protostars, see e.g.\ Andr{\'e} \etal\ 2000 for a discussion) should
show no signatures of infall motions, and that protostellar cores at
later stages of evolution should exhibit collapse motions confined to
their central regions. This can be tested by mapping molecular cloud
cores at the same time in optically thin and thick lines. Inward
motions can be inferred from asymmetry of optically thick lines, if
the zero point of the velocity frame is determined from the optically
thin lines.  This procedure allows the separation of infall signatures
from those of rotation and possible outflows
(Myers \etal\ 1996).

One of the best studied examples is the apparently starless core L1544, which
exhibits infall asymmetries (implying velocities up to
$0.1\,$km$\,$s$^{-1}$) that are too extended ($\sim 0.1\,$pc) to be
consistent with inside-out collapse (Tafalla \etal\ 1998, Williams
\etal\ 1999). Similar conclusions can be derived for a variety of
other sources (see the review by Myers, Evans, \& Ohashi 2000; or the
extended survey for infall motions in prestellar cores by Lee, Myers,
\& Tafalla 1999, 2001). Typical observed contraction velocities in the
prestellar phase are between $0.05$ and $0.1\,$km$\,$s$^{-1}$,
corresponding to mass infall rates ranging from a few $10^{-6}$ to a
few $10^{-5}\,$M$_{\odot}\,$yr$^{-1}$. The sizes of the infalling
regions (e.g.\ as measured in CS) typically exceed the sizes of the
corresponding cores as measured in high-density tracers like
N$_2$H$^+$ by a factor of 2--3.  Even the dark globule B335, which was
considered ``a theorists dream'' (Myers \etal\ 2000) and which was
thought to match standard theory very well (Zhou \etal\ 1993) is also
consistent with Larson-Penston collapse (e.g.\ Masunaga and Inutsuka
2001) when analyzed using improved radiation transfer techniques but
relying on single dish data only. The core however exhibits
considerable sub-structure and complexity (clumps, outflows, etc.)
when observed with high spatial and spectral resolution using
interferometry (Wilner \etal\ 2001). This raises questions about the
applicability of {\em any} 1D, isothermal collapse model
to real cloud cores.

Extended inward motions are a common feature in prestellar cores, and
appear a necessary ingredient for the formation of stars as predicted
by dynamical theories (Sections \ref{sub:classical} and
\ref{sub:new}).

\paragraph{Density profiles}
\label{para:density}
Observed prestellar cores have flat inner density profiles, as
described in \S~\ref{sub:cores}.  The basis of the Shu (1977) model is
the singular isothermal sphere: the theory assumes radial density
profiles $\rho \propto 1/r^2$ at all radii $r$ as starting conditions
of protostellar collapse.  High-resolution mapping of the density
profiles of prestellar cores provides the most direct evidence against
singular isothermal spheres as initial conditions of protostellar
collapse.

\begin{figure}[th]
\begin{center}
\unitlength1cm
\begin{picture}( 8.00, 7.00)
\put( 0.00, 0.00){\epsfxsize=9cm \epsfbox{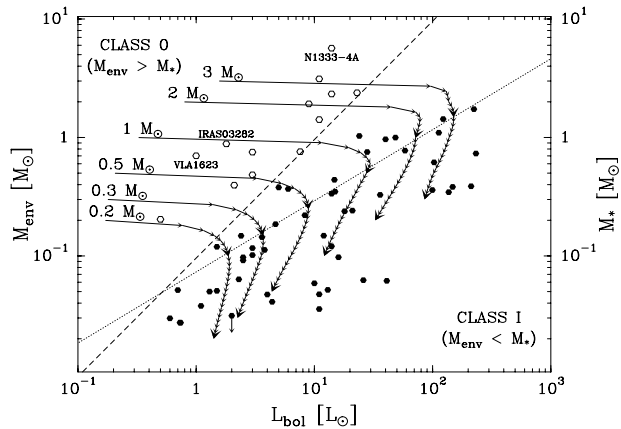}}
\end{picture}
\end{center}
\caption{\label{fig:L_bol-M_env} Plot of envelope mass $M_{\rm env}$
versus bolometric luminosity $L_{\rm bol}$ for a sample of cores
containing protostars in the main accretion phase with masses $M_{\rm
\star}$, from Andr{\'e} \& Montmerle (1994) and Saraceno \etal\
(1996). Open circles are objects with $M_{\rm env} > M_{\rm \star}$)
and filled circles are in the later evolutionary stage where $M_{\rm
env} < M_{\rm \star}$.  The evolutionary tracks shown assume a bound
initial configuration of finite mass that has $L_{\rm bol} =
GM_{\star} \dot{M_{\star}}/R_{\star} + L_{\star}$ with $L_{\star}$
from Stahler (1988), and assume that both $M_{\rm env}$ and
$\dot{M_{\star}}=M_{\rm env}/\tau$ ($\tau \approx 10^5\,$yr), decline
exponentially with time (Bontemps \etal\ 1996, also Myers \etal\
1998). Exponentially declining $\dot{M_{\star}}$ shows better agreement
with the data than do constant accretion rates. Small arrows are
plotted on the tracks every $10^4\,$yr, and large arrows when 50\% and
90\% of the total mass is accreted onto the central protostar. The dashed
and dotted lines indicate the transition from $M_{\rm env}> M_{\star}$
to $M_{\rm env}< M_{\star}$ using two different relations,
$M_{\star}\propto L_{\rm bol}$ and $M_{\star}\propto L_{\rm
bol}^{0.6}$, respectively, indicating the range proposed in the
literature (e.g.\ Andr{\'e} \& Montmerle 1994, or Bontemps \etal\
1996). The latter relation is suggested by the accretion scenario
adopted in the tracks. The figure is adapted from Andr{\'e} \etal\
(2000).}
\end{figure}

\paragraph{Chemical ages}
\label{para:chemical}
The chemical age of substructure in molecular clouds, as derived from
observations of chemical abundances (also see \S~\ref{sub:clouds}), is
much smaller than the ambipolar diffusion time. This poses a timescale
argument against magnetically regulated star formation.
The comparison of multi-molecule observations of cloud cores with
time-dependent chemical models indicates typical ages of about
$10^5\,$years
(see the reviews by van~Dishoeck \etal\ 1993, van~Dishoeck \& Black
1998, and Langer \etal\ 2000). This is orders of magnitude shorter
than the timescales of up to $10^7$~yr required for ambipolar
diffusion to become important as required by the standard model.

\subsubsection{Protostars and young stars}
\label{subsub:obs-stars}
\paragraph{Accretion rates}
\label{para:accretion}
Observed protostellar accretion rates decline with time, in
contradiction to the constant rates predicted by the standard model.
As matter falls onto the central protostar it goes through a shock and
releases energy that is radiated away giving rise to a luminosity
$L_{\rm acc} \approx GM_{\star}\dot{M}_{\star}/R_{\star}$ (Shu \etal\
1987, 1993). The fact that most of the matter first falls onto a
protostellar disk, where it gets transported inwards on a viscous
timescale before it is able to accrete onto the star does not alter
the expected overall luminosity by much (see e.g.\ Hartmann 1998).

During the early phases of
protostellar collapse, while the mass $M_{\rm env}$ of the
infalling envelope exceeds the mass $M_{\star}$ of the central
protostar, the accretion luminosity $L_{\rm acc}$ far exceeds the
intrinsic luminosity $L_{\star}$ of the young star. Hence the observed
bolometric luminosity $L_{\rm bol}$ of the object is a direct measure
of the accretion rate as long as reasonable estimates of $M_{\star}$
and $R_{\star}$ can be obtained. Determinations of bolometric
temperature $T_{\rm bol}$ and luminosity $L_{\rm bol}$ therefore
should provide a fair estimate of the evolutionary stage of a
protostellar core (e.g.\ Chen \etal\ 1995, Myers \etal\
1998). Scenarios in which the accretion rate decreases with time and
increases with total mass of the collapsing cloud fragment yield
qualitatively better agreement with the observations than do models
with constant accretion rate (Andr{\'e} \etal\ 2000, see however
Jayawardhana, Hartmann, \& Calvet 2001, for an alternative
interpretation based on environmental conditions).  A comparison of
observational data with theoretical models where $\dot{M}_{\star}$
decreases exponentially with time is shown in Figure
\ref{fig:L_bol-M_env}.  

\begin{figure}[ht]
\begin{center}
\unitlength1cm
\begin{picture}( 8.00, 8.00)
\put( 0.00, 0.10){\epsfxsize=9cm \epsfbox{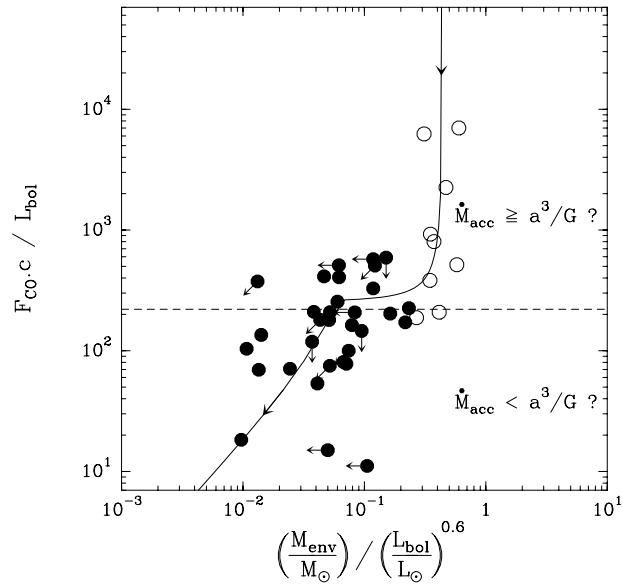}}
\end{picture}
\end{center}
\caption{\label{fig:decreasing-dM/dt}
Outflow momentum flux $F_{\rm CO}$ versus envelope mass
$M_{\rm env}$, normalized to the bolometric luminosity $L_{\rm bol}$,
using the relations $M_{\rm env}\propto L_{\rm bol}^{0.6}$ and $F_{\rm
CO}c\propto L_{\rm bol}$. Protostellar cores with $M_{\rm env}>
M_{\star}$ are shown by open circles, and $M_{\rm env}> M_{\star}$
by filled circles (data from Bontemps \etal\ 1996). $F_{\rm
CO}c/L_{\rm bol}$ is an empirical tracer for the accretion rate; the
speed of light $c$ is invoked in order to obtain a dimensionless quantity. 
$M_{\rm env}/L_{\rm bol}^{0.6}$ is an evolutionary indicator that
decreases with time. The abscissa therefore corresponds to a time axis,
with early times at the right and later times to the
left. Overlayed on the data is a evolutionary model that assumes a
flat inner density profile (for details see Henriksen
\etal\ 1997, where the figure was published originally).
}
\end{figure}

A closely related method to estimate the accretion rate
$\dot{M}_{\star}$ is determination of protostellar outflow strengths
(Bontemps \etal\ 1996). Most embedded young protostars have powerful
molecular outflows (Richer \etal\ 2000), while outflow strength
decreases towards later evolutionary stages.  At the end of the main
accretion phase, the bolometric luminosity of protostars $L_{\rm bol}$
strongly correlates with the momentum flux $F_{\rm CO}$ (e.g.\ Cabrit
\& Bertout 1992). Furthermore, $F_{\rm CO}$ correlates well with
$M_{\rm env}$ for all protostellar cloud cores (Bontemps \etal\ 1996,
Hoherheijde \etal\ 1998, Henning and Launhardt 1998). This result is
independent of the $F_{\rm CO}-L_{\rm bol}$ relation and most likely
results from a progressive decrease of outflow power with time during
the main accretion phase. With the linear correlation between outflow
mass loss and protostellar accretion rate (Hartigan \etal\ 1995) these
observations therefore suggest stellar accretion rates
$\dot{M}_{\star}$ that decrease with time. This is illustrated in
Figure \ref{fig:decreasing-dM/dt} which compares the observed values
of the normalized outflow flux and the normalized envelope mass for a
sample of $\sim40$ protostellar cores with a simplified dynamical
collapse model with decreasing accretion rate $\dot{M}_{\star}$
(Henriksen \etal\ 1997). The model describes the data relatively well,
as opposed to models of constant $\dot{M}_{\star}$.

\paragraph{Embedded Objects}
The fraction of protostellar cores with embedded protostellar objects
is very high. Further indication that the standard theory may need to
be modified comes from estimates of the time spent by protostellar
cores during various stages of their evolution. As the standard model
assumes that cloud cores in the prestellar phase evolve on ambipolar
diffusion timescales, which are an order of magnitude longer than the
dynamical timescales of the later accretion phase, one would expect a
significantly larger number of starless cores than cores with embedded
protostars.  

For a uniform sample of protostars, the relative numbers of objects in
distinct evolutionary phases roughly correspond to the relative time
spent in each phase. Beichman \etal\ (1986) used the ratio of numbers
of starless cores to the numbers of cores with embedded objects
detected with the {\em Infrared Astronomical Satellite} (IRAS) and
estimated that the duration of the prestellar phase is about equal to
the time needed for a young stellar object to completely accrete its
protostellar envelope.  Millimeter continuum mapping of pre-stellar
cores gives similar results (Ward-Thompson \etal\ 1994, 1999), leading
Andr{\'e} \etal\ (2000) to argue that the timespan of cores to
increase their central density $n({\rm H}_2)$ from $\sim10^4$ to
$\sim10^5\,$cm$^{-3}$ is about the same as to go from $n({\rm
H}_2)\approx 10^5\,$cm$^{-3}$ to the formation of the central
protostar. This clearly disagrees with standard ambipolar diffusion
models (e.g.\ Ciolek \& Mouschovias 1994), which predict a duration six times
longer.  Ciolek \& Basu (2000) were indeed able to accurately
model infall in L1544 using an ambipolar diffusion model, but they did
so by using initial conditions that were already almost supercritical,
so that very little ambipolar diffusion had to occur before dynamical
collapse would set in.  Ciolek
\& Basu (2001) quantify the central density required to match the
observations, and conclude that observed pre-stellar cores are either
already supercritical or just about to be. 
These observations that already in the prestellar phase the timescales
of core contraction are determined by fast dynamical processes rather
than by slow ambipolar diffusion.

\paragraph{Stellar Ages}
If the contraction time of individual cloud cores in the prestellar
phase is determined by ambipolar diffusion, then
the age spread in a newly formed group or cluster 
should considerably exceed the dynamical timescale. Within a
star-forming region, high-density protostellar cores will evolve and
form central protostars faster than their low-density counterparts,
so the age distribution is roughly determined by the evolution time
of the lowest-density condensation. 

However, the observed age spread in star clusters is very short. For
example, in the Orion Trapezium cluster, a dense cluster of a few
thousand stars, it is less than $10^6\,$years (Prosser \etal\ 1994,
Hillenbrand 1997, Hillenbrand \& Hartmann 1998), and the same holds
for L1641 (Hodapp \& Deane 1993). These age spreads are comparable to
the dynamical time in these clusters. Similar conclusions can be
obtained for Taurus (Hartmann 2001), NGC$\,$1333 (Bally \etal\ 1996,
Lada \etal\ 1996), NGC$\,$6531 (Forbes 1996), and a variety of other
clusters (see Elmegreen \etal\ 2000, Palla \& Stahler 1999, Hartmann
2001).

Larger regions form stars for a
longer time. This correlation 
suggests that typical star formation times
correspond to about 2--3 turbulent crossing times in that region
(Efremov \& Elmegreen 1998a).  This is very fast compared to the
ambipolar diffusion timescale, which is about 10 crossing times in a
uniform medium with cosmic ray ionization (Shu \etal\ 1987) and is
even longer if stellar UV sources contribute to the ionization (Myers
\& Khersonsky 1995) or if the cloud is very clumpy (Elmegreen \&
Combes 1992). Magnetic fields, therefore, appear unable regulate star
formation on scales of stellar clusters.

\section{TOWARD A NEW PARADIGM}
\label{sec:paradigm}

In this section we suggest that self-gravity acting in a supersonic,
turbulent flow can lead to behavior consistent with observations of
star formation, returning to the pioneering ideas of Larson (1981) in
a more quantitative fashion.  
We first examine whether magnetic fields
can maintain the supersonic motions observed in molecular clouds in
\S~\ref{sub:motions}.  We then study the behavior of self-gravitating,
turbulent gas, beginning in \S~\ref{sub:self-grav}.  We conclude that
turbulence often inhibits collapse without preventing it entirely in
\S~\ref{sub:promotion}, and discuss the relation between turbulence
and star formation in the subsequent subsections.  Finally, we outline
our conclusions in \S~\ref{sub:new}.

\begin{figure}[bt]
\begin{center}
\includegraphics[width=.6\textwidth]{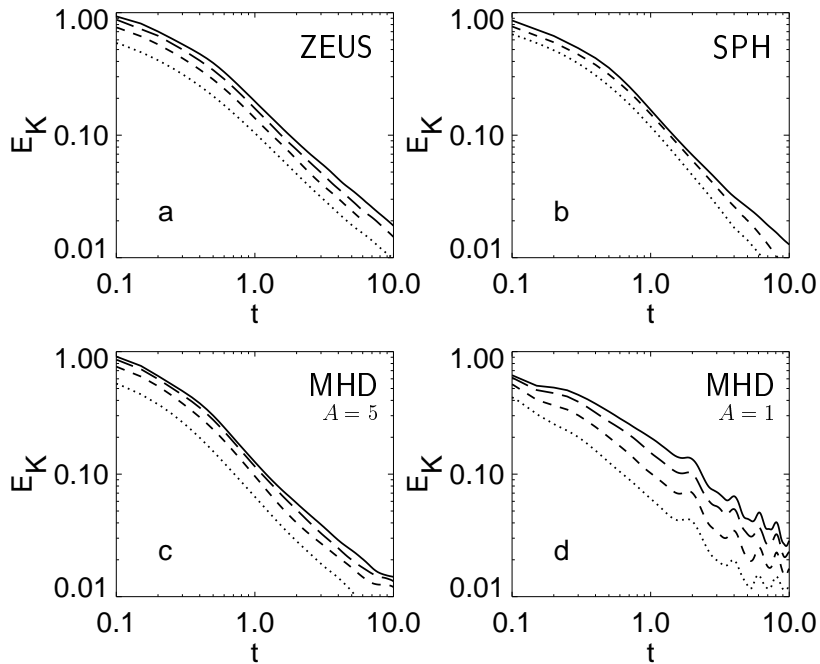}
\end{center}
\caption[Kinetic energy vs.\ time in decaying
turbulence]{Decay of supersonic turbulence. The plots show the time
  evolution of the total kinetic energy $E_K$ in a variety of 3D
  numerical calculations of decaying supersonic turbulence in
  isothermal ideal gas with initial rms Mach number ${\cal M}=5$.
  Grid-based ZEUS models have $32^3$ ({\em dotted}), $64^3$ ({\em
    short dashed}), $128^3$ ({\em long dashed}), or $256^3$ ({\em
    solid}) zones, while the particle-based SPH models have 7000 ({\em
    dotted}), 50,000 ({\em short dashed}), or 350,000 ({\em solid})
  particles.  The panels show {\em (a)} gas dynamic runs with ZEUS,
  {\em (b)} gas dynamic runs with SPH, {\em (c)} MHD
  ZEUS calculations with Alfv\'enic Mach number $A = 5$ MHD, and {\em
    d)} $A=1$ MHD runs with ZEUS, where $A = v_{\rm rms}/v_A = v_{\rm
    rms}\sqrt{4\pi\rho}/|B|$.  (From Mac Low \etal\ 1998).}
\label{fig:prlres}
\end{figure}

\subsection{Maintenance of supersonic motions}
\label{sub:motions}
We first consider the question of how to maintain observed
supersonic motions.  As described above in \S~\ref{sub:standard},
magnetohydrodynamic waves were generally thought to provide the means
to prevent the dissipation of interstellar turbulence.  However,
numerical models have now shown that they do not.  One-dimensional
simulations of decaying, compressible, isothermal, magnetized
turbulence by Gammie \& Ostriker (1996) showed quick decay of kinetic
energy $E_{\rm kin}$ in the absence of driving, but found that the
quantitative decay rate depended strongly on initial and boundary
conditions because of the low dimensionality.  Mac Low \etal\ (1998),
Stone, Ostriker \& Gammie (1998), and Padoan
\& Nordlund (1999) measured the decay rate in direct numerical
simulations in three dimensions, using a number of different numerical
methods.  They uniformly found rather faster decay, with Mac Low
\etal\ (1998) characterizing it as $E_{\rm kin} \propto t^{-\eta}$,
with $0.85 < \eta < 1.1$. A resolution and algorithm study is shown in
Figure~\ref{fig:prlres}.  Magnetic fields with strengths ranging up to
equipartition with the turbulent motions (ratio of thermal to magnetic
pressures as low as $\beta = 0.025$) do indeed reduce $\eta$ to the
lower end of this range, but not below that, while unmagnetized
supersonic turbulence shows values of $\eta \simeq$~1--1.1.

Stone \etal\ (1998) and Mac Low (1999) showed that supersonic
turbulence decays in less than a free-fall time under molecular cloud
conditions, regardless of whether it is magnetized or unmagnetized.
The gas dynamical result agrees with the high-resolution, transsonic,
decaying models of Porter \& Woodward (1992) and Porter \etal\
(1994).  Mac Low (1999) showed that the formal dissipation time
$\tau_{\rm d} = E_{\rm kin}/\dot{E}_{\rm kin}$, scaled in units of the
free fall time $t_{\rm ff}$, is
\begin{equation} \label{eqn:decay}
\tau_{\rm d}/\tau_{\rm ff} = \frac{1}{4 \pi \xi} \left(\frac{32}{3}\right)^{1/2}
\frac{\kappa}{{\cal M}_{\rm rms}} \simeq \,3.9 \,\frac{\kappa}{{\cal
M}_{\rm rms}},
\end{equation}
where $\xi = 0.21/\pi$ is the energy-dissipation coefficient, ${\cal M}_{\rm
rms} = v_{\rm rms}/c_{\rm s}$ is the rms Mach number of the turbulence, and
$\kappa$ is the ratio of the driving wavelength to the Jeans
wavelength $\lambda_{\rm J}$.  In molecular clouds, ${\cal M}_{\rm rms}$ is
typically observed to be of order 10 or higher.  The value of $\kappa$
is less clear.  Molecular clouds do appear to be driven from large
scales (\S~\ref{sub:LSS}), while the effective value of $\lambda_{\rm
J}$ is not entirely clear in a strongly inhomogeneous medium.  
If $\kappa < $3--4, then undriven turbulence will decay in a free-fall
time.  As we discuss in \S~\ref{sub:clouds}, the observational
evidence suggests that clouds are a few free-fall times old, on
average, though perhaps not more than two or three, so there may be
continuing energy input into the clouds.
This energy input may come from the same compressive motions that have
been suggested to form the clouds (Ballesteros-Paredes, Hartmann, \&
V\'azquez-Semadeni 1999).

\subsection{Turbulence in self-gravitating gas}
\label{sub:self-grav}

This leads to the question of what effect supersonic turbulence has
on self-gravitating clouds.  Can turbulence alone delay gravitational
collapse beyond a free-fall time?  In {\S}s \ref{sub:turbulence} and
\S~\ref{sub:classical}, we summarized analytic approaches to this
question, and pointed out that, aside from Sasao (1973), they were all
based on the assumption that the turbulent flow is close to
incompressible.  Sasao (1973) concluded that compressible turbulence
with a Kolmogorov spectrum would show collapse at roughly the Jeans
scale. 

Numerical models of highly compressible, self-gravitating turbulence
have shown the importance of density fluctuations generated by the
turbulence to understanding support against gravity.  Early models by
Bonazzola \etal\ (1987), Passot \etal\ (1988), and L{\'e}orat, Passot
\& Pouquet (1990) used low resolution calculations ($32^2$ to
$64^2$ collocation points) with a two-dimensional (2D) spectral code to
support their analytical results.  The gas dynamical studies by
V{\'a}zquez-Semadeni, Passot, \& Pouquet (1995) and
Ballesteros-Paredes \etal~(1999b), were also restricted to two
dimensions, and were focussed on the ISM at kiloparsec
scales rather than molecular clouds, although they were performed with
far higher resolution (up to $800 \times 800$ points).  Magnetic
fields were introduced in these models by Passot, V\'azquez-Semadeni,
\& Pouquet~(1995), and extended to three dimensions with self-gravity
(though at only $64^3$ resolution) by V\'azquez-Semadeni, Passot, \&
Pouquet~(1996).  One-dimensional computations focussed on molecular
clouds, including both MHD and self-gravity, were presented by Gammie
\& Ostriker~(1996) and Balsara, Crutcher \& Pouquet (1999).  Ostriker,
Gammie, \& Stone (1999) extended their work to 2.5 dimensions.

These early models at low resolution, low dimension, or both,
suggested two conclusions important for this review. First,
gravitational collapse, even in the presence of magnetic fields, does
not generate sufficient turbulence to markedly slow further
collapse. Second, turbulent support against gravitational collapse may
act globally, while still allowing local collapse.  More recent
three-dimensional (3D), high-resolution computations by Klessen \etal\
(1998, 2000), Klessen (2000), Klessen \& Burkert (2000, 2001), and
Heitsch \etal\ (2001a) have now confirmed
and extended 
these earlier results. In the following, we use these calculations to
draw consequences for the theory of star formation.

\subsection{A numerical approach}
\label{sub:numerics}

Klessen \etal\ (2000) and Heitsch \etal\ (2001a) applied two different
numerical methods: ZEUS-3D, an Eulerian MHD code (Stone \& Norman
1992ab, Clarke 1994, Hawley \& Stone 1995), and an implementation of
smoothed particle hydrodynamics (SPH), a Lagrangian hydrodynamics
method using particles as an unstructured grid (Benz 1990, Monaghan
1992).  Klessen \etal\ (1998), Klessen (2000), and Klessen \& Burkert
(2000, 2001) used only SPH computations.  

SPH can resolve very high density contrasts because it increases the
particle concentration in regions of high density, and thus the
effective spatial resolution, making it well suited for computing
collapse problems.  By the same token, though, it resolves low-density
regions poorly. Shock structures tend to be broadened by the averaging
kernel in the absence of adaptive techniques.  The correct numerical
treatment of gravitational collapse requires the resolution of the
local Jeans mass at every stage of the collapse (Bate \& Burkert
1997).  In the models described here, once an object with density
beyond the resolution limit of the code has formed in the center of a
collapsing gas clump, it is replaced by a `sink' particle (Bate,
Bonnell, \& Price 1995).  Replacing high-density cores and keeping
track of their further evolution in a consistent way prevents the time
step from becoming prohibitively small. This allows modeling of the
collapse of a large number of cores until the overall gas reservoir
becomes exhausted.

ZEUS-3D, conversely, gives equal resolution in all regions,
resolving shocks equally well everywhere, as well as allowing the inclusion
of magnetic fields (see \S~\ref{sub:MHD}).  On the other hand,
collapsing regions cannot be followed to scales less than one or two
grid zones.  Once again the resolution required to follow
gravitational collapse must be considered.  For a grid-based
simulation, the criterion given by Truelove \etal\ (1997) holds.
Equivalent to the SPH resolution criterion, the mass contained in one
grid zone has to be rather smaller than the local Jeans mass
throughout the computation.  In the models described here, this
criterion is satisfied until gravitational collapse is underway.

The computations discussed here are done on periodic cubes, with an
isothermal equation of state, using up to $256^3$ zones (with one
model at $512^3$ zones) or $80^3$ SPH particles. To generate turbulent
flows, Gaussian velocity fluctuations are introduced with power only in
a narrow interval $k-1 \leq |\vec{k}| \leq k$, where $k =
L/\lambda_{\rm d}$ counts the number of driving wavelengths
$\lambda_{\rm d}$ in the box (Mac Low~\etal\ 1998).  This offers a
simple approximation to driving by mechanisms that act on a single
scale. To drive the turbulence, this fixed pattern is normalized to
maintain constant kinetic energy input rate $\dot{E}_{\rm in} = \Delta
E / \Delta t$ (Mac~Low 1999).  Self-gravity is turned on only after
the turbulence reaches a state of dynamical equilibrium.

\begin{figure}[th]
\unitlength1.0cm
\begin{picture}(16,14.4)
\put(3.5,0.0){\epsfxsize=10.7cm \epsfbox{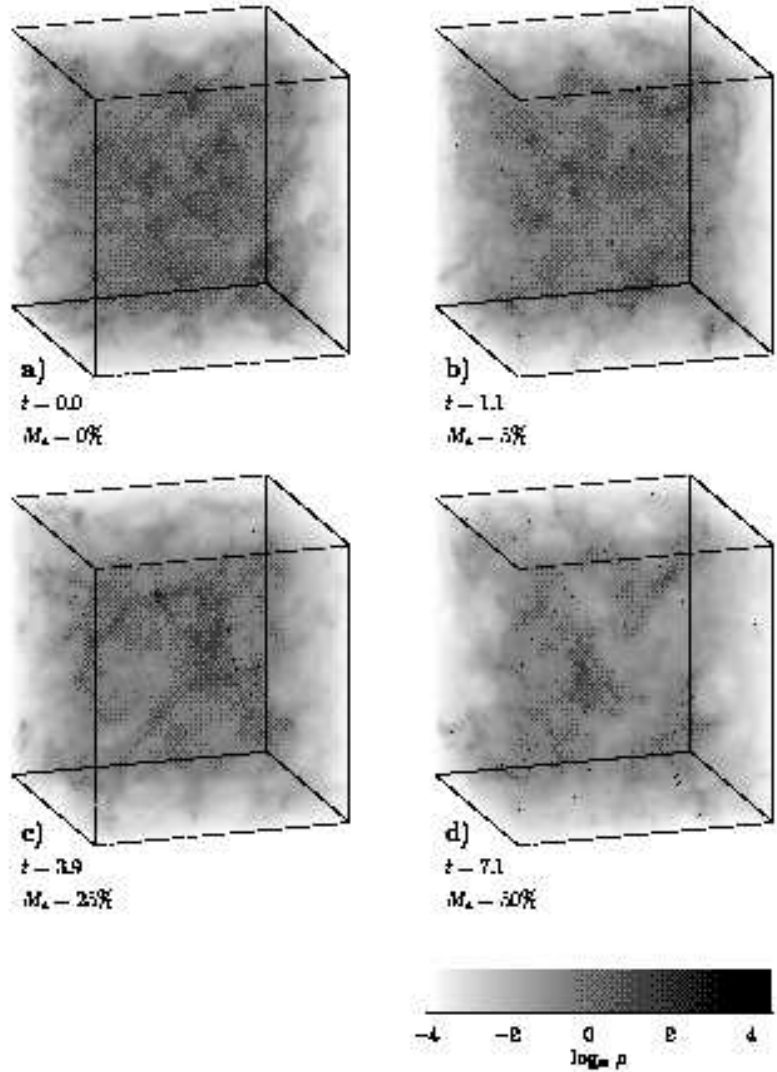}}
\end{picture}
\caption{\label{fig:3D-cubes} Density cubes in a model of supersonic
turbulence driven on intermediate scales, with wavenumbers $k= 3 - 4$,
from Klessen \etal\ (2000). Four different evolutionary stages are
shown: (a) just before gravity is turned on, (b) when the first
collapsed cores are formed and have accreted $M_* = 5$\% of the mass,
(c) when the mass in dense cores is $M_* = 25$\%, and (d) when $M_* =
50$\%. Time is measured in units of the global system free-fall
timescale $\tau_{\rm ff}$, and dark dots indicate the locations of
protostars. }
\end{figure}

\subsection{Global collapse}
\label{sub:global}

First, we examine the question of whether gravitational collapse can itself
generate enough turbulence to prevent further collapse. Gas dynamical
SPH models initialized at rest with Gaussian density perturbations
show fast collapse, with the first collapsed objects forming in a
single free-fall time (Klessen, Burkert, \& Bate 1998; Klessen \&
Burkert 2000, 2001; Bate \etal\ 2002a,b). Models set up with a freely decaying turbulent
velocity field behave similarly (Klessen 2000).  Further accretion of
gas onto collapsed objects then occurs over the next free-fall time,
defining the predicted spread of stellar ages in a freely-collapsing
system. The turbulence generated by collapse (or virialization)
fails to prevent further collapse, contrary to previous suggestions (e.g.\ by
Elmegreen 1993). Such a mechanism only works for thermal pressure
support in systems such as galaxy cluster halos where energy is lost
inefficently, while turbulence dissipates energy quite efficiently
(Eq.~\ref{eqn:decay}).

Models of freely collapsing, magnetized gas remain to be done, but models of
self-gravitating, decaying, magnetized turbulence by Balsara, Ward-Thompson,
\& Crutcher (2001) using an MHD code incorporating a Riemann solver suggest
that the presence of magnetic fields
does not
markedly extend collapse
timescales.  They further show that accretion down filaments aligned with
magnetic field lines onto cores occurs readily.  This allows high mass-to-flux
ratios to be maintained even at small scales, which is necessary for
supercritical collapse to continue 
in a magnetized medium
after fragmentation occurs.

\begin{figure}[th]
\begin{center}
\unitlength1.0cm
\begin{picture}( 8, 8.0)
\put( 0.0, 1.3){\epsfxsize=8cm \epsfbox{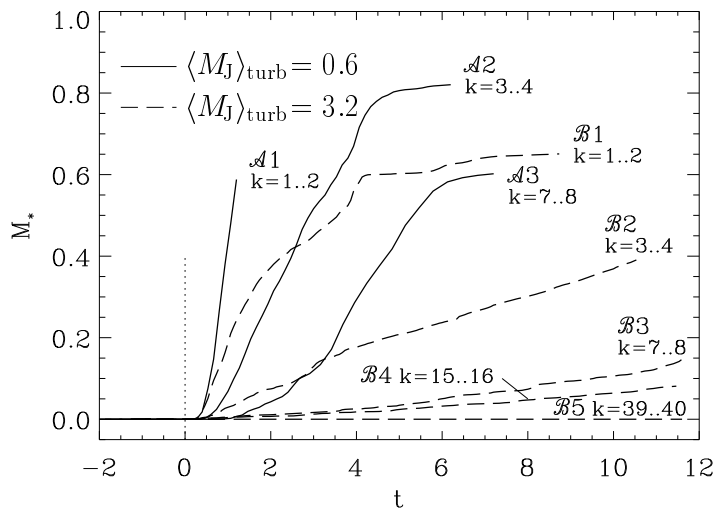}}
\end{picture}
\end{center}
\caption{\label{fig:accretion-history} Fraction $M_{\rm core}$ of mass
accreted onto protostars as function of time for different models of
self-gravitating supersonic turbulence. The models differ by driving
strength and driving wavenumber, as indicated in the figure. The mass
in the box is initially unity, so the $\cal A$-models (solid curves)
are formally unsupported, while the $\cal B$-models (dashed lines) are
formally supported.  This is indicated by the effective turbulent
Jeans mass $\langle M_{\rm J}\rangle_{\rm turb}$ defined at the mean
density. This number has to be compared with the total mass in the
cube, which is unity.  The figure shows that the efficiency of local
collapse depends on the scale and strength of turbulent driving.  Time
is measured in units of the global system free-fall timescale
$\tau_{\rm ff}$. Only the model ${\cal B}5$ which is driven strongly
at scales smaller than the Jeans wavelength $\lambda_J$ in
shock-compressed regions shows no collapse. (From Klessen \etal\
2000.) }
\end{figure}

\subsection{Local collapse in globally stable regions}
\label{sub:local}
Second, we examine whether continuously driven turbulence can support
against gravitational collapse.  The models of driven,
self-gravitating turbulence by Klessen \etal\ (2000) and Heitsch
\etal\ (2001a) described in \S~\ref{sub:numerics} show that
{\em local} collapse occurs even when the turbulent velocity field
carries enough energy to counterbalance gravitational contraction on
global scales.  This idea was first suggested by Hunter (1979), who
used the virial theorem to make his case. Later support for it was
offered by 2D gas dynamical computations by
L\'eorat \etal\ (1990).  
An example of local collapse in a globally supported
cloud is given in Figure~\ref{fig:3D-cubes}. A hallmark of global
turbulent support is isolated, inefficient, local collapse.

Local collapse in a globally stabilized cloud is not predicted by any
of the analytic models for turbulent, self-gravitating gas, as
discussed in Klessen \etal\ (2000). The resolution to this apparent
paradox lies in the requirement that any substantial turbulent support
must come from supersonic flows, as otherwise pressure support would
be at least equally important.  However, supersonic flows compress the
gas in shocks. In isothermal gas with density $\rho$ the postshock gas
has density $\rho' = {\cal M}^2 \rho$, where ${\cal M}$ is the Mach
number of the shock.  The turbulent Jeans length $\lambda_{\rm J}
\propto \rho'^{-1/2}$ in these density enhancements, so it drops by a
factor of ${\cal M}$ in isothermal shocks.

Klessen \etal\ (2000) demonstrated that supersonic turbulence can
completely prevent collapse only when it can support not just the
average density, but also the shocked, high-density regions, as shown
in Figure~\ref{fig:accretion-history}. This basic point was earlier
made by Elmegreen (1993) and V\'azquez-Semadeni \etal\ (1995).  Two
criteria must be fulfilled in these regions. The rms velocity must be
high enough,
and the driving wavelength $\lambda_{\rm d} < \lambda_{\rm J}(\rho')$
small enough.  If these two criteria are not met, the localized high-density
regions collapse, although the surrounding flow remains turbulently
supported.

\begin{figure}[h]
\begin{center}
\unitlength1.0cm
\begin{picture}(15,9.0)
\put( 2.0, 0.0){\epsfxsize=11.7cm \epsfbox{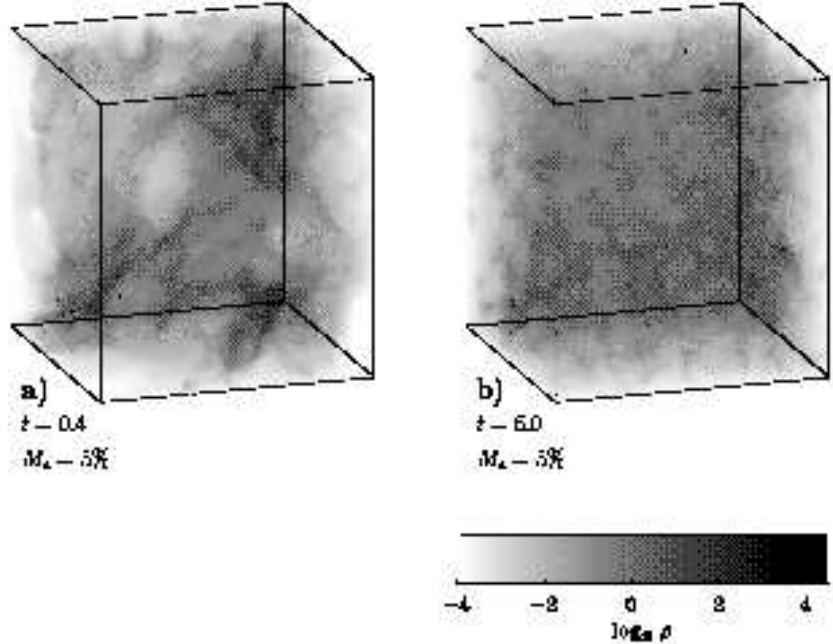}}
\end{picture}
\end{center}
\caption{\label{fig:3D-cubes-1..2+7..8} Density cubes for (a) a model
  of large-scale driven turbulence and (b) a model of small-scale
  driven turbulence at an evolutionary phase when $M_* = 5$\% of the
  total mass has been accumulated by protostars. Compare with
  Figure~\ref{fig:3D-cubes}b. Together they illustrate the influence
  of different driving wavelengths for otherwise identical physical
  parameters. Larger-scale driving results in a more organized
  distribution of protostars (black dots), while smaller-scale driving
  results in a more random structure. Note the different times at
  which $M_* = 5\%$ is reached. (From Klessen \etal\ 2000.)}
\end{figure}

The {\em length scale} and {\em strength} of energy injection into the
system determine the structure of the turbulent flow, and therefore
the locations at which stars are most likely to form.  Large-scale
driving leads to large coherent shock structures
(Figure~\ref{fig:3D-cubes-1..2+7..8}a). Local collapse occurs
predominantly in these filaments and layers of shocked gas (Klessen
\etal\ 2000).  Increasing the driving scale produces larger and more
massive structures that can become gravitationally unstable. Hence,
the star formation efficiency (eq.~\ref{eqn:sfe}) increases.  The same
is true for weaker driving.  Reducing the turbulent kinetic energy
means that more and larger volumes exceed the Jeans criterion for
gravitational instability. The more massive the unstable region is,
the more stars it will form.  Dense clusters or associations of stars
build up, either in the complete absence of energy input, or when
small-scale turbulence is too weak to support large volumes, or when
large-scale turbulence sweeps up large masses of gas that collapse. In
all of these cases star formation is high efficiency and proceeds on a
free-fall timescale (Klessen \etal~1998, Klessen \& Burkert 2000,
2001).

\begin{figure}[htp]
\begin{center}
\unitlength1.0cm
\begin{picture}(14,16.0)
\put( 0.0, 1.0){\epsfxsize=14cm \epsfbox{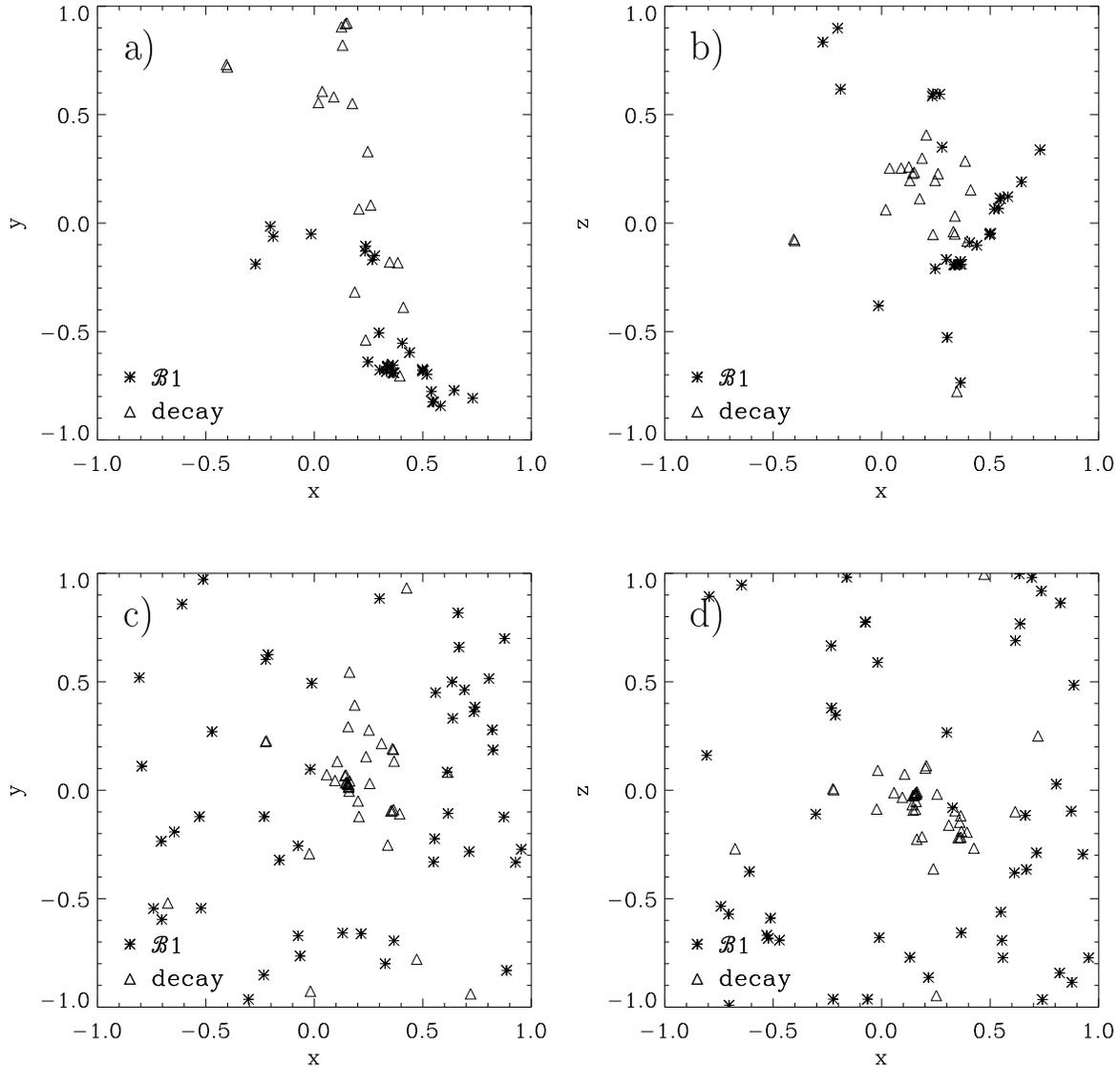}}
\end{picture}
\end{center}
\caption{\label{fig:2D-projection} Projected positions of protostars
in models of turbulence driven at large scale (model ${\cal B}1$ with
driving wavenumber $k=$~1--2) and freely decaying turbulence at two
different times.  The upper panels show projections into (a) the x-y
plane and (b) the x-z plane at an evolutionary stage when $M_*
\approx 20$\% of the total mass has been accreted.  The
gravitational potential is still dominated by non-accreted gas, and
the two models are statistically indistinguishable. In both cases
protostars form in a group. However, only decaying turbulence leads to a
bound cluster. In the model of driven turbulence, the group disperses
quickly, and stars end up widely distributed throughout the
computational volume.  This is illustrated in the lower panels (c) and
(d), which show the same projections of the system at later times when
$M_* \approx 65$\%.  (From Klessen \etal\ 2000.)  }
\end{figure}
These points are illustrated in Figure~\ref{fig:2D-projection}, which
compares the distribution of protostars in a model of freely-decaying
turbulence to the distribution in a model of turbulence driven at
large scale strongly enough to formally support against collapse.
Both scenarios lead to star formation in aggregates and
clusters. However, the figure suggests a possible way to distinguish
between the two.  Decaying turbulence typically results in a bound
cluster of stars, while stellar aggregates associated with
large-scale, coherent, shock fronts often have super-virial velocity
dispersions that result in their quick dispersal.  The numerical
models discussed here do not include feedback from the newly formed
stars, however.  Ionization and outflows may retard or impede further
star formation or gas accretion, possibly preventing a bound cluster
from forming even in the case of freely decaying turbulence.

The efficiency of collapse depends on the properties of the supporting
turbulence. It could be regulated by the amount of gas available for
collapse on scales where turbulence turns from supersonic to subsonic
(V\'azquez-Semadeni, Ballesteros-Paredes, \& Klessen 2003).
Sufficiently strong driving on short enough scales can prevent local
collapse for arbitrarily long periods of time, but such strong driving
may be rather difficult to arrange in a real molecular cloud.  If we
assume that stellar driving sources have an effective wavelength close
to their separation, then the condition that driving acts on scales
smaller then the Jeans wavelength in `typical' shock generated gas
clumps requires the presence of an extraordinarily large number of
stars evenly distributed throughout the cloud, with typical separation
0.1 pc in Taurus, or only 350 AU\footnote{One astronomical unit is the
mean radius of Earths orbit around the Sun, $1\,{\rm
AU}=1.5\times10^{13}\,{\rm cm}$.} in Orion.  This is not
observed. Very small driving scales seem to be at odds with the
observed large-scale velocity fields in at least some molecular clouds
(e.g.\ Ossenkopf \& Mac~Low 2002).

Triggering of star formation by compression has been discussed at
least since the work of Elmegreen \& Lada (1977) on star formation in
the gas swept up by expanding H~{\sc ii} regions.  The turbulent
compressions described here do indeed trigger local star formation in
globally supported regions.  However, in the absence of the flow, 
global collapse would cause more vigorous star formation.
Indeed, Elmegreen \& Lada (1977) noted themselves that the time for
gravitational instability to occur in the shocked, compressed layer
was actually somewhat longer than the Jeans time in the undisturbed
cloud.  We discuss this issue further in \S~\ref{subsub:disks}.

The domination of cloud structure by large-scale modes leads to the
formation of stars in groups and clusters.  When stars form in groups,
their velocities initially reflect the turbulent velocity field of the
gas from which they formed.  However, as more and more mass
accumulates in protostars, their mutual gravitational interaction
becomes increasingly important, beginning to determine the dynamical
state of the system, which then behaves more and more like a collisional
$N$-body system, where close encounters occur frequently (see
\S~\ref{sub:clusters}).

\subsection{Effects of magnetic fields}
\label{sub:MHD}

So far, we have concentrated on the effects of purely gas dynamical
turbulence. How does the picture discussed here change if we consider
the presence of magnetic fields?  Magnetic fields have been suggested
to support molecular clouds well enough to prevent gravitationally
unstable regions from collapsing (McKee 1999), either magnetostatically
or dynamically through MHD waves.

Assuming ideal MHD, a self-gravitating cloud of mass $M$ permeated by
a uniform flux $\Phi$ is stable unless the mass-to-flux ratio exceeds
the value given by Eq.~(\ref{eqn:crit-phi}).  Without any other
mechanism of support, such as turbulence acting along the field lines,
a magnetostatically supported cloud collapses to a sheet which is then
supported against further collapse. Fiege \& Pudritz (1999) found an
equilibrium configuration of helical field that could support a
filament, rather than a sheet, from fragmenting and collapsing. Such
configurations do not appear in numerical models of turbulent
molecular clouds, however, suggesting that reaching this
stable equilibrium is difficult.

Investigation of support by MHD waves concentrates mostly on the
effect of Alfv\'{e}n waves, as they (1) are not as subject to damping
as magnetosonic waves and (2) can exert a force {\em along} the mean
field, as shown by Dewar (1970) and Shu \etal\ (1987). This is because
Alfv\'{e}n waves are {\em transverse} waves, so they cause
perturbations $\delta \vec{B}$ perpendicular to the mean magnetic
field $\vec{B}$. McKee \& Zweibel (1994) argue that Alfv\'{e}n waves
can even lead to an isotropic pressure, assuming that the waves are
neither damped nor driven. However, in order to support a region
against self-gravity, the waves would have to propagate outwardly,
because inwardly propagating waves would only further compress the
cloud. Thus, this mechanism requires a negative radial gradient in
wave sources in the cloud (Shu \etal\ 1987).

\begin{figure}[htp]
\begin{center}
\unitlength1.0cm
\begin{picture}(8,15.0)
\put(-1.0,-1.5){\epsfxsize=13cm \epsfbox{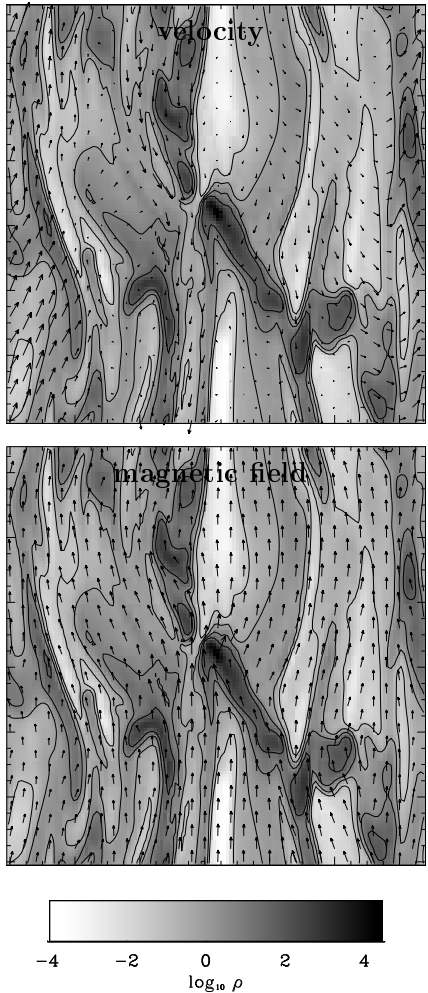}}
\end{picture}
\end{center}
\caption{\label{fig:magstatsup}
  Two dimensional slice through a cube of magnetostatically supported,
  self-gravitating turbulence driven at large scale from Heitsch
  \etal\ (2001a).  The upper panel shows  velocity vectors
  and the lower panel magnetic field vectors.  The initial
  magnetic field is along the $z$-direction, i.e.\ vertically oriented
  in all plots presented.  The field is strong enough in this case not
  only to prevent the cloud from collapsing perpendicular to the field
  lines, but even to suppress turbulent motions in the cloud. The
  turbulence barely affects the mean field. The density greyscale is
given in the colorbar, in model units. The time shown is
  $t=5.5t_{\rm ff}$. (From Heitsch \etal\ 2001a.)}
\end{figure}
It can be demonstrated (e.g.\ Heitsch \etal\ 2001a) that supersonic
turbulence does not cause a magnetostatically supported region to
collapse, and vice versa, that in the absence of magnetostatic
support, MHD waves cannot completely prevent collapse, although they
can retard it to some degree.  The case of a subcritical region with
$M < M_{cr}$ is illustrated in Figure \ref{fig:magstatsup}.
Indeed, sheets form, roughly
perpendicular to the field lines.  This is because the
turbulent driving can shift the sheets along the field lines without
changing the mass-to-flux ratio. The sheets do not collapse further,
because the shock waves cannot sweep gas across field lines and the
entire region is initially supported magnetostatically.

\begin{figure}
\begin{center}\unitlength1.0cm
\begin{picture}(14,14.0)
\put( 0.0,  0.0){\epsfxsize=14cm \epsfbox{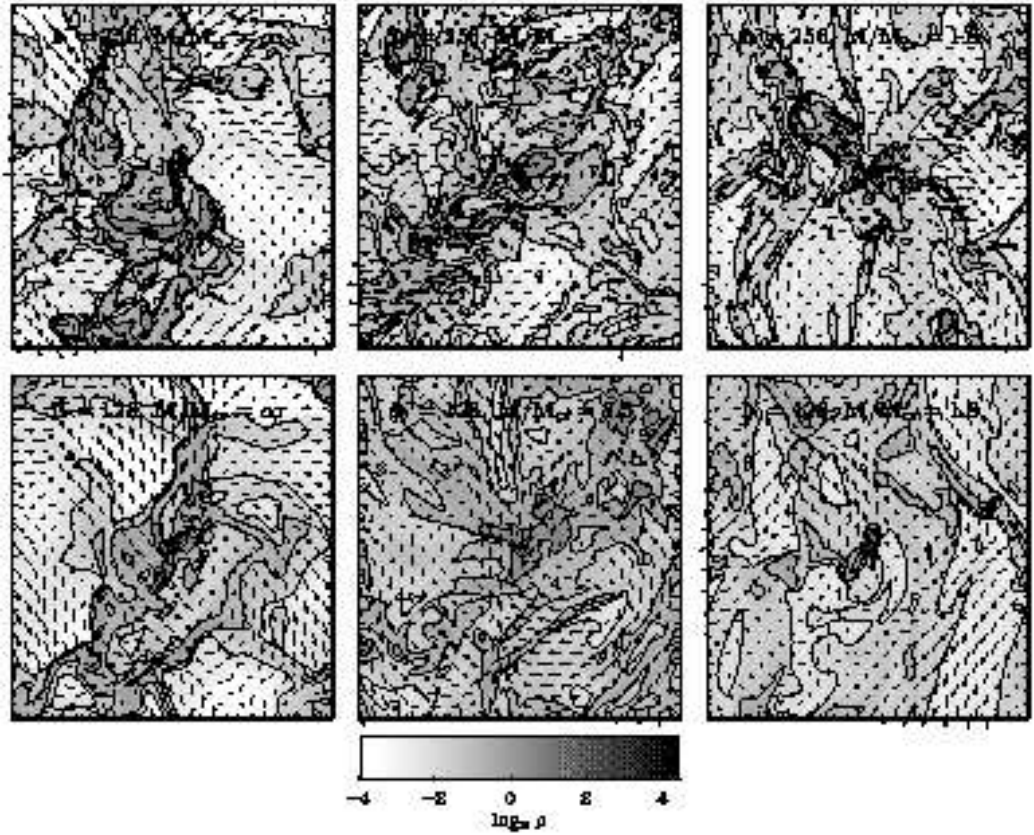}}
\end{picture}
\end{center}\caption{\label{fig:2Dslices}Two dimensional slices of $128^3$ and
$256^3$ models from Heitsch \etal\ (2001a) driven at large scales with
wavenumbers $k=1-2$ and high enough energy input that the mass in the
box represents only 1/15 $\langle M_{\rm J}\rangle_{\rm turb}$. The
magnetic field is initially vertical and strong enough to give
critical mass fractions as shown.  The slices are shown at the
location of the zone with the highest density, at the time when $10$\%
of the total mass has been accreted onto dense cores. The plot is
centered on this zone using the periodic boundary conditions of the
models.  Arrows show velocities in the plane. The longest arrows
corresponds to a velocity of $v \sim 20 c_s$. The density greyscale is
given in the colorbar, in model units. As magnetic fields become
stronger, they influence the flow more, producing anisotropic
structure. (From Heitsch \etal\ 2001a.)
}
\end{figure}
\begin{figure}[htp]
\begin{center}
\unitlength1.0cm
\begin{picture}(9.0, 10.0)
\put( 0.0, 0.0){\epsfxsize=8cm \epsfbox{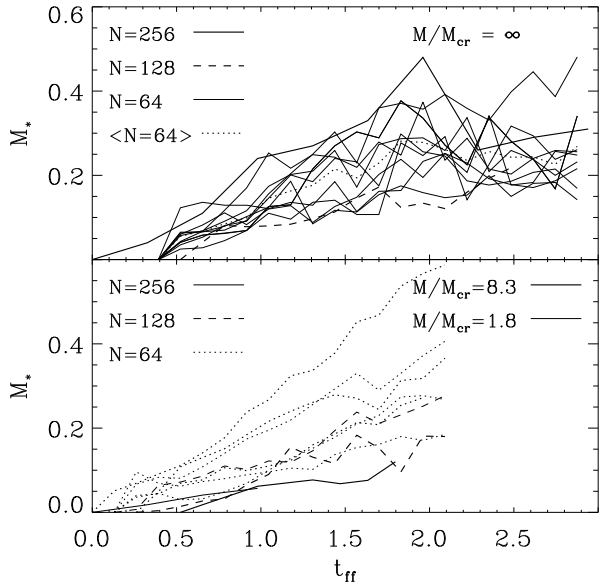}}
\end{picture}
\end{center}
\caption{\label{fig:variance}
  {\em Upper panel:} Core-mass accretion rates for $10$ different
  low-resolution models ($64^3$ zones) of purely gas dynamic
  turbulence with identical parameters but different realizations of
  the random turbulent driving. The dotted line shows a mean accretion
  rate calculated from averaging over the sample. For comparison,
  higher-resolution runs with identical parameters but $128^3$ (dashed
  line) and $256^3$ (thick solid line) zones are shown as well.
{\em Lower panel:} Mass accretion rates for various models with
  different magnetic field strength and resolution.  Common to all
  models is the occurrence of local collapse and star formation
  regardless of the detailed choice of parameters, as long as the
  system is magnetostatically supercritically.  (For further details
  see Heitsch \etal\ 2001a.)}
\end{figure}

A supercritical cloud with $M > M_{cr}$
could only be stabilized by MHD wave pressure.
This is insufficient to completely prevent gravitational collapse, as
shown
in Figure \ref{fig:2Dslices}.
Collapse occurs in all models of unmagnetized and magnetized
turbulence regardless of the numerical resolution and magnetic field
strength as long as the system is magnetically supercritical. This is
shown quantitatively in Figure~\ref{fig:variance}. Increasing the
resolution makes itself felt in different ways in gas dynamical and
MHD models.  In the gas dynamical case, higher resolution results in
thinner shocks and thus higher peak densities.  These higher density
peaks form cores with deeper potential wells that accrete more mass
and are more stable against disruption.  Higher resolution in the MHD
models, on the other hand, better resolves short-wavelength MHD waves,
which apparently can delay collapse, but not prevent it.  This result
extends to models with $512^3$ zones (Heitsch \etal\ 2001b, Li \etal\
2001).  The delay of local collapse seen in our magnetized simulations is
caused mainly by weakly magnetized turbulence acting to decrease
density enhancements due to shock interactions.

\subsection{Promotion and prevention of local collapse}
\label{sub:promotion}

Highly compressible turbulence both promotes and prevents collapse.
Its net effect is to inhibit collapse globally, while perhaps
promoting it locally.  This can be seen by examining the dependence of
the Jeans mass $M_{\rm J} \propto \rho^{-1/2} c_{\rm s}^3$, Eq.\
(\ref{eqn:jeans-mass}), on the rms turbulent velocity $v_{\rm rms}$.
If we follow the classical picture that treats turbulence as an
additional pressure (Chandrasekhar 1951a,b), then we define $c_{{\rm
s},e\!\!\!\;f\!\!f}^2 = c_{\rm s}^2 + v_{\rm rms}^2/3$, giving the
Jeans mass a dependence on velocity of $v_{\rm rms}^3$.  However,
compressible turbulence in an isothermal medium causes local density
enhancements that increase the density by ${\cal M}^2
\propto v_{\rm rms}^2$, adding a dependence $1/v_{\rm rms}$.
Combining these two effects, we find that 
\begin{equation}
M_{\rm J} \propto v_{\rm rms}^2
\end{equation}
for $v_{\rm rms} \gg c_{\rm s}$, so that ultimately turbulence does
inhibit collapse.  However, there is a broad intermediate region,
especially for long wavelength driving, where local collapse occurs
despite global support, as shown in
Figure~\ref{fig:3D-cubes-1..2+7..8}, which can be directly compared
with Figure~\ref{fig:3D-cubes}b).

The total mass and lifetime of a Jeans-unstable fluctuation determine
whether it will actually collapse.  Roughly speaking, the lifetime of
an unstable clump is determined by the interval between two successive
passing shocks: the first creates it, while the second one, if strong
enough, may disrupt the clump again (Klein, McKee \& Colella 1994, Mac
Low \etal\ 1994).  If the time interval between the two shocks is
sufficiently long, however, an unstable clump can contract to high
enough densities to effectively decouple from the ambient gas flow and
becomes able to survive the encounter with further shock fronts (e.g.\
Krebs \& Hillebrandt 1983). Then it continues to accrete from the
surrounding gas, forming a dense core.

A more detailed understanding of how local collapse proceeds comes
from examining the full time history of accretion for different models
(Figure \ref{fig:accretion-history}). 
The cessation of strong accretion onto cores occurs long before all
gas has been accreted, with the mass fraction at which this occurs
depending on the properties of the turbulence (V\'azquez-Semadeni
\etal\ 2003).  This is because the
time that dense cores spend in shock-compressed, high-density regions
decreases with increasing driving wave number and increasing driving
strength.  In the case of long wavelength driving, cores form
coherently in high-density regions associated with one or two large
shock fronts that can accumulate a considerable fraction of the total
mass of the system, while
in the case of short wavelength driving, the network of shocks is
tightly knit, and cores form in smaller clumps and remain in them for
shorter times.

\subsection{The timescales of star formation}

\begin{figure}[h]
\begin{center}
\unitlength1.0cm
\begin{picture}(12,12.0)
\put( 0.0, 0.0){\epsfxsize=13cm \epsfbox{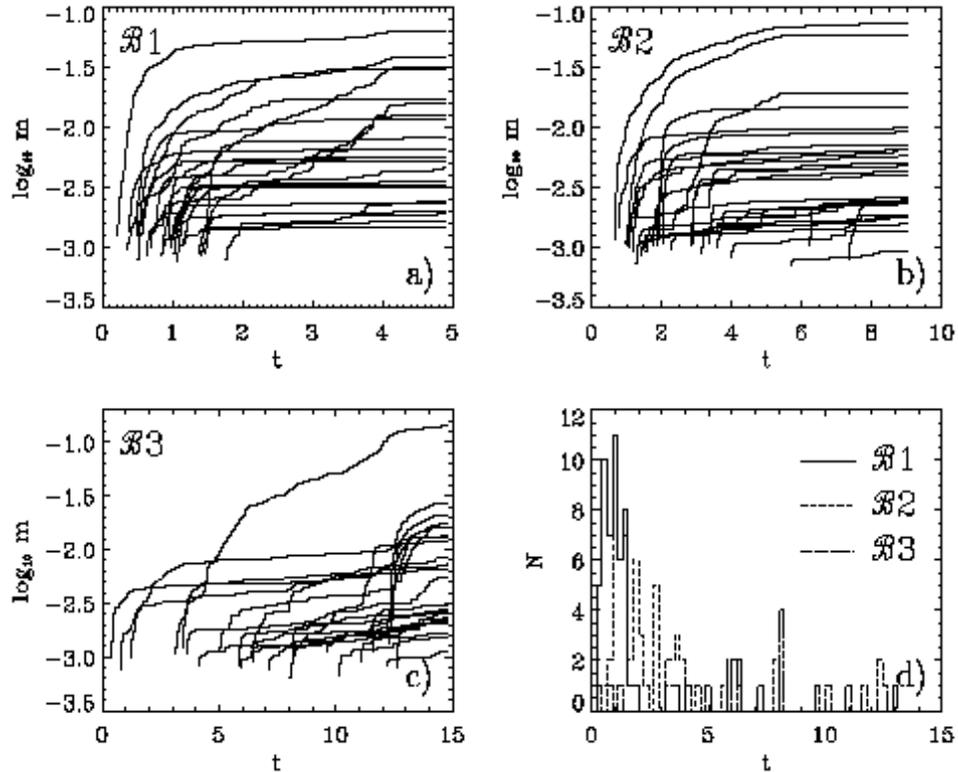}}
\end{picture}
\end{center}
\caption{\label{fig:core-formation-histogram} 
  Masses of individual protostars as function of time in SPH models
  (a) ${\cal B}1$ driven at large scales with driving wavenumber $k=1-2$, (b)
  ${\cal B}2$ with intermediate scale driving ($k=3-4$), and
  (c) ${\cal B}3$ with small scale driving ($k=7-8$).  
For the sake of clarity, only every other object is shown in (a) and
(b), whereas in (c) the evolution of every single object is plotted.
Time is given in units of the global free-fall time $\tau_{\rm
ff}$. Note the different timescale in each plot. In the depicted time
interval models ${\cal B}1$ and ${\cal B}2$ reach a core mass fraction
$M_* =70$\%, and both form roughly 50 cores.  Model ${\cal B}3$
reaches $M_* =35$\% and forms only 25 cores.  Figure (d) compares the
distributions of formation times. The age spread increases with
decreasing driving scale, showing that clustered core formation should
lead to a coeval stellar population, whereas a distributed stellar
population should exhibit considerable age spread. (From Klessen
\etal\ 2000.)  }
\end{figure}

Turbulent control of star formation predicts that stellar clusters
form predominantly in regions that are insufficiently supported by
turbulence or where only large-scale driving is active.  In the
absence of driving, molecular cloud turbulence decays on order of the
free-fall timescale $\tau_{\rm ff}$ (Eq.~\ref{eqn:decay}), which is
roughly the timescale for dense star clusters to form.  Even in the
presence of support from large-scale driving, substantial collapse
still occurs within a few free-fall timescales
(Figures~\ref{fig:accretion-history}
and~\ref{fig:core-formation-histogram}a).  If the dense cores followed
in these models continue to collapse to quickly build up stellar
objects in their centers, then this directly implies the star
formation timescale.  Therefore the age distribution will be roughly
$\tau_{\rm ff}$ for stellar clusters that form coherently with high
star formation efficiency.  When scaled to low densities, say $n({\rm
H}_2) \approx 10^2\,{\rm cm}^{-3}$ and $T\approx10\,$K, the global
free-fall timescale (eq.~\ref{eqn:free-fall-time}) in the models is
$\tau_{\rm ff} = 3.3$ Myr.

If star forming clouds such as Taurus indeed have ages of order
$\tau_{\rm ff}$, as suggested by Ballesteros-Paredes \etal\ (1999a),
then the star formation timescale computed here is quite consistent
with the low star formation efficiencies seen in Taurus (e.g.\
Leisawitz \etal\ 1989, Palla \& Stahler 2000, Hartmann 2001), as the
cloud simply has not had time to form many stars.  In the case of
high-density regions, $n({\rm H}_2) \approx 10^5\,{\rm cm}^{-3}$ and
$T\approx10\,$K, the dynamical evolution proceeds much faster and the
corresponding free-fall timescale drops to $\tau_{\rm ff} \sim
10^5\,$years.  These values are indeed supported by observational data
such as the formation time of the Orion Trapezium cluster.  It is
inferred to have formed from gas of density $n({\rm H}_2) \sil
10^5\,{\rm cm}^{-3}$, and is estimated to be less than $10^6$ yr old
(Hillenbrand \& Hartmann 1998).  The age spread in the models
increases with increasing driving wave number $k$ and increasing
effective turbulent Jeans mass $\langle M_{\rm J} \rangle _{\rm
turb}$, as shown in Figure~\ref{fig:core-formation-histogram}. Long
periods of core formation for globally supported clouds appear
consistent with the low efficiencies of star-formation in regions of
isolated star formation, such as Taurus, even if they are rather young
objects with ages of order $\tau_{\rm ff}$.

\subsection{Scales of interstellar turbulence}
\label{sub:scales}

Turbulence only has self-similar properties on scales between the
driving and dissipation scales.  What are these scales for
interstellar turbulence?
The driving scale is determined by what stirs the turbulence.  We
discuss different driving mechanisms in \S~\ref{sub:driving}, where we
conclude that supernovae are likely dominant in star-forming galaxies.
Norman \& Ferrara (1996) attempted to estimate from general analytic
arguments the driving scale of an ensemble of blast waves from
supernovae and superbubbles.  However, their description remains to be
tested against nonlinear models.  An outer limit to the turbulent
cascade in disk galaxies is given by the scale height. If indeed
molecular clouds are created at least in part by converging
large-scale flows generated by the collective influence of recurring
supernovae explosions in the gaseous disk of our Galaxy, as we argue
in \S~\ref{sub:clouds}, then the extent of the Galactic disk is indeed
the true upper scale of turbulence in the Milky Way. For individual
molecular clouds this means that turbulent energy is fed in at scales
well above the size of the cloud itself.  This picture of molecular
cloud turbulence being driven by large-scale, external sources is
supported by the observation that density and velocity structure shows
power-law scaling extending up to the largest scales observed in all
clouds that have been analyzed (Ossenkopf \& Mac Low 2002).

In a purely gas dynamic system the dissipation scale is the scale
where molecular viscosity becomes important.  In interstellar clouds
the situation may be different.  Zweibel \& Josafatsson (1983) showed
that ambipolar diffusion (discussed in \S~\ref{sub:standard}) is the
most important dissipation mechanism in typical molecular clouds with
very low ionization fractions $x = \rho_i/\rho_n$, where $\rho_i$ is
the density of ions, $\rho_n$ is the density of neutrals, and the
total density $\rho = \rho_i + \rho_n$.  An ambipolar diffusion
coefficient with units of viscosity can be defined as
\begin{equation}
\lambda_{\rm AD} = v_{\rm A}^2 / \nu_{ni},
\end{equation}
where $v_{\rm A}^2 = B^2/4\pi\rho_n$ approximates the effective
Alfv\'en speed for the coupled neutrals and ions if $\rho_n \gg
\rho_i$, and $\nu_{ni} = \alpha \rho_i$ is the rate at which each
neutral is hit by ions, where the coupling constant $\alpha$ is given
by equation~(\ref{eqn:couple}).  Zweibel \& Brandenburg (1997) define
an ambipolar diffusion Reynolds number by analogy with the normal
viscous Reynolds number as
\begin{equation}
R_{\rm AD} = \tilde{L}\tilde{V} / \lambda_{\rm AD} = {\cal M}_{\rm A} \tilde{L} \nu_{ni}/v_{\rm A},
\end{equation}
which must fall below unity for ambipolar diffusion to be important
(also see Balsara 1996), where $\tilde{L}$ and $\tilde{V}$ are the
characteristic length and velocity scales, and ${\cal M}_{\rm A} =
\tilde{V}/v_{\rm A}$ is the characteristic Alfv\'en Mach number. In
our situation we again can take the rms velocity as a typical value for
$\tilde{V}$.  By setting $R_{\rm AD} = 1$, we can derive a critical
lengthscale below which ambipolar diffusion is important
\begin{equation} \label{lcrit1}
\tilde{L}_{\rm cr} = \frac{v_{\rm A}}{{\cal M}_{\rm A} \nu_{ni}} 
         \approx (0.041 \mbox{ pc})\left(\frac{B}{10\,\mu{\rm G}}\right) {\cal M_{\rm
         A}}^{-1} \left(\frac{x}{10^{-6}}\right)^{-1}
         \left(\frac{n_n}{10^3\,{\rm cm}^{-3}}\right)^{-3/2},
\end{equation}
with the magnetic field strength $B$, the ionization fraction $x$, the
neutral number density $n_n$, and where we have taken $\rho_n = \mu
n_n$, with $\mu = 2.36\,m_{\rm H}$.  This is consistent with typical
sizes of protostellar cores (e.g.\ Bacmann \etal\ 2000), if we assume
that ionization and magnetic field both depend on the density of the
region and follow the empirical laws $n_i = 3 \times 10^{-3}\,{\rm
cm}^{-3}\,(n_n / 10^5\,{\rm cm}^{-3})^{1/2}$ (e.g.\ Mouschovias 1991b)
and $B \approx 30\,\mu{\rm G}\, (n_n/10^3\,{\rm cm}^{-3})^{1/2}$
(e.g.\ Crutcher 1999).  Balsara (1996) notes that there are wave
families that can survive below $L_{\rm cr}$ that resemble
gas dynamic sound waves.  This means that this scale may determine
where the magnetic field becomes uniform, but not necessarily where
the gas dynamic turbulent cascade cuts off.

\subsection{Termination of local star formation}
\label{sub:termination}
It remains quite unclear what terminates stellar birth on scales of
individual star forming regions, and even whether these processes are
the primary factor determining the overall efficiency of star
formation in a molecular cloud 
(eq.~\ref{eqn:sfe}).  
Three main possibilities exist.
First, feedback from the stars themselves in the form of ionizing
radiation and stellar outflows may heat and stir surrounding gas up
sufficiently to prevent further collapse and accretion.  Second,
accretion might peter out either when all the high density,
gravitationally unstable gas in the region has been accreted in
individual stars, or after a more dynamical period of competitive
accretion, leaving any remaining gas to be dispersed by the background
turbulent flow.  Third, background flows may sweep through, destroying
the cloud, perhaps in the same way that it was created. Most likely
the astrophysical truth lies in some combination of all three
possibilities.

If a stellar cluster formed in a molecular cloud contains OB stars,
then the radiation field and stellar winds from these high-mass stars
strongly influence the surrounding cloud material. The UV flux ionizes
gas out beyond the local star forming region. Ionization heats the
gas, raising its Jeans mass, and possibly preventing further
protostellar mass growth or new star formation.  The termination of
accretion by stellar feedback has been suggested at least since the
calculations of ionization by Oort \& Spitzer (1955). Whitworth (1979)
and Yorke \etal\ (1989) computed the destructive effects of individual
blister H{\sc ii} regions on molecular clouds, while in a series of
papers, Franco \etal\ (1994), Rodriguez-Gaspar \etal\ (1995), and
Diaz-Miller \etal\ (1998) concluded that indeed the ionization from
massive stars may limit the 
star formation efficiency (eq.~\ref{eqn:sfe}) of molecular clouds to
about 5\%.  Matzner (2002) analytically modeled the effects of
ionization on molecular clouds, concluding as well that turbulence
driven by H{\sc ii} regions could support and eventually destroy
molecular clouds.  The key question facing these models is whether
H{\sc ii} region expansion couples efficiently to clumpy,
inhomogeneous molecular clouds, a question probably best addressed
with numerical simulations.

Bipolar outflows are a different manifestation of protostellar
feedback, and may also strongly modify the properties of star forming
regions (Norman \& Silk 1980, Lada \& Gautier 1982, Adams \& Fatuzzo
1996).  Matzner \& McKee (2000) modeled the ability of
bipolar outflows to terminate low-mass star formation, finding that
they can limit star formation efficiencies to 30--50\%, although they
are ineffective in more massive regions.  How important these
processes are compared to simple exhaustion of available reservoirs of
dense gas (Klessen \etal\ 2000, V{\'a}zquez-Semadeni \etal\ 2003)
remains an important question.

The models relying on exhaustion of the reservoir of dense gas argue
that only dense gas will actually collapse, and that only a small
fraction of the total available gas reaches sufficiently high
densities, due to cooling (Elmegreen \& Parravano 1994, Schaye 2002),
gravitational collapse and turbulent triggering (Elmegreen 2002), or
both (Wada, Meurer, \& Norman 2002). This of course pushes the
question of local star formation efficiency up to larger scales, which
may indeed be the correct place to ask it.

Other models focus on competitive accretion in local star formation,
showing that the distribution of masses in a single group or cluster
can be well explained by assuming that star formation is fairly
efficient in the dense core, but that stars that randomly start out
slightly heavier tend to fall towards the center of the core and
accrete disproportionately more gas (Bonnell \etal\ 1997; 2001a).
These models have recently been called into question by the
observation that the stars in lower density young groups in Serpens
simply have not had the time to engage in competitive accretion, but
still have a normal IMF (Olmi \& Testi 2002).

Finally, star formation in dense clouds created by turbulent flows may
be terminated by the same flows that created them.  Ballesteros-Pardes
\etal\ (1999a) suggested that the coordination of star formation over
large molecular clouds, and the lack of post-T Tauri stars with ages
greater than about 10$\,$Myr tightly associated with those clouds, could
be explained by their formation in a larger-scale turbulent flow.
Hartmann \etal\ (2001) make the detailed argument that these flows may
disrupt the clouds after a relatively short time, limiting their star
formation efficiency that way. 
The extensive evidence for short sequences of cluster ages
(e.g. Blaauw 1964, Walborn \& Parker 1992, Efremov \& Elmegreen 1998b)
is often attributed to sequential triggering (Elmegreen \& Lada 1977)
by shock fronts expanding into existing clouds.  An additional
mechanism for producing such sequences may be the sequential formation
of clouds by the larger scale flow.  Below, in \S~\ref{sub:driving} we
argue that field supernovae are the most likely driver for the
background turbulence, at least in the star-forming regions of
galaxies.  Supernovae associated with any particular star-forming
region will not be energetically important, although they may produce
locally significant compressions.

\subsection{Outline of a new theory of star formation}
\label{sub:new}

The support of star-forming clouds by supersonic turbulence can
explain many of the same observations successfully explained by the
standard theory, while also addressing the inconsistencies between
observation and the standard theory described in the previous section.
The key point that is new in our argument is that supersonic
turbulence produces strong density fluctuations in the interstellar
gas, sweeping gas up from large regions into dense sheets and
filaments, and does so even in the presence of magnetic fields.
Supersonic turbulence decays quickly, but so long as it is maintained
by input of energy from some driver it can support regions against
gravitational collapse.

Such support comes at a cost, however.  The very turbulent flows that
support the region produce density enhancements in which the Jeans
mass drops as $M_{\rm J} \propto \rho^{-1/2}$ (Eq.\
\ref{eqn:jeans-mass}), and the magnetic critical mass above which
magnetic fields can no longer support against that collapse drops even
faster, as $M_{\rm cr} \propto \rho^{-2}$ (Eq.\ \ref{eqn:crit-rho}).
For local collapse to actually result in the formation of stars,
Jeans-unstable, shock-generated, density fluctuations must collapse to
sufficiently high densities on time scales shorter than the typical
time interval between two successive shock passages.  Only then can
they decouple from the ambient flow and survive subsequent shock
interactions.  The shorter the time between shock passages, the less
likely these fluctuations are to survive. Hence, the timescale and
efficiency of protostellar core formation depend strongly on the
wavelength and strength of the driving source, and the accretion
histories of individual protostars are strongly time varying
(\S~\ref{sub:accretion}).  Global support by supersonic turbulence
thus tends to produce local collapse
and low rate star formation,
exactly as seen in low-mass star
formation regions characteristic of the disks of spiral galaxies.
Conversely, lack of turbulent support results in regions that collapse
freely.  In gas dynamic simulations,
freely collapsing gas forms a web of density
enhancements in which star formation can proceed efficiently, as seen
in regions of massive star formation and starbursts.

The regulation of the star formation rate then occurs not just at the
scale of individual star-forming cores through ambipolar diffusion
balancing magnetostatic support, but rather at all scales 
via the dynamical processes that determine whether regions of
gas become unstable to prompt gravitational collapse.  Efficient star
formation occurs in collapsing regions; apparent inefficiency occurs
when a region is turbulently supported and only small subregions get
compressed sufficiently to collapse.  The star formation rate is
determined by the balance between turbulent support and local density,
and is a continuous function of the strength of turbulent support for
any given region.   Fast and efficient star formation is the natural
behavior of gas lacking sufficient turbulent support for its local
density.

Regions that are gravitationally unstable in this picture collapse
quickly, on the free-fall timescale.
They never pass
through a quasi-equilibrium state as envisioned by the standard model.
Large-scale density enhancements
such as molecular clouds could be caused either by gravitational
collapse, or by ram pressure from turbulence. 
If collapse does not succeed, the same large-scale
turbulence that formed molecular clouds can destroy them again.

\section{LOCAL STAR FORMATION}
\label{sec:local}

In this section we apply the theoretical picture of
\S~\ref{sec:paradigm} to observations of individual star forming
regions. We show how the efficiency and time and length scales of star
formation depend on the properties of turbulence
(\S~\ref{sub:multiple}), followed by a discussion of the properties of
protostellar cores (\S~\ref{sub:cores}), the immediate progenitors of
individual stars.  We then speculate about the formation of binary
stars (\S\ \ref{sub:binary}), and stress the importance of the dynamical
interaction between protostellar cores and their competition for mass
growth in dense, deeply embedded clusters
(\S~\ref{sub:clusters}). This implies strongly time-varying
protostellar mass accretion rates (\S~\ref{sub:accretion}).  Finally,
we discuss the consequences of the probabilistic processes of
turbulence and stochastic mass accretion for the resulting stellar
inital mass function (\S~\ref{sub:imf}).

\subsection{Star formation in molecular clouds}
\label{sub:multiple}

Not only does all star formation occur in molecular clouds, but 
all giant molecular clouds appear to form stars.  At least, all those
surveyed within distances less than 3$\,$kpc form stars (Blitz 1993,
Williams \etal\ 2000), except possibly the Maddalena \& Thaddeus
(1985) cloud (Lee, Snell, \& Dickman 1996; Williams \& Blitz 1998),
and this last cloud may have formed just recently.

The star formation process in molecular clouds appears to be fast.
Once the collapse of a cloud region sets in, it rapidly forms an
entire cluster of stars within $10^6$ years or less. This is indicated
by the young stars associated with star forming regions,
typically T~Tauri stars with ages less than $10^6$ years (e.g.\ Gomez
\etal\ 1992, Green \& Meyer 1996, Carpenter \etal\ 1997, Hartmann
2001), and by the small age spread in more evolved stellar clusters
(e.g.\ Hillenbrand 1997, Palla \& Stahler 1999, 2001). 
Star clusters in the Milky Way also exhibit an amazing degree of
chemical homogeneity (in the case of the Pleiades, see Wilden \etal\
2002), implying that the gas
out of which these stars formed must have been chemically well-mixed
initially (see also Avillez \& Mac~Low 2002, Klessen \& Lin 2003).   

Star-forming molecular clouds in our Galaxy vary enormously
in size and mass. In small, low-density, clouds stars form with low
efficiency, more or less in isolation or scattered around in small
groups of up to a few dozen members. Denser and more massive clouds
may build up stars in associations and clusters of a few hundred
members.  This appears to be the most common mode of star formation in
the solar neighborhood (Adams \& Myers 2001). Examples of star
formation in small groups and associations are found in the
Taurus-Aurigae molecular cloud (e.g.\ Hartmann 2002). Young stellar
groups with a few hundred members form in the Chamaeleon I dark cloud
(e.g.\ Persi \etal\ 2000) or $\rho$-Ophiuchi (Bontemps \etal\
2001). Each of these clouds is at a distance of about 130 to 160$\,$pc
from the Sun.  Many nearby star forming regions have been associated
with a ring-like structure in the Galactic disk called Gould's belt
(P{\"o}ppel 1997), although its reality remains uncertain.

The formation of dense rich clusters with thousands of stars is rare.
The closest molecular cloud where this happens 
is the Orion Nebula Cluster in L1641 (Hillenbrand 1997; Hillenbrand \&
Hartmann 1998), which lies at a distance of $\sim 450\,$pc.  A rich
cluster somewhat further away is associated with the Monoceros R2
cloud (Carpenter \etal\ 1997) at a distance of $\sim 830\,$pc.  The
cluster NGC~3603 is roughly ten times more massive than the Orion
Nebula Cluster.  It lies in the Carina region, at about $7\,$kpc
distance. It contains about a dozen O stars, and is the nearest object
analogous to a starburst knot (Brandl \etal\ 1999, Moffat \etal\
2002). To find star-forming regions building up hundreds
of O stars one has to look towards giant extragalactic 
H{\sc ii}-regions, the nearest of which is 30 Doradus in the Large 
Magellanic Cloud, a satellite galaxy of our Milky Way at a distance at
55$\,$kpc (for an overview see the book edited by Chu \etal\ 1999). The giant
star forming region 30 Doradus is thought to contain up to a hundred
thousand young stars, including more than 400 O stars (Hunter \etal\
1995; Walborn \etal\ 1999). Even more massive star forming regions are
associated with tidal knots in interacting galaxies, as observed in
the Antennae (NGC~4038/8, see e.g.\ Zhang, Fall, \& Whitmore 2001) or
as inferred for starburst galaxies at high redshift (Sanders \&
Mirabel 1996).

This sequence demonstrates that the star formation process spans many
orders of magnitude in scale, ranging from isolated single
stars ($M\approx 1$~M$_{\odot}$) to ultra-luminous starburst galaxies
with masses of several $10^{11}$M$_{\odot}$ and star formation rates of
$10^2$--$10^3$~M$_{\odot}$~yr$^{-1}$; for comparison the
present-day rate in the Milky Way is about 1~M$_{\odot}$~yr$^{-1}$.
This enormous variety of star
forming regions appears to be controlled by 
the competition between self-gravity and the turbulent
velocity field in interstellar gas (see \S~\ref{sub:applications}). 

The control of star formation by supersonic turbulence gives rise to a
continuous but articulated picture. There may not be physically
distinct modes of star formation, but qualitatively different
behaviors do appear over the range of possible turbulent flows. The
apparent dichotomy between a clustered and an isolated mode of star
formation, as discussed by Lada (1992) for L1630 and Strom, Strom, \&
Merrill (1993) for L1941, disappears, if a different balance between
turbulent strength and gravity holds at the relevant length scales in
these different clouds.

Turbulent flows tend to have hierarchical structure (e.g.\ She \&
Leveque 1994), which may explain the hierarchical distribution of stars
in star forming regions shown by statistical studies of the
distribution of neighboring stars in young stellar clusters (e.g.\
Larson 1995; Simon 1997; Bate, Clarke, \& McCaughrean 1998, Nakajima
\etal\ 1998; Gladwin \etal\ 1999; Klessen \& Kroupa 2001).
Hierarchical clustering seems to be a common feature of all star
forming regions (e.g.\ Efremov \& Elmegreen 1998a). It may be a natural
outcome of turbulent fragmentation 
(e.g.\ Padoan \& Nordlund 2002).

\begin{figure}[ht]
\unitlength1.0cm
\begin{picture}(16,10.0)
\put(0.0,-2.0){\epsfxsize=9.5cm \epsfbox{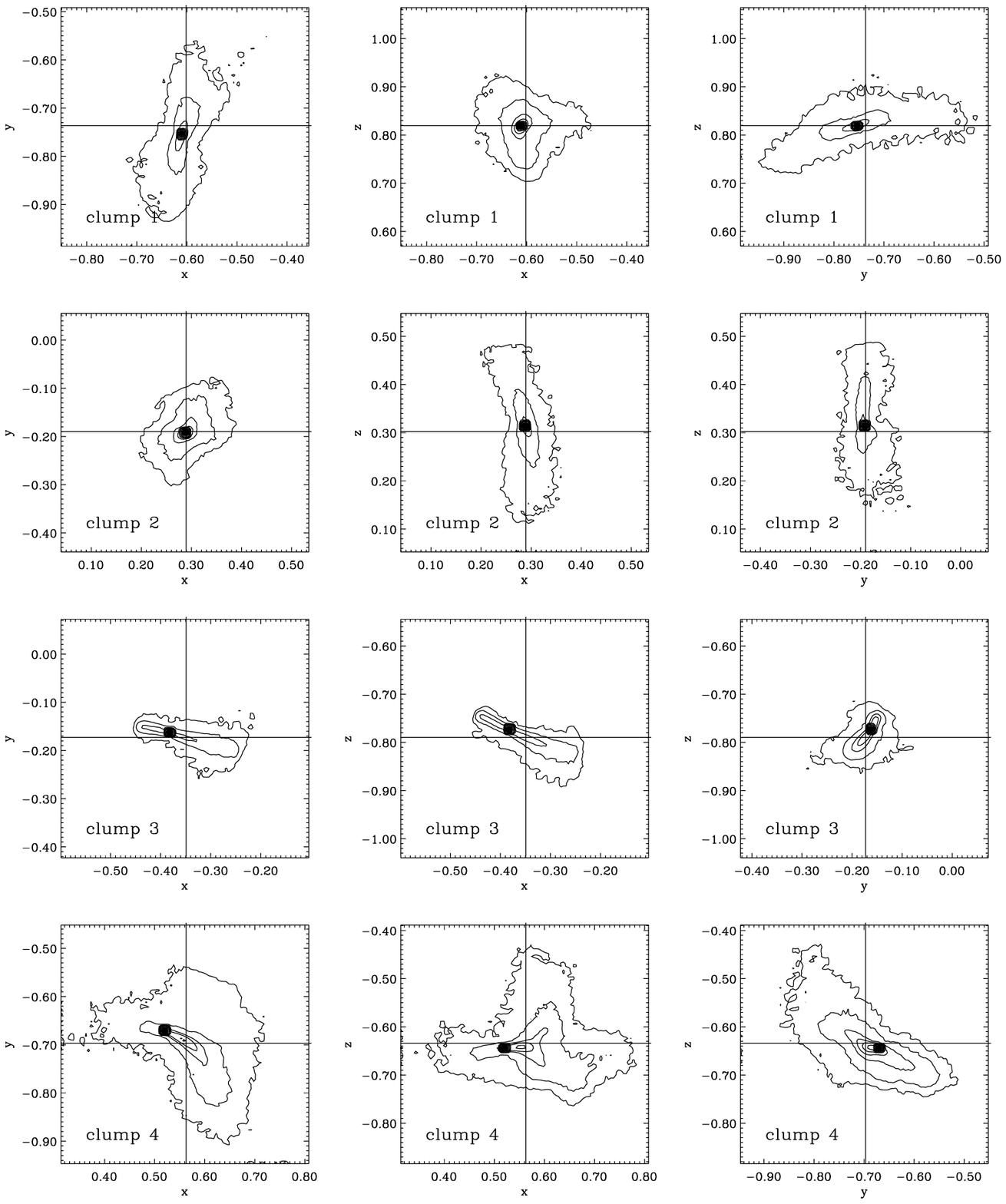}}
\put(8.0,-2.0){\epsfxsize=9.5cm \epsfbox{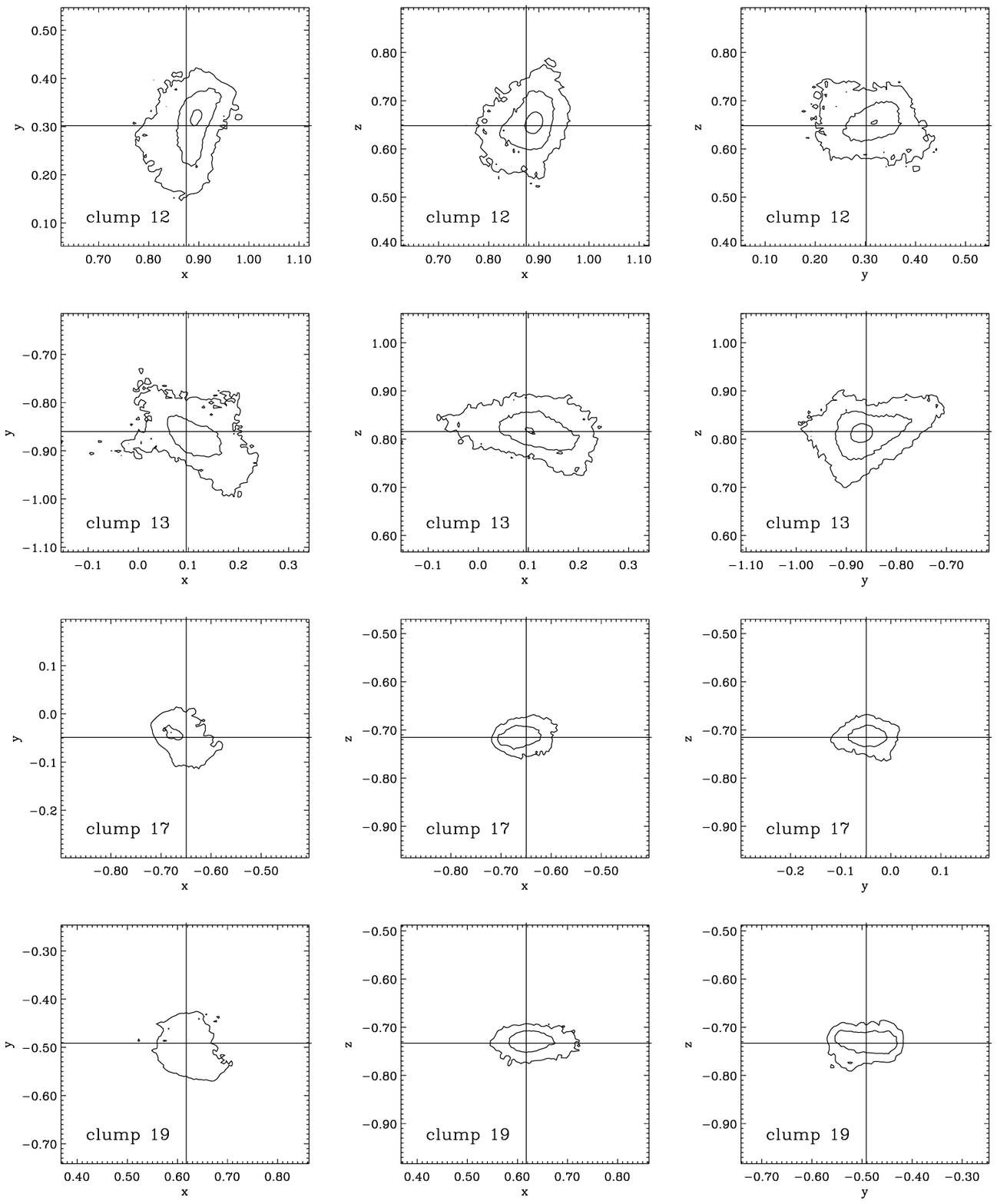}}
\end{picture}
\caption{\label{fig:ind-clumps}Protostellar cores from a model of
  clustered star formation. The left side shows protostellar cores
  with collapsed central objects (indicated by a black dot), the right
  side ``starless'' cores without central protostars.  Cores are numbered
  accoring to their peak density. Surface density contours are spaced
  logarithmically with two contour levels spanning one decade,
  $\log_{10}\Delta \rho = 0.5$. The lowest contour is a factor of
  $10^{0.5}$ above the mean density. (From Klessen \& Burkert 2000.)
  }
\end{figure}

\subsection{Protostellar core models}
\label{sub:core-models}

The observed properties of molecular cloud cores as discussed in
Section \ref{subsub:core-obs} can be compared with gas clumps
identified in numerical models of interstellar cloud turbulence. Like
their observed counterparts, the model cores are generally highly
distorted and triaxial.  Depending on the projection angle, they often
appear extremely elongated, being part of a filamentary structure
that may connect several objects. Figure \ref{fig:ind-clumps} plots a
sample of model cores from Klessen \& Burkert (2000). Those which have
already formed a protostellar object in their interior are shown on
the left, while starless cores without central protostars are shown on
the right. Note the similarity to the appearance of observed
protostellar cores (Figure \ref{fig:core-shapes}). The model clumps
are clearly elongated.  The ratios between the semi-major and the
semi-minor axis measured at the second contour level are typically
between 2:1 and 4:1. However, there are significant deviations from
simple triaxial shapes.  As a general trend, high density contour
levels typically are regular and smooth, because there the gas is
mostly influenced by pressure and gravitational forces.  On the other
hand, the lowest contour level samples gas that is strongly influenced by
environmental effects. Hence, it appears patchy and irregular. The
location of the protostar is not necessarily identical with the
center of mass of the core, especially when it is irregularly shaped.

\begin{figure}[ht]
\unitlength01.0cm
\begin{picture}(16,6)
\put( 2.500, 0.300){\epsfxsize=12.0cm \epsfbox{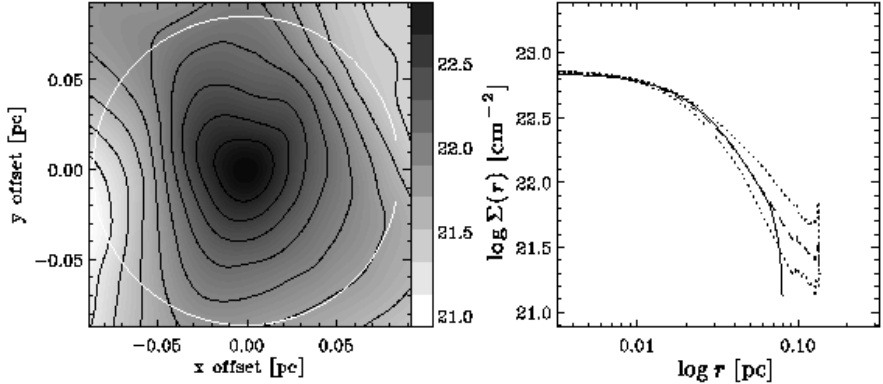}}
\end{picture}
\caption[F13]{\label{fig:BE-fit} Typical prestellar core in numerical
model of turbulent molecular cloud fragmentation. The projected
column density distribution is shown in the left panel, while the resulting
radial profile $\Sigma(r)$ on the right-hand side is plotted as a dashed
line. The maximum deviations from spherical symmetry are given by the
dotted lines. The best-fit Bonnor Ebert profile within the area
enclosed by the white circle in the left image is given by the solid
line.  For details see Ballesteros-Paredes \etal\ (2003).  Compare
with the observed core L1689B in Figure \ref{fig:L1689B}. }
\end{figure}
The surface density profiles of 
observed protostellar cores are
often interpreted in terms of equilibrium Bonnor-Ebert spheres (Ebert
1955, Bonnor 1956). The best example is the Bok globule B68 (Alves
\etal\ 2001).  However, it is a natural prediction of turbulent
fragmentation calculations that stars form from cloud cores with
density profiles that are constant in the inner region, then exhibit
an approximate power-law fall-off, and that appear finally truncated
at some maximum radius. Such a configuration can easily be
misinterpreted as Bonnor-Ebert sphere, as illustrated in Figure
\ref{fig:BE-fit}.  Indeed, Ballesteros-Paredes, Klessen, \&
V\'azquez-Semadeni (2003) argue that turbulent fragmentation produces
cores that in about 60\% of all cases fit well to Bonnor-Ebert
profiles, of which most imply stable equilibrium conditions.
However, supersonic turbulence 
does not
create hydrostatic equilibrium configurations. Instead, the density
structure is transient and dynamically evolving, as the different
contributions to virial  
equilibrium do not balance
(V{\'a}zquez-Semadeni \etal\ 2003).  None of the cores analyzed by
Ballesteros-Paredes \etal\ (2002) 
are in true equilibrium, and the physical properties inferred from
fitting Bonnor-Ebert profiles generally deviate from the real values.

Besides the direct comparison of projected surface density maps, there
is additional evidence supporting the idea of the turbulent origin of
the structure and kinematics of molecular cloud cores and clouds as a
whole.  This includes comparisons of numerical models of supersonic
turbulence to stellar extinction measurements (Padoan \etal\ 1997),
Zeeman splitting measurements (e.g.\ Padoan \& Nordlund 1999),
polarization maps (e.g.\ Padoan \etal\ 2001a, Heitsch \etal\ 2001b,
Ostriker, Stone, \& Gammie 2001), determination of the velocity
structure of dense cores and their immediate environment (e.g.\ Padoan
\etal\ 2001b), and various other statistical measures of structure and
dynamics of observed clouds (as mentioned in \S~\ref{sub:LSS}).

The density profiles of observed cores can also be fit by invoking
confinement by helical magnetic fields (Fiege \& Pudritz 2000a,b).
Helical field structures unwind, though, as magnetic tension forces
straighten field lines.  Therefore these models require external
forces to continuously exert strong torques in order to twist the
field lines.  These would necessarily drive strong flows, making it
impossible to achieve the static equilibrium configurations required
by the model.

Other models that have been proposed to describe the properties of
protostellar cores are based on quasi-static equilibrium conditions
(e.g.\ with a composite polytropic equation of state, as discussed by
Curry \& McKee 2000); invoke thermal instability (e.g.\ Yoshii \&
Sabano 1980, Gilden 1984b, Graziani \& Black 1987, Burkert \& Lin
2000); gravitational instability through ambipolar diffusion (e.g.\
Basu \& Mouschovias 1994, Nakamura, Hanawa, \& Nakano 1995, Indebetouw
\& Zweibel 2000, Ciolek \& Basu 2000); nonlinear Alfv{\'e}n waves
(e.g.\ Carlberg \& Pudritz 1990, Elmegreen 1990, 1997a, 1999b); or
rely on clump collisions (e.g., Gilden 1984a, Murray \& Lin 1996,
Kimura \& Tosa 1996). Models based on supersonic turbulence
as discussed here appear to be
most consistent with observational data.

\subsection{Binary formation}
\label{sub:binary}
In order to study binary formation, stellar collapse has to be studied
in multiple dimensions. Dynamical modeling in 2D has the advantage of
speed compared to 3D simulations, and therefore allows for the
inclusion of a larger number of physical processes while reaching
higher spatial resolution.  Early 2D calculations were reported by
Larson (1972), Tscharnuter (1975), Black \& Bodenheimer (1976),
Fricke, Moellenhoff, \& Tscharnuter (1976), Nakazawa, Hayashi, \&
Takahara (1976), Bodenheimer \& Tscharnuter (1979), Boss (1980a), and
Norman, Wilson, \& Barton (1980).  The disadvantage of 2D models is
that only axisymmetric perturbations can be studied.  Initial attempts
to study collapse in 3D were reported by Cook \& Harlow (1978),
Bodenheimer \& Boss (1979), Boss (1980b), Rozyczka \etal\ (1980), and
Tohline (1980). Since these early studies, numerical simulations of
the collapse of isolated isothermal objects have been extended, for
example, to include highly oblate cores (Boss, 1996), elongated
filamentary cloud cores (e.g.\ Bastien \etal\ 1991; Tomisaka 1995,
1996a; Inutsuka \& Miyama, 1997), differential rotation (Boss \&
Myhill, 1995), and different density distributions for the initial
spherical cloud configuration with or without bar-like perturbations
(Burkert \& Bodenheimer 1993; Klapp, Sigalotti, \& de~Felice 1993;
Burkert \& Bodenheimer 1996; Bate \& Burkert 1997; Burkert, Bate, \&
Bodenheimer 1997; Truelove \etal\ 1997, 1978; Tsuribe \& Inutsuka
1999a; Klein 1999; Boss \etal\ 2000).  Whereas 1D spherical collapse
models could only treat the formation of single stars, the 2D and 3D
calculations show that the formation of binary and higher-order
multiple stellar systems can be described in terms of the classical
dynamical theory.  Indeed, they show that it is a likely outcome of
protostellar collapse and molecular cloud fragmentation. For a
comprehensive overview see Bodenheimer \etal\ (2000).

The observed fraction of binary and multiple stars relative to single
stars is about 50\% for the field star population in the solar
neighborhood. This has been determined for all known F7--G9 dwarf
stars within 22 pc from the Sun by Duquennoy \& Mayor (1991) and for M
dwarfs out to similar distances by Fischer \& Marcy (1992; also
Leinert \etal\ 1997).  The binary fraction for pre-main sequence stars
appears to be at least equally high (see e.g.\ K\"ohler \& Leinert
1998, or Table 1 in Mathieu \etal\ 2000).  These findings put strong
constraints on the theory of star formation, as {\em any} reasonable
model needs to explain the observed high number of binary and multiple
stellar systems. It has long been suggested that sub-fragmentation and
multiple star formation is a natural outcome of isothermal collapse
(Hoyle 1954). However, stability analyses show that the growth time of
small perturbations in the isothermal phase is typically small
compared to the collapse timescale itself (e.g.\ Silk \& Suto 1988;
Hanawa \& Nakayama 1997).

Hence, in order to form multiple stellar systems, either perturbations
to the collapsing core must be external and strong, or
subfragmentation must occur at a later, non-isothermal phase of
collapse, after a protostellar disk has formed. This disk may become
gravitationally unstable if the surface density exceeds a critical
value given by the epicyclic frequency and the sound speed (Safranov,
1960; Toomre 1964; 
see derivation in \S~\ref{subsub:disk-grav}),
allowing fragmentation into multiple objects (as summarized by
Bodenheimer \etal\ 2000).

Contracting gas clumps with strong external perturbation occur
naturally in turbulent molecular clouds or when stars form in
clusters.  While collapsing to form or feed protostars, clumps may
lose or gain matter from interaction with the ambient turbulent flow
(Klessen \etal\ 2000). In a dense cluster environment, collapsing
clumps may merge to form larger clumps containing multiple
protostellar cores that subsequently compete with each other for
accretion from the common gas environment (Murray \& Lin, 1996;
Bonnell \etal\ 1997, Klessen \& Burkert, 2000, 2001). Strong external
perturbations and capture through clump merger leads to {\em wide}
binaries or multiple stellar systems. Stellar aggregates with more
than two stars are dynamically unstable, so protostars may
be ejected again from the gas rich environment they accrete from.
This not only terminates their mass growth, but leaves the remaining
stars behind more strongly bound. These dynamical effects can
transform the original wide binaries into close binaries (see also
Kroupa 1995a,b,c). Binary stars that form through disk fragmentation
are close binaries right from the beginning, as typical sizes of
protostellar disks are of order of a few hundred$\,$AU (e.g.\ Bate, Bonnell, \&
Bromm 2002b).

Magnetic fields above all influence the development of close binaries,
as magnetic braking acts during the collapse of protostellar cores
(\S~\ref{sub:standard}). Preliminary 3D models of the
collapse of a magnetized, rotating cloud, described in Balsara (2001),
already demonstrate that braking can drain enough angular momentum to
prevent the formation of a binary from a core that would form a binary
in the absence of magnetic field.  This effect is not captured by the
simple inclusion of an additional pressure term to model the magnetic
field, as done by Boss (2000, 2002), as it is the magnetic tension
that brakes 
rotation.  
The observed prevalence of binaries
suggests that the magnetic field must indeed decouple from collapsing
cores at some stage, presumably by ambipolar diffusion.  However,
quantitative models of this process remain to be done.

\subsection{Dynamical interactions in clusters}
\label{sub:clusters}

Star forming regions can differ enormously in scale and density as a
consequence of supersonic turbulence (as discussed in
\S~\ref{sub:multiple}). Stars almost never form in isolation, but
instead in groups and clusters.  The number density of protostars and
protostellar cores in rich compact clusters can be high enough for
mutual dynamical interactions to become important.  This introduces a
further degree of stochasticity to the star formation process in dense
clusters beyond the statistical chaos associated with turbulence and
turbulent fragmentation in the first place.

\begin{figure}[htp]
\unitlength1cm
\begin{picture}(16,5.5)
%
 \put( 0.00,0.50){\epsfxsize=16cm  \epsfbox{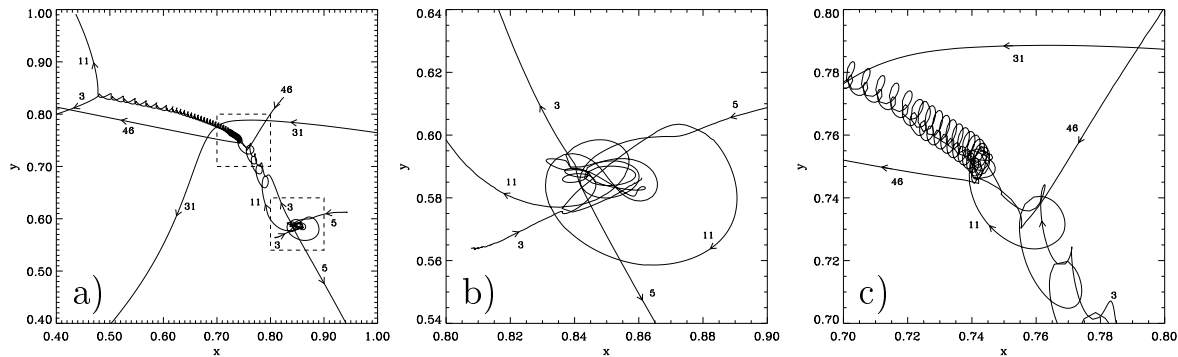}}
\end{picture}
\caption{\label{fig:trajectory} Example of protostellar interactions
in an embedded nascent star cluster.  (a) The projected trajectories
of five accreting cores in a numerical model of star cluster formation
by Klessen \& Burkert (2000).  We highlight two events in the
evolutionary sequence: (b) the formation of an unstable triple system
at the beginning of cluster formation with the lowest mass member
being expelled from the cluster, and (c) binary hardening in a close
encounter together with subsequent acceleration of the resulting close
binary due to another distant encounter during late evolution. The
corresponding parts of the orbital paths are enlarged by a factor of
six. For simplicity, neither the trajectories of other cores in the
cluster nor the distribution of gas are shown. Numbers next to the
trajectories identify the protostellar cores. For further detail see
Klessen \& Burkert (2000).  }
\end{figure}

When a molecular cloud region of a few hundred solar masses or more
coherently becomes gravitationally unstable, it contracts and builds
up a dense cluster of embedded protostars within one or two free-fall
timescales. While contracting individually to build up a star in their
interior, protostellar gas clumps still follow the global flow
patterns. They stream towards a common center of attraction, undergo
further fragmentation, or merge together. The timescales for clump
mergers and clump collapse are comparable. Merged clumps therefore can
contain multiple protostars that now compete with each other for
further accretion. They are embedded in the same limited and rapidly
changing reservoir of contracting gas. As the cores are dragged along
with the global gas flow, a dense cluster of accreting protostellar
cores quickly builds up.  Analogous to dense stellar clusters, the
dynamical evolution is subject to the complex gravitational
interaction between the cluster members. Close encounters or even
collisions occur, drastically altering the protostellar orbits. Triple
or higher-order systems often form. They are generally unstable, so a
large fraction of protostellar cores are expelled from the parental
cloud. The expected complexity of protostellar dynamics already in the
deeply embedded phase of evolution is illustrated in Figure
\ref{fig:trajectory}, which shows trajectories of five accreting
protostars in a calculation of molecular cloud fragmentation and
clustered star formation by Klessen \& Burkert (2000).

The effects of mutual dynamical interaction of protostellar cores in
the embedded phase of star cluster formation have been investigated by
a variety of authors. Here, we list some basic results:

{\em (a)} Close encounters in nascent star clusters strongly influence
the accretion disk expected to surround every protostar. These disks
can be tidally truncated or even disrupted. This influences mass
accretion through the disk, modifies the ability to subfragment and
form a binary star, and the probability of planet formation (e.g.\
Clarke \& Pringle 1991; Murray \& Clarke 1993 ; McDonald \& Clarke
1995; Hall \etal\ 1996; Scally \& Clarke 2001; Kroupa \& Burkert 2001;
Smith \& Bonnell 2001; Bonnell \etal\ 2001c). In particular, Ida,
Larwood, \& Burkert (2000) note that an early stellar encounter may
explain features of our own solar system, namely the high
eccentricities and inclinations observed in the outer part of the
Edgeworth-Kuiper Belt at distances larger than $42\,$AU.

{\em (b)} Stellar systems with more than two members are in general
unstable. In a triple system, for example, the lowest-mass member has
the highest probability to be expelled. If this happens during the
embedded phase, the protostar leaves a region of high-density gas.
This terminates further mass growth and sets its final mass. Thus, the
dynamical processes have important consequences for the resulting
stellar mass spectrum in dense stellar clusters (see
\S~\ref{sub:imf}).  Ejected objects can travel quite far, and indeed
this has been suggested to account for the so called ``run away''
T-Tauri stars found in X-ray observation in the vicinities of
star-forming molecular clouds (e.g.\ Sterzik \& Durison 1995, 1998,
Smith \etal\ 1997; Klessen \& Burkert 2000; or for observations e.g.\
Neuh{\"a}user \etal\ 1995; or Wichmann \etal\ 1997).  However, many of
these stars could not have traveled to their observed positions in
their own lifetimes if they were formed in the currently star-forming
cloud It appears more likely that the observed extended stellar
population is associated with clouds that dispersed long ago.
 
{\em (c)} Dynamical interaction leads to mass segregation. Star
clusters evolve towards equipartition. For massive stars this means
that they have on average smaller velocities than low-mass stars (in
order to keep the kinetic energy $E_{\rm kin} = 1/2\, m v^2$ roughly
constant). Thus, massive stars sink towards the cluster center,
while low-mass stars predominantly populate large cluster radii
(e.g.\ Kroupa 1995a,b,c). This holds already for nascent star
clusters in the embedded phase (e.g.\ Bonnell \& Davis 1998). 

{\em (d)} Dynamical interaction and competition for mass accretion
lead to highly time-variable protostellar mass growth rates. This is 
discussed in more detail in \S~\ref{sub:accretion}.

{\em (e)} The radii of stars in the pre-main sequence contraction
phase are several times larger than stellar radii on the main sequence
(for a review on pre-main sequence evolution see, e.g., Palla 2000,
2002).  Stellar collisions are therefore more likely to occur during the
very early evolution of star clusters. During the embedded phase the
encounter probability is further increased by gas drag and dynamical
friction. 

Collisions in dense protostellar clusters have 
been proposed as mechanism to produce massive stars (Bonnell, Bate, \&
Zinnecker 1998; Stahler, Palla, \& Ho 2000). The formation of massive
stars has long been considered a puzzle in theoretical astrophysics,
because 1D calculations predict for stars above $\sim
10\,$M$_{\odot}$ the radiation pressure acting on the infalling dust
grains to be strong enough to halt or even revert further mass
accretion (e.g.\ Yorke \& Kr{\"u}gel 1977; Wolfire \& Cassinelli 1987;
or Palla 2000, 2001). However, detailed 2D calculations
by Yorke \& Sonnhalter (2002) demonstrate that in the more realistic
scenario of mass growth via an accretion disk the radiation barrier
may be overcome. Mass can accrete from the disk onto the star along
the equator while radiation is able to escape along the polar direction.
Massive stars, thus, may form via the same processes as ordinary
low-mass stars 
(also see McKee \& Tan 2002). 
Collisional processes need not be invoked.

\subsection{Accretion rates}
\label{sub:accretion}

When a gravitationally unstable gas clump collapses 
onto a
central star, it follows the observationally well-determined sequence
described in \S~\ref{sub:cores}.  
In the main accretion phase (class 0), the release of gravitational energy by
accretion dominates the energy budget. Hence, protostars exhibit large
IR and sub-millimeter luminosities and drive powerful outflows.  Both
phenomena can be used to estimate the protostellar mass accretion rate
$\dot{M}$ (e.g.\ Andr{\'e} \& Montmerle 1994, Bontemps \etal\ 1996,
Henriksen, Andr{\'e}, \& Bontemps 1997). These observations suggest
that $\dot{M}$ varies strongly, 
and declines with time after the class 0 phase (\S~\ref{para:accretion}). 
The estimated lifetimes are a few tens of thousands of years for the
class 0 and a few hundreds of thousands of years for the class I
phase.

\begin{figure}[tp]
\unitlength1cm
\begin{center}
\begin{picture}(10.0,10.0)
\put(  0.0,  0.0){\epsfxsize=9.2cm \epsfbox{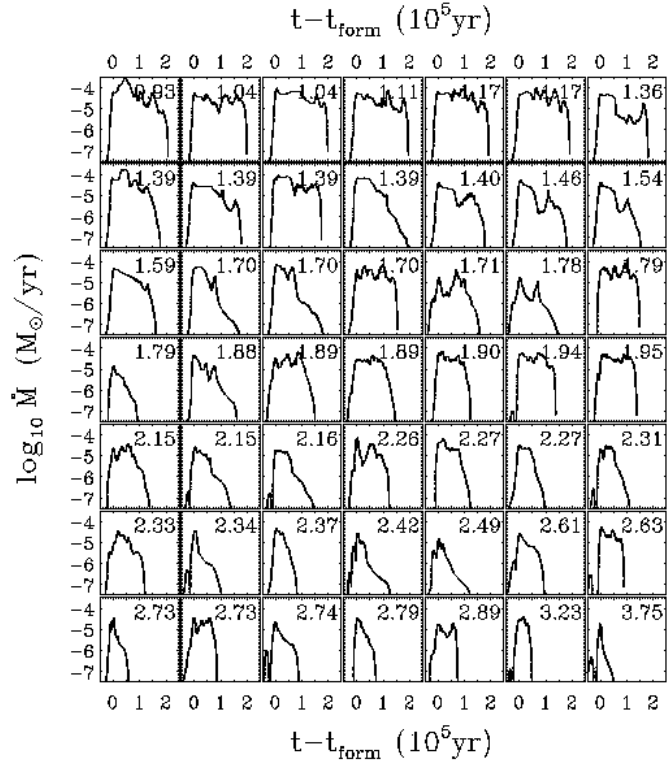}}
\end{picture}
\end{center}
\caption{\label{fig:accretion-rates} Examples of time varying mass
accretion rates for protostellar cores forming in a dense cluster
environment. The figure shows accretion rate $\dot{M}$ versus time
after formation $t-t_{\rm form}$ for 49 randomly selected protostellar
cores in a numerical model of molecular cloud fragmentation from
Klessen \& Burkert (2000), with $t_{\rm form}$ being the formation
time of the protostar in the center of the collapsing gas clump. To
link individual accretion histories to the overall cluster evolution,
$t_{\rm form}$ is indicated in the upper right corner of each plot and
measures the elapsed time since the start of the simulation. The
free-fall timescale of the considered molecular region is $\tau_{\rm
ff} \approx 10^5\,$years. High-mass stars tend to form early in the
dynamical evolution and are able to maintain high accretion rates
throughout the entire simulation. On the contrary, low-mass stars tend
to form later in the cluster evolution and $\dot{M}$ declines strongly
after the short initial peak accretion phase.  The accretion histories
of cores, even those with similar masses, differ dramatically from
each other because of the stochastic influence of the cluster
environment, as clumps merge and protostellar cores compete for
accretion from a common gaseous environment.  (From Klessen 2001a.)}
\end{figure}

Most models of protostellar core collapse concentrate on isolated
objects, whether the models are analytic (e.g.\ Larson 1969, Penston
1969a, Hunter 1977, Henriksen \etal\ 1997, Basu 1997) or numerical
(e.g.\ Foster \& Chevalier 1993, Tomisaka 1996, Ogino, Tomisaka, \&
Nakamura 1999, Wuchterl \& Tscharnuter 2002).  A typical accretion
history is shown in Figure~\ref{fig:larson-accretion-rate}.  However,
stars predominantly form in groups and clusters. Numerical studies
that investigate the effect of the cluster environment on protostellar
mass accretion rates have been reported by Bonnell \etal\ (1990,
2001a,b), Klessen \& Burkert (2000, 2001), Klessen \etal\ (2001),
Heitsch \etal\ (2001a), and Klessen (2001a).  These numerical models
suggest the following predictions about protostellar accretion in
dense clusters:

{\em(a)} Protostellar accretion rates from turbulent fragmentation in
a dense cluster environment are strongly time variable. This is
illustrated in Figure \ref{fig:accretion-rates} for 49 randomly
selected cores from the model by Klessen (2001a).

{\em(b)} The typical density profiles of gas clumps that give birth to
protostars indeed exhibit a flat inner core, followed by a density
fall-off $\rho \propto r^{-2}$, and are truncated at some finite
radius, which in the dense centers of clusters often is due to tidal
interaction with neighboring cores (see \S~\ref{sub:cores} and
\S~\ref{sub:clusters}).  As a result, the modeled accretion rates
agree well with the observations.  A short-lived initial phase of
strong accretion occurs when the flat inner part of the pre-stellar
clump collapses, corresponding to class 0.
If the cores remain isolated and unperturbed, the mass growth rate
gradually declines in time as the outer envelope accretes, giving
class I.  Once the truncation radius is reached, accretion fades and
the object enters class II.
However, collapse does not start from rest for the density
fluctuations considered here, so accretion rates exceed the values
predicted by models of isolated objects, even for objects in the
simulations far from their nearest neighbors.

{\em (c)} The mass accretion rates of protostellar cores in a dense
cluster also deviate strongly from the rates of isolated cores because of
mergers and competition between cores, as discussed above
(\S~\ref{sub:clusters}).
Mergers drastically change the density
and velocity structure of cores, so
the predictions for isolated cores no longer hold. Furthermore
these new, larger cores contain multiple protostars that
subsequently compete with each other for accretion from a common
gas reservoir.  The most massive protostar in a clump 
accretes more matter than its competitors (see also Bonnell \etal\ 1997,
Klessen \& Burkert 2000, Bonnell \etal\ 2001a,b). Its accretion rate
is enhanced through the clump merger, whereas the accretion rate of
low-mass cores typically decreases.  Many temporary accretion peaks in the
wake of clump mergers are visible in Figure~\ref{fig:accretion-rates}.
The small aggregates of cores 
that build up are dynamically unstable, so low-mass cores may be
ejected. As they leave the high-density environment, accretion
terminates and their final mass is reached.
  
{\em (d)} The most massive protostars begin to form first and continue
to accrete at high rate throughout the entire cluster evolution. As
the most massive gas clumps tend to have the largest density contrast,
they are the first to collapse, forming the center of the
nascent cluster.  These protostars are fed at high rate and
gain mass very quickly.  
McKee \& Tan (2002) argue on analytical grounds that they may form in
as little as $10^5$\,yr, even in the absence of competitive accretion.  
As their parental cores merge with others, more gas is fed into their
`sphere of influence'. They are able to maintain or even increase the
accretion rate when competing with lower-mass objects (e.g.\ core 1
and 8 in Figure~\ref{fig:accretion-rates}).  Low-mass stars tend to
form somewhat later in the dynamical evolution of the system (as
indicated by the absolute formation times in
Figure~\ref{fig:accretion-rates}; also Figure 8 in Klessen \& Burkert
2000), and typically have only short periods of high accretion.

{\em (e)} As high-mass stars form in massive cores, while low-mass
stars form in less massive cores, the stellar population in clusters
should be mass segregated right from the start. High-mass stars form
in the center, lower-mass stars tend to form towards the cluster
outskirts. This agrees with recent observations of the cluster NGC330
in the Small Magellanic Cloud (Sirianni \etal\ 2002). Dynamical
effects during the embedded phase of star cluster evolution enhance
this initial segregation even further (see \S~\ref{sub:clusters}.c).

{\em(f)} Individual cores in a cluster environment form and evolve
through a sequence of highly probabilistic events, so their accretion
histories differ even if they accumulate the same final
mass. Accretion rates for protostars of a certain mass can only be
determined in a statistical sense. Klessen (2001a) suggests that an
exponentially declining rate with a peak value of a few
$10^5\,$M$_{\odot}$yr$^{-1}$, a time constant in the range $0.5$ to
$2.5\times 10^5\,$yr, and a cut-off related to gas dispersal from the
cluster offers a reasonable fit to the average protostellar mass
growth in dense embedded clusters, but with large variations.

\subsection{Initial mass function}
\label{sub:imf}

Knowledge of the distribution of stellar masses at birth, described by
the initial mass function (IMF), is necessary to understand many
astrophysical phenomena, but no analytic derivation of the observed
IMF has yet stood the test of time. In fact, it appears likely that a
fully deterministic theory for the IMF does not exist.  Rather, any
viable theory must take into account the probabilistic nature of the
turbulent process of star formation, which is inevitably highly
stochastic and indeterminate.
We gave a brief overview
of the observational constraints on the IMF in
\S~\ref{sub:IMF-observed}, and we here review
models for it.

\subsubsection{Models of the IMF}
\label{subsub:IMF-models}

Existing models to explain the IMF can be divided into five major
groups.  In the first group feedback from the stars themselves
determines their masses. Silk (1995) suggests that stellar masses are
limited by the feedback from both ionization and protostellar
outflows. Nakano, Hasegawa, \& Norman (1995) describe a model in which
stellar masses are sometimes limited by the mass scales of the
formative medium and sometimes by stellar feedback. 
Adams \& Fatuzzo (1996) 
apply the central limit theorem to the hypothesis that many
independent physical variables contribute to the stellar masses to
derive a log-normal IMF regulated by protostellar feedback. However,
for the overwhelming majority of stars with masses $M\sil
5\,M_{\odot}$, protostellar feedback (i.e.\ winds, radiation and
outflows) are unlikely to be strong enough to halt mass accretion, as
shown by detailed protostellar collapse calculations (e.g.\ Wuchterl
\& Klessen 2001, Wuchterl \& Tscharnuter 2003).

In the second group of models, initial and environmental conditions
determine the IMF. In this picture, the structural properties of
molecular clouds determine the mass distribution of Jeans-unstable gas
clumps, and the clump properties determine the mass of the stars that
form within. If one assumes a fixed star formation efficiency for
individual clumps, there is a one-to-one correspondence between the
molecular cloud structure and the final IMF.  The idea that
fragmentation of clouds leads directly to the IMF dates back to Hoyle
(1953) and later Larson (1973). More recently, this concept has been
extended to include the observed fractal and hierarchical structure of
molecular clouds Larson (1992, 1995). Indeed random sampling from a
fractal cloud seems to be able to reproduce the basic features of the
observed IMF (Elmegreen \& Mathieu 1983, Elmegreen 1997, 1999,
2000a,c, 2002). A related approach is to see the IMF as a domain
packing problem (Richtler 1994).

The hypothesis that stellar masses are determined by clump masses in
molecular clouds is supported by observations of the dust continuum
emission of protostellar condensations in the Serpens, $\rho$ Ophiuchi,
and Orion star forming regions (Testi \& Sargent 1998; Motte \etal\
1998, 2001; Johnstone \etal\ 2000, 2001). These protostellar cores are
thought to be in a phase immediately before 
a star forms 
in their interior. Their mass distribution resembles the stellar IMF
reasonably well, suggesting a close correspondence between
protostellar clump masses and stellar masses, leaving little room for
stellar feedback processes, competitive accretion, or collisions to act
to determine the stellar mass spectrum.

A third group of models relies on competitive coagulation or accretion
processes to determine the IMF.  This has a long tradition and dates
back to investigations by Oort (1954) and Field \& Saslaw (1965), but
the interest in this concept continues to the present day (e.g.\ Silk
\& Takahashi 1979; Lejeune \& Bastien 1986; Price \& Podsiadlowski
1995; Murray \& Lin 1996; Bonnell \etal\ 2001a,b; Durisen, Sterzik, \&
Pickett 2001).  Stellar collisions require very high stellar
densities, however, for which observational evidence and theoretical
mechanisms remain scarce.

Fourth, there are models that connect the supersonic turbulent motions
in molecular clouds to the IMF. In particular there are a series of
attempts to find an analytical relation between the stellar mass
spectrum and statistical properties of interstellar turbulence (e.g.\
Larson 1981, Fleck 1982, Hunter \& Fleck 1982, Elmegreen 1993, 
Padoan 1995, Padoan \etal\ 1997, Myers 2000, Padoan \& Nordlund
2002).  However, properties such as the probability distribution of
density in supersonic turbulence in the absence of gravity have never
successfully been shown to have a definite relationship to the final
results of gravitational collapse (Padoan \etal\ 1997).  Even the more
sophisticated model of Padoan \& Nordlund (2002) 
does not take into account
that not single but multiple compressions and rarefactions determine
the density structure of supersonic turbulence (Passot \&
V\'azquez-Semadeni 1998, 2003), and that higher mass clumps will fragment
into multiple protostars.  Furthermore, such models neglect the
effects of competitive accretion in dense cluster environments
(\S~\ref{sub:accretion}), which may be important for determining the
upper end of the IMF.  

Finally there is a more statistical approach.  Larson (1973) and
Zinnecker (1984, 1990) argued that whenever a large set of parameters
is involved in determining the masses of stars, invoking the central
limit theorem of statistics naturally leads to a log-normal stellar
mass spectrum (Adams \& Fatuzzo 1996 made similar arguments).

Regardless of the detailed physical processes involved, the common
theme in all of these models is the probabilistic nature of star
formation.  It appears impossible to predict the formation of specific
individual objects.  Only the fate of an ensemble of stars can be
described {\em ab initio}.  The implication is that the star formation
process can only be understood within the framework of a probabilistic
theory.

 \begin{figure}[ht]
 \unitlength1.0cm
 \begin{picture}(16.0,10.9)
 \put( 2.3,-0.0){\epsfxsize=11.7cm\epsfbox{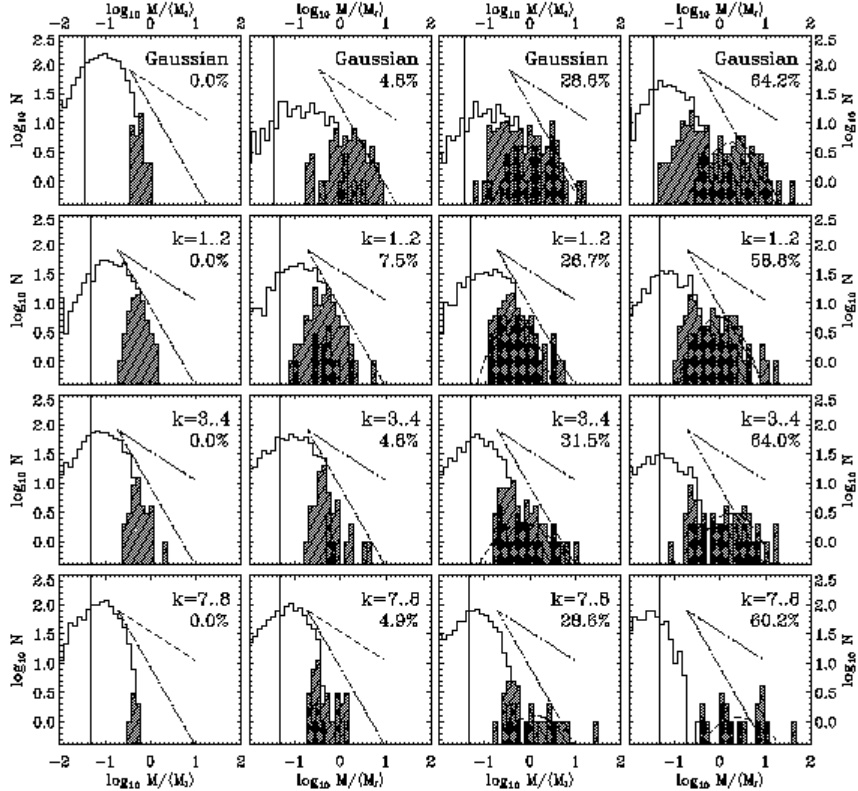}}
 \end{picture}
 \caption{\label{fig:massspectra} Mass spectra of gas clumps
   (thin lines), of the subset of Jeans unstable clumps (thin
   lines, hatched distribution), and of 
   protostars 
   (hatched thick-lined histograms), for four different
   models.  The decaying model started with Gaussian density
   perturbations and no turbulence, while the other three models were
   nominally supported by turbulence driven at long, intermediate, or
   short scales as indicated by the driving wavenumbers $k$.  Masses
   are binned logarithmically and normalized to the average Jeans mass
   $\langle M_{\rm J}\rangle$. The left column gives the initial state
   of the system when the turbulent flow has reached equilibrium but
   gravity has not yet been turned on, the second column shows the
   mass spectra when $M_{\rm *} \approx 5$\% of the mass is accreted
   onto dense cores, the third column shows $m_{\rm *} \approx 30$\%,
   and the last one $M_{\rm *} \approx 60$\%. For comparison with
   power-law spectra ($dN/dM \propto M^{\alpha}$), a slope $\alpha =
   -1.5$ typical for the observed clump mass distribution, and the
   Salpeter slope $\alpha=-2.33$ for the IMF, are indicated by the
   dotted lines.  The vertical line shows the resolution limit of the
   numerical model. In columns 3 and 4, the long dashed curve shows
   the best log-normal fit to the protostars. (From Klessen 2001b.) } 
\end{figure}

\subsubsection{Turbulent fragmentation example}
\label{subsub:mass-spectra}

To illustrate some of the issues discussed above, we examine the mass
spectra of gas clumps and collapsed cores from models of
self-gravitating, isothermal, supersonic turbulence driven with
different wavelengths (Klessen 2001b).  In the absence of magnetic
fields and more accurate equations of state, these models can only be
illustrative, not definitive, but nevertheless they offer insight into
the processes acting to form the IMF.  Figure~\ref{fig:massspectra}
shows that before local collapse begins to occur, the clump mass
spectrum is not well described by a single power law.  During
subsequent evolution, as clumps merge and grow larger, the mass
spectrum extends towards higher masses, approaching a power law with
slope $\alpha \approx -1.5$.  Local collapse sets in, resulting in the
formation of dense cores, most quickly in the freely collapsing model.
The influence of gravity on the clump mass distribution weakens when
turbulence dominates over gravitational contraction on the global
scale, as in the other three models. The more the turbulent energy
dominates over gravity, the more the spectrum resembles the initial
case of pure gas dynamic turbulence. This suggests that the clump mass
spectrum in molecular clouds will be shallower in regions where
gravity dominates over turbulent energy. This may explain the observed
range of slopes for the clump mass spectrum in different molecular
cloud regions (\S~\ref{sub:LSS}).

Like the distribution of Jeans-unstable clumps, the mass spectrum of
dense protostellar cores resembles a log-normal in the models without
turbulent support and with long-wavelength turbulent driving, with a
peak at roughly the average thermal Jeans mass $\langle m_{\rm
J}\rangle$ of the system. These models also predict initial mass
segregation (Section \ref{sub:accretion}.{\em e}).  
The protostellar clusters discussed here only contain between 50 and
100 cores. This allows for comparison with the IMF only around the
characteristic mass scale, typically about 1~M$_{\odot}$, since the
numbers are too small to study the very low- and high-mass end of the
distribution. Focusing on low-mass star formation, however, Bate,
Bonnell, \& Bromm (2002a) demonstrate that brown dwarfs are a natural
and frequent outcome of turbulent fragmentation. In this model, brown
dwarfs form when dense molecular gas fragments into unstable multiple
systems that eject their smallest members from the dense gas before
they have been able to accrete to stellar masses.  Numerical models
with sufficient dynamic range to treat the full range of stellar
masses (Eq.\ \ref{eqn:mass-range}) remain to be done.

\section{GALACTIC SCALE STAR FORMATION}
\label{sec:galactic}

How do the mechanisms that control local star formation determine the
global rate and distribution of star formation in galaxies? 
In this section we begin by examining how molecular clouds form from
the interstellar medium in \S~\ref{sub:clouds}.  We then outline what
determines the efficiency of star formation 
in \S~\ref{sub:efficient}.  We argue that the balance between the density
of available gas and its turbulent velocity determines where star
formation will occur, and how efficiently.  Even when the turbulent
velocity in a region is relatively high, if the density in that region
is also high, the region may still not be supported against
gravitational collapse and prompt star formation.  Therefore, any
mechanism that increases the local density without simultaneously
increasing the turbulent velocity sufficiently can lead to star
formation, via molecular cloud formation.
Most mechanisms that increase local density
appear to be external to the star formation process, however.
Accretion during initial galaxy formation, interactions and collisions
between galaxies, spiral gravitational instabilities of galactic
disks, and bar formation are major examples. In this review we cannot
do justice to the vast literature on galactic dynamics and
interactions that determine the density distribution in galaxies.  We
do, however, examine what physical mechanisms control the turbulent
velocity dispersion in \S~\ref{sub:driving}.  Finally, in
\S~\ref{sub:applications} we briefly speculate on how turbulent
control of star formation may help explain objects with very different
star formation properties, including low surface brightness galaxies,
normal galactic disks, globular clusters, galactic nuclei, and
primordial dwarf galaxies.

\subsection{Formation and lifetime of molecular clouds}
\label{sub:clouds}

How do molecular clouds form?  Any explanation must account for their
low star-formation efficiencies and broad linewidths.
Molecular hydrogen forms on dust grains at a rate calculated by Hollenbach,
Werner, \& Salpeter (1971) to be 
\begin{equation}
t_{\rm form} = (1.5 \times 10^9 \mbox{ yr}) \left(\frac{n}{1\,{\rm cm}^{-3}}\right)^{-1},
\end{equation}
where $n$ is the number density of gas particles.  Recent experimental work
by Piranello \etal\ (1997a, 1997b, 1999) on molecular hydrogen
formation on graphite and olivine suggests that these rates may be strongly
temperature dependent, so that the Hollenbach \etal\ (1971) result
may be a lower limit to the formation time. However, the same group
reports that molecule formation is rather more efficient on amorphous
ices (Manic\'o \etal\ 2001) such as would be expected on grain
surfaces deep within dark clouds, so that the rates computed by
Hollenbach \etal\ (1971) may be reached after all.  Further
experimental investigation of molecule formation appears necessary.

When molecular clouds were first discovered, they were thought to have
lifetimes of over 100$\,$Myr (e.g.\ Scoville \& Hersh 1979) because of
their apparent predominance in the inner galaxy.  These estimates were
shown to depend upon too high a conversion factor between CO and H$_2$
masses by Blitz \& Shu (1980).  They revised the estimated lifetime
down to roughly 30$\,$Myr based on the association of clouds with spiral arms,
apparent ages of associated stars, and overall star formation rate in
the Galaxy.  

Chemical equilibrium models of dense cores in molecular clouds (as
reviewed, for example, by Irvine, Goldsmith, \& Hjalmarson 1986)
showed disagreements with observed abundances in a number of
molecules. These cores would take as long as 10$\,$Myr to reach
equilibrium, which could still occur in the standard model. However,
Prasad, Heere, \& Tarafdar (1991) demonstrated that the abundances of
the different species agreed much better with the results at times of
less than 1$\,$Myr from time-dependent models of the chemical
evolution of collapsing cores (also see \S~\ref{para:chemical}).
Bergin \& Langer (1997), Bergin \etal\ (1997), Pratap \etal\ (1997),
and Aikawa \etal\ (2001) came to similar conclusions from careful
comparison of several different cloud cores to extensive chemical
model networks. Aikawa \etal\ (2001) and Saito
\etal\ (2002) also studied deuterium fractionation, again finding
short lifetimes.

Ballesteros-Paredes \etal\ (1999a) argue for a lifetime of less than
10$\,$Myr for molecular clouds as a whole.  They base their argument
on the notable lack of a population of 5--20~Myr old stars in
molecular clouds.  Stars in the clouds typically have ages under
3--5~Myr, judging from their position on pre-main-sequence
evolutionary tracks in a Hertzsprung-Russell diagram (D'Antona \&
Mazzitelli 1994; Swenson \etal\ 1994; with discrepancies resolved by
Stauffer, Hartmann, \& Barrado y Navascues 1995).  Older weak-line T
Tauri stars identified by X-ray surveys with {\em Einstein} (Walter
\etal\ 1988) and {\em ROSAT} (Neuh\"auser \etal\ 1995) are dispersed
over a region as much as 70$\,$pc away from molecular gas, suggesting
that they were not formed in the currently observed gas (Feigelson
1996).  Leisawitz, Bash, \& Thaddeus (1989), Fukui \etal\ (1999) and
Elmegreen (2000) have made similar arguments based on the observation
that only stellar clusters with ages under about 10$\,$Myr are
associated with substantial amounts of molecular gas in the Milky Way
and the LMC.

\begin{figure}[tbhp]
\begin{center}
\includegraphics[width=0.6\textwidth,angle=270]{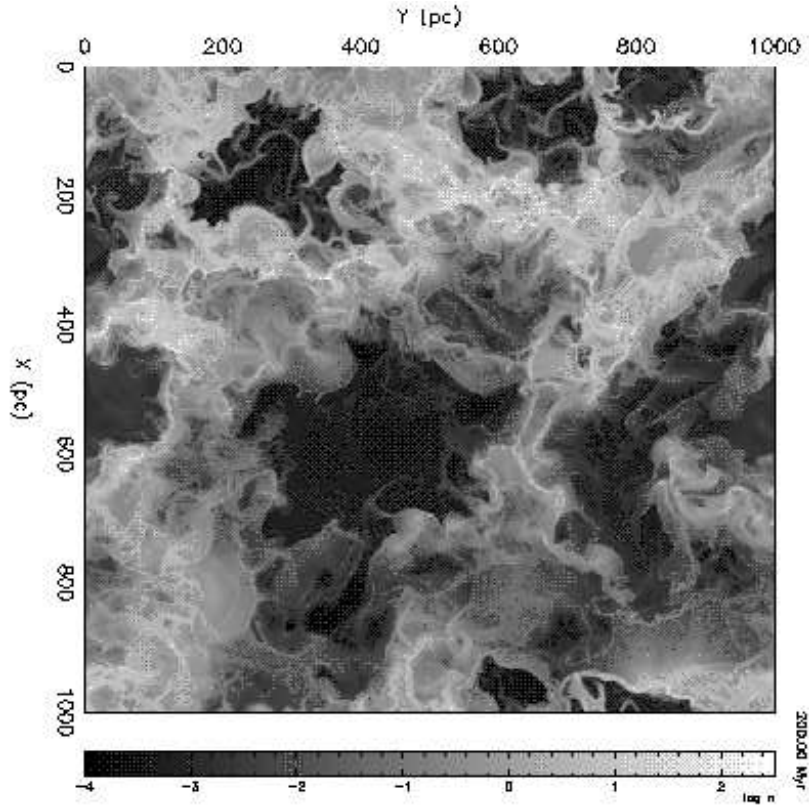}
\end{center}
\caption{\label{fig:mbak-1a} 
Log of number density in a cut through the galactic plane from a 3D
SN-driven model of the ISM with resolution of 1.25~pc, including
radiative cooling and the vertical gravitational field of the stellar
disk, as described by Mac Low \etal\ (2003) and Avillez \&
Breitschwerdt (2003).  High-density, shock-confined regions are
naturally produced by intersecting SN-shocks from field SNe.  }
\end{figure}
For these short lifetimes to be plausible, either molecule formation
must proceed quickly, and therefore at high densities, or observed
molecular clouds must be formed from preexisting molecular gas, as
suggested by Pringle, Allen, \& Lubow (2001).  A plausible place for
fast formation of H$_2$ at high density is the shock compressed layers
naturally produced in a supernova-driven ISM, as shown in Figure\
\ref{fig:mbak-1a} from simulations similar to those described in Mac
Low \etal\ (2003) and Avillez \& Breitschwerdt (2003).  Similar
morphologies have been seen in many other global simulations of the
ISM, including Rosen, Bregman, \& Norman (1993), Rosen \& Bregman
(1995), Rosen, Bregman, \& Kelson (1996), Korpi \etal\ (1999), Avillez
(2000), and Wada \& Norman (1999, 2001). Mac Low (2000) reviews these earlier
simulations.  Mac Low \etal\ (2003) showed that pressures in the ISM
are broadly distributed, with peak pressures in cool gas ($T < 10^3$
K) as much as an order of magnitude above the average because of shock
compressions
(also see Passot \& V\'azquez-Semadeni 2003).  
This gas is swept up from ionized $10^4\,$K gas, so between cooling
and compression its density has already been raised up to two orders
of magnitude from $n\approx 1\,$cm$^{-3}$ to $n\approx
100\,$cm$^{-3}$.  These simulations did not include a correct cooling
curve below $10^4\,$K, so further cooling could not occur even if
physically appropriate, but it would be expected.

We can understand this compression quantitatively.  The sound speed in
the warm gas is $(8.1\,$km$\,$s$^{-1})(T/10^4\mbox{ K})^{1/2}$, taking
into account the mean mass per particle $\mu = 2.11 \times 10^{-24}\,$g
for gas 90\% H and 10\% He by number.  The typical velocity dispersion
for this gas is 10--12~km~s$^{-1}$ (e.g.\ Dickey \& Lockman 1990,
Dickey, Hanson, \& Helou 1990), so that shocks with Mach numbers
${\cal M} =$ 2--3 are moderately frequent.  Temperatures in these
shocks reach values $T \le 10^5\,$K, which is close to the peak of
the interstellar cooling curve (e.g.\ Dalgarno \& McCray 1972;
Raymond, Cox, \& Smith 1976), so the gas cools quickly back to
$10^4\,$K.  The density behind an isothermal shock is $\rho_1 = {\cal
M}^2 \rho_0$, where $\rho_0$ is the pre-shock density, so order of
magnitude density enhancements occur easily.  The optically-thin
radiative cooling rate $\Lambda(T)$ drops off at $10^4\,$K as H~atoms
no longer radiate efficiently (Dalgarno \& McCray 1972; Spaans \&
Norman 1997), but the radiative cooling $L \propto n^2 \Lambda(T)$.
Therefore 
density enhancements strongly increase the ability to cool.  Hennebelle \&
P\'erault (1999) show that such shock compressions can trigger the
isobaric thermal instability (Field \etal\ 1969; Wolfire \etal\ 1995),
reducing temperatures to of order 100$\,$K or less.  Heiles (2001)
observes a broad range of temperatures for neutral hydrogen from below
100$\,$K to a few thousand$\,$K.  The reduction in temperature by two
orders of magnitude from $10^4\,$K to 100$\,$K raises the density
correspondingly.  Combined with the initial isothermal shock
compression, this results in a total of as much as three orders of
magnitude of compression.  Gas that started at densities somewhat
higher than average, say at 10$\,$cm$^{-3}$, can be compressed to
densities of $10^4\,$cm$^{-3}$, enough to reduce H$_2$ formation times
to a few hundred thousand years.

\begin{figure}[tbhp]
\begin{center}
\includegraphics[width=0.4\textwidth,angle=270]{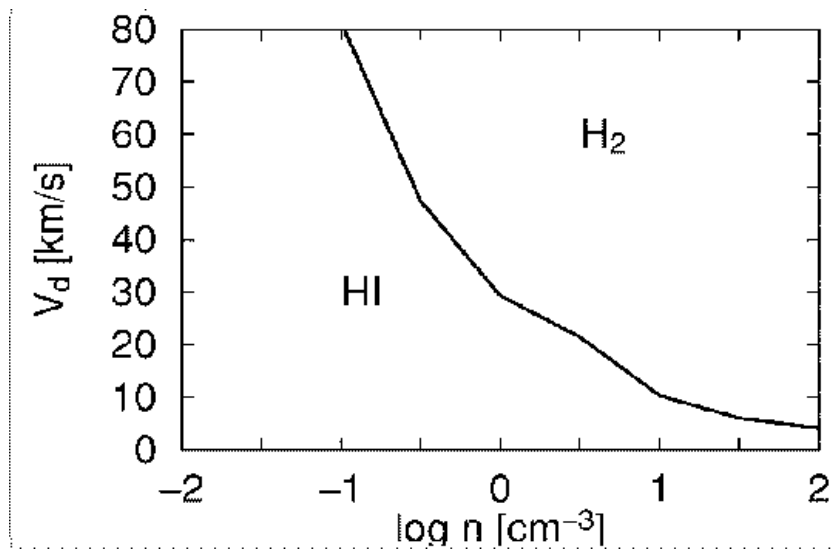}
\end{center}
\caption{ \label{fig:ko-in-10}
Shock velocities $V_d$ and pre-shock number densities $n$ at which
the cold post-shock layer is more than 8\% molecular, taken from
1D simulations by Koyama \& Inutsuka (2000) that include
H$_2$ formation and dissociation, and realistic heating and cooling
functions from Wolfire \etal\ (1995).  
}
\end{figure}
Koyama \& Inutsuka (2000) have demonstrated numerically that
shock-confined layers do indeed quickly develop high enough densities
to form H$_2$ in under a million years, using 1D
computations including heating and cooling rates from Wolfire \etal\
(1995) and H$_2$ formation and dissociation.  In Figure\ \ref{fig:ko-in-10}
we show the parameter space in which they find H$_2$ formation is
efficient.  Hartmann, Ballesteros-Paredes, and Bergin (2001) make a
more general argument for rapid H$_2$ formation, based in part on
lower-resolution, 2D simulations described by Passot,
V\'azquez-Semadeni, \& Pouquet (1995) that could not fully resolve
realistic densities like those of Koyama \& Inutsuka (2000), but do
include larger-scale flows showing that the initial conditions for the
1D models are quite reasonable.  Hartmann \etal\ (2001)
further argue that the self-shielding against the background UV field
also required for H$_2$ formation will become important at
approximately the same column densities required to become
gravitationally unstable.

\begin{figure}[htp]
\begin{center}
\unitlength1cm
 \begin{picture}(8.0,13.9)
 \put( -2.5,0.3){\epsfxsize=13cm\epsfbox{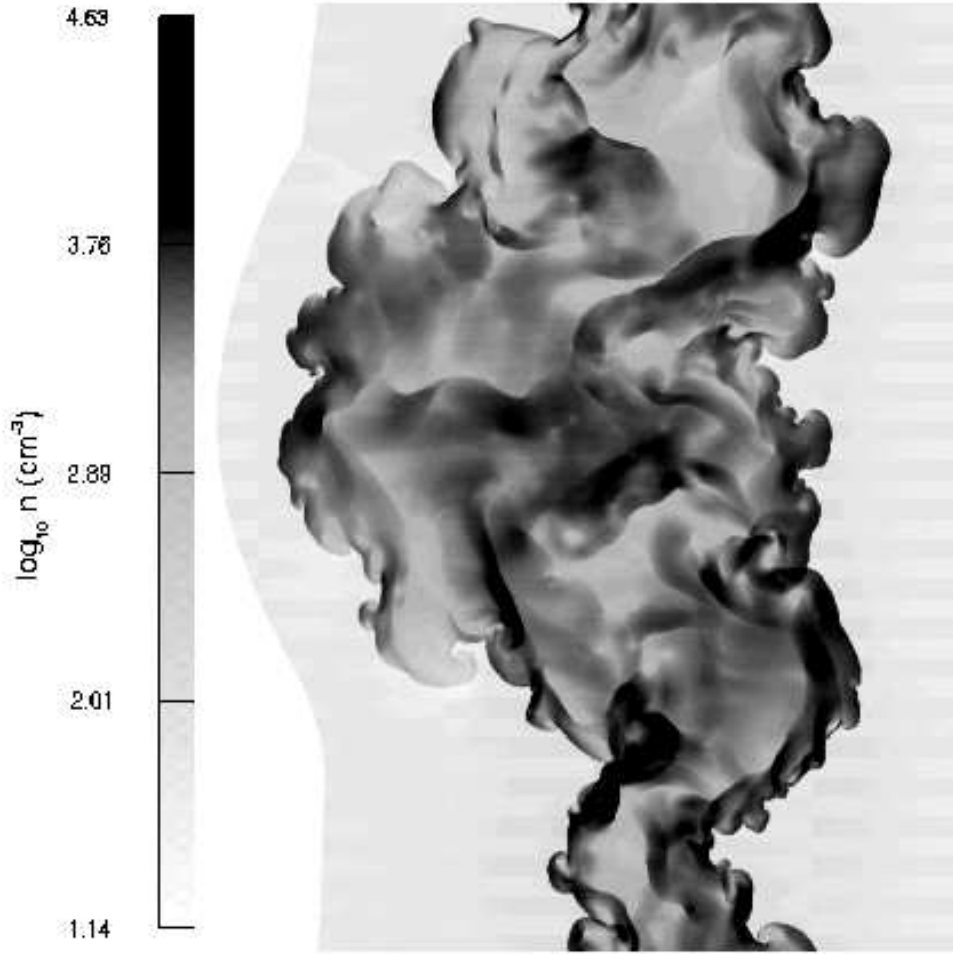}}
 \end{picture}
\end{center}
\caption{\label{fig:wal-fol-2} Instability of radiatively cooled layer
confined from left and right by strong shocks with
Mach number ${\cal M} =16.7$, computed in two dimensions with an
adaptive mesh refinement technique by Walder \& Folini (2000).
White regions have densities of 14~cm$^{-3}$, while the darkest
regions have densities over $10^4$ cm$^{-3}$.}
\end{figure}

Shock-confined layers were shown numerically to be unstable by Hunter
\etal\ (1986) in the context of colliding spherical density
enhancements, and by Stevens, Blondin, \& Pollack (1992) in the
context of colliding stellar winds.  Vishniac (1994) demonstrated
analytically that isothermal, shock-confined layers are subject to a
nonlinear thin shell instability.  The physical mechanism can
be seen by considering a shocked layer perturbed sinusoidally.  The
ram pressure on either side of the layer acts parallel to the incoming
flow, and thus at an angle to the surface of the perturbed layer.
Momentum is deposited in the layer with a component parallel to the
surface, which drives material towards extrema in the layer, causing
the perturbation to grow.  A numerical study by Blondin \&
Mark (1996) in two dimensions demonstrated that the nonlinear thin
shell instability saturates in
a thick layer of transsonic turbulence when the flows become
sufficiently chaotic that the surface no longer rests at a substantial
angle to the normal of the incoming flow.

Thermal instability will act in conjunction with shock confinement
(see \S~\ref{subsub:thermal}).  Goldsmith (1970) and Schwarz, McCray,
\& Stein (1972) first computed the nonlinear development of the
thermal instability, demonstrating that shock waves form during the
dynamical collapse of nonlinear regions. Hennebelle \&
P\'erault (1999) demonstrated that shock compression can trigger
thermal instability in otherwise stable regions in the diffuse ISM,
even in the presence of magnetic fields (Hennebelle \& P\'erault
2000), so that compressions much greater than the isothermal factor of
${\cal M}^2$ can occur (see the quantitative discussion by
V\'azquez-Semadeni \etal\ 1996).

These cold, dense layers are themselves subject to dynamical
instabilities, as has recently been shown in 2D computations by Koyama
\& Inutsuka (2002).  The instabilities they found are caused by some
combination of thermal instability and mechanisms very similar to the
nonlinear thin shell instability (Vishniac 1994) for the isothermal
case. Figure\ \ref{fig:wal-fol-2} shows another example of these
instabilities from a numerical study by Walder \& Folini (2000).
These dynamical instabilities can drive strongly supersonic motions in
the cold, dense layer.  If that layer is dense enough for molecule
formation to proceed quickly, those molecules will show strongly
supersonic linewidths on all but the very smallest scales, as seen in
the models of Koyama \& Inutsuka (2002), in agreement with the
observations of molecular clouds.  It remains to be shown whether this
scenario can quantitatively explain the full ensemble of molecular
clouds observed in the solar neighborhood, or elsewhere in our own and
external galaxies.

The final destruction of molecular clouds then proceeds from a
combination of several effects.  First, once the external turbulent
compression has passed, they will begin to freely expand
(V\'azquez-Semadeni \etal\ 2002), but only at the sound speed of the
cold gas of 0.2 km~s$^{-1}$, or roughly a parsec every 5~Myr.  Second,
the same turbulent flows that formed them may again tear them
apart. As the density decreases in either of these cases, background
dissociating radiation will tend to destroy the molecules (McKee
1989).  Third, radiation from stars forming in the cloud may heat and
dissociate the molecular gas, reducing its density and preventing it
from forming further stars (Matzner 2002 and references therein).

\subsection{When is star formation efficient?}
\label{sub:efficient}

\subsubsection{Overview}
Observers have documented a surprisingly strong connection between the
star formation rate and the local velocity dispersion, column density
and rotational velocity of disk galaxies (Kennicutt 1998a, Martin \&
Kennicutt 2001).  A global Schmidt (1959) law relates star formation
rate per unit area to gas surface density as
\begin{equation}
\Sigma_{\rm SFR} = A \Sigma_{\rm gas}^N,
\end{equation}
where a value of $N = 1.4 \pm 0.05$ can be derived from the
observations (Kennicutt 1989, 1998b).  Star formation also cuts off
sharply at some radius in most star-forming galaxies (Kennicutt 1989,
Martin \& Kennicutt 2001), which also appears related to the gas
surface density.  The Schmidt law can be interpreted as reflecting
star formation on a free-fall timescale, so that (following Wong \&
Blitz 2001 for example) the star formation rate per unit volume of gas
with density $\rho$ is
\begin{equation}
\dot{\rho}_{\rm SF} =\epsilon_{\rm SFR} \frac{\rho}{\tau_{\rm ff}} = 
\epsilon_{\rm SF} \frac{\rho}{(G\rho)^{-1/2}} \propto \rho^{1.5},
\end{equation}
where $\epsilon_{\rm SFR}$ is an efficiency factor observed to be
substantially less than unity. (It differs from the star formation
efficiency $\epsilon_{SF}$ defined in equation~(\ref{eqn:sfe}) only in
that we here compare to the free-fall lifetime rather than the total
lifetime of the system.)

The connection between magnetically controlled small-scale star
formation and large-scale star formation is not clear in the standard
theory. Shu, Adams, \& Lizano (1987) did indeed suggest that OB
associations were formed by freely collapsing gas that had overwhelmed
the local magnetic field, but that still implied that the star
formation rate was controlled by the details of the magnetic field
structure, which in turn is presumably controlled by the galactic
dynamo.  The connection appears clearer, though, if turbulence, as
represented by the velocity dispersion, controls the star formation
rate.  The same physical mechanisms control star formation at all
scales.  Regions that are globally supported by turbulence still
engage in inefficient star formation, but the frequency of regions of
efficient star formation determines the overall star formation rate in
a feedback loop.

The big open question in this area remains the importance of radiative
cooling for efficient star formation, either on its own or induced by
turbulent compression.  Is cooling, and indeed molecule formation,
necessary for gravitational collapse to begin, or is it rather a
result of already occurring collapse in gravitationally unstable gas?
Certainly there are situations where cooling will make the difference
between gravitational stability and instability, but are those just
marginal cases or the primary driver for star formation in galaxies?

\begin{figure}[bt]
\begin{center}
\includegraphics[width=0.5\textwidth,angle=270]{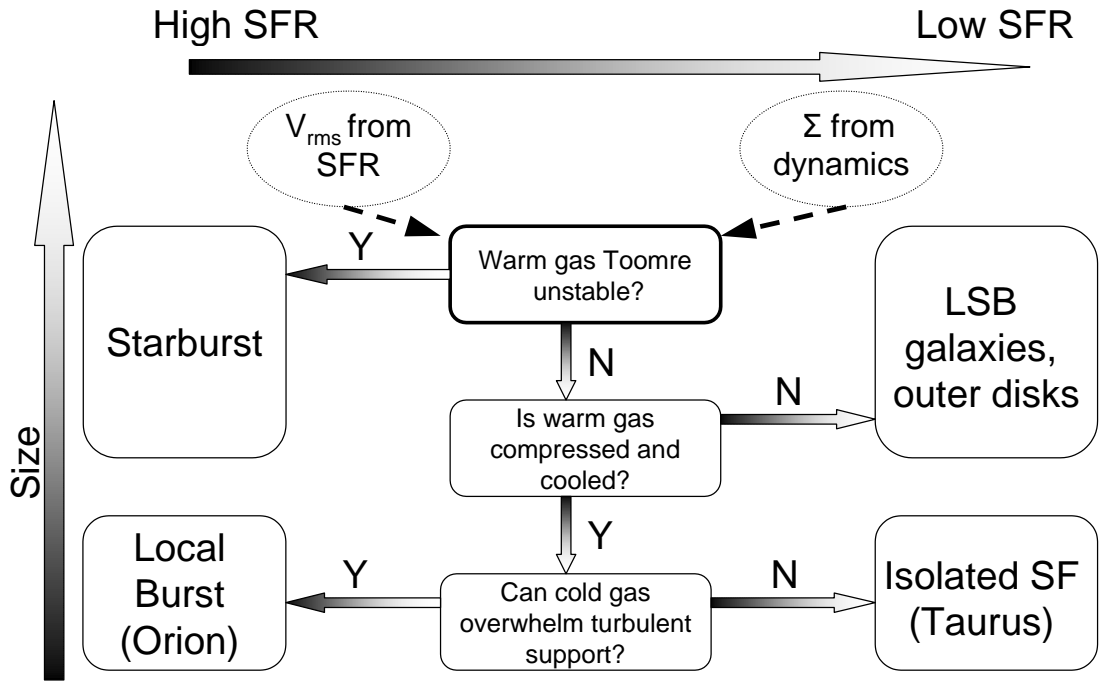}
\end{center}
\caption[Star formation in galaxies]{
\label{fig:sf-flowchart}
Criteria for different regimes of star formation efficiency in
galaxies.  See text for further details.}
\end{figure}

In Figure~\ref{fig:sf-flowchart} we outline a unified picture that
depends on turbulence and cooling to control the star formation rate.  
After describing the different elements of this picture, we 
discuss the steps that we think will be needed to move from this
cartoon to a quantitative theory of the star formation rate.
The factor that determines the star formation rate above
any other is whether the gas is sufficiently dense to be
gravitationally unstable without additional cooling.  Galactic
dynamics, and interactions with other galaxies and the surrounding
intergalactic gas determine the average gas densities in different
regions of a galaxy.  The gravitational instability criterion here
includes both turbulent motions and galactic shear, as well as
magnetic fields.  If gravitational instability sets in at large
scale, collapse will continue so long as sufficient cooling mechanisms
exist to prevent the temperature of the gas from rising (effective
adiabatic index $\gamma_{\rm eff} \leq 1$).  Molecular
clouds can form in less than $10^5$ yr, as the gas passes through
densities of $10^4$~cm$^{-3}$ or higher, as an incidental effect of
the collapse.  A starburst results, with stars forming efficiently in
compact clusters.  The size of the gravitationally unstable region
determines the size of the starburst.

If turbulent support, rather than thermal support, prevents the gas
from immediately collapsing, compression-induced cooling can become
important.  Supersonic turbulence compresses some fraction of the gas
strongly.  As most cooling mechanisms depend on the gas density
non-linearly, the compressed regions cool quickly.  When these regions
reach densities of order $10^4$~cm$^{-3}$, again molecule
formation occurs, allowing the gas to cool to even lower
temperatures 
(see \S~\ref{sub:clouds}).
These cold regions then can become gravitationally unstable and collapse,
if allowed by the local turbulence. 
Triggering by nearby star formation events (Elmegreen \& Lada 1977)
represents a special case of this mode (see \S~\ref{subsub:disks}).
This mechanism is less efficient than prompt gravitational
instability, as much of the gas is not compressed enough to form
molecules. It is, however, more efficient in regions of higher average
density.  Galactic dynamics again determines the local average
density, and so, in the end, the star formation efficiency in this
regime as well.

If the turbulence even in the cooled regions supports the gas against
general gravitational collapse, isolated, low-rate star formation can
still occur locally in regions further compressed by the turbulence.
This may describe regions of low-mass star formation like the Taurus
clouds.  On the other hand, if the cooled gas begins to collapse
gravitationally, locally efficient star formation can occur.  The size
of the gravitationally unstable region then really determines whether
a group, OB association, or bound cluster eventually forms.  Star
formation in regions like Orion may result from this branch.

\subsubsection{Gravitational instabilities in galactic disks}
\label{subsub:disk-grav}
Now let us consider the conditions under which gravitational instability will
set in.  On galactic scales, the Jeans instability criterion for gravitational
instability must be modified to include the additional support offered by the
shear coming from differential rotation, as well as the effects of magnetic
fields.  The gravitational potential of the stars can also contribute to
gravitational instability on large scales.  Which factor determines the onset
of gravitational instability remains unknown.  Five that have been proposed
are the temperature of the cold phase, the surface density, the local shear,
the presence of magnetic fields, and the velocity dispersion, in different
combinations.

We can heuristically derive the Toomre (1964) criterion for stability
of a rotating, thin disk with uniform velocity dispersion $\sigma$ and
surface density $\Sigma$ using timescale arguments (Schaye 2003).
First consider the Jeans criterion for instability in a thin disk,
which requires that the timescale for collapse of a perturbation of
size $\lambda$
\begin{equation}
t_{\rm coll} = \sqrt{\lambda / G \Sigma}
\end{equation}
be shorter than the time required for the gas to respond to the
collapse, the sound crossing time
\begin{equation}
t_{\rm sc} = \lambda / c_{\rm s}.
\end{equation}
This implies that gravitational stability requires perturbations
with size
\begin{equation} \label{eqn:press}
\lambda < c_{\rm s}^2 / G \Sigma.
\end{equation}
Similarly, in a disk rotating differentially, a perturbation will
spin around itself, generating centrifugal motions that can also
support against gravitational collapse.  This will be effective if the
collapse timescale $t_{\rm coll}$ exceeds the rotational period
$t_{\rm rot} = 2\pi/\kappa$, where $\kappa$ is the epicyclic
frequency, so that stable perturbations have
\begin{equation} \label{eqn:rot}
\lambda > 4 \pi^2 G \Sigma / \kappa^2.
\end{equation}
A regime of gravitational instability occurs if there are
wavelengths that lie between the regimes of pressure and rotational
support, with 
\begin{equation}
\frac{c_{\rm s}^2}{G\Sigma} < \lambda < \frac{4 \pi^2 G\Sigma}{\kappa^2}.
\end{equation}
This will occur if 
\begin{equation} \label{eqn:Toomre}
Q = c_{\rm s}\kappa / 2\pi G \Sigma < 1,
\end{equation}
which is the Toomre criterion for gravitational instability to within
a factor of two.  The full criterion from a linear analysis of
the equations of motion of gas in a shearing disk gives a factor of
$\pi$ in the denominator (Safronov 1960, Goldreich \& Lynden-Bell
1965), while a kinetic theory approach appropriate for a collisionless
stellar system gives a factor of 3.36 (Toomre 1964).

Kennicutt (1989) and Martin \& Kennicutt (2001) have demonstrated that
the Toomre criterion generally can explain the location of the edge of
the star-forming disk in galaxies, although they must introduce a
correction factor $\alpha = 0.69\pm 0.2$ into the left-hand-side of
equation~(\ref{eqn:Toomre}). Schaye (2003) notes that this factor
should be corrected to $\alpha = 0.53$ to account for the use of
the velocity dispersion rather than sound velocity, and the exact
Toomre criterion for a stellar rather than a gas disk.

The Toomre criterion given in Eq.\ (\ref{eqn:Toomre}) was derived
for a pure gas disk with uniform temperature and velocity dispersion,
and no magnetic field.  Relaxation of each of these assumptions
modifies the criterion, and indeed each has been argued to be the
controlling factor in determining star formation thresholds by
different authors.

Stars in a gas disk respond as a collisionless fluid to density
perturbations large compared to their mean separation.  Jog \& Solomon
(1984a) computed the Toomre instability in a disk composed of gas
and stars, and found it to always be more unstable than either
component considered individually.  Both components contribute to the
growth of density perturbations, allowing gravitational collapse to
occur more easily.   Taking into account both
gas (subscript $g$) and stars (subscript $r$), instability occurs when
\begin{equation}
2\pi G k \left(\frac{\Sigma_r}{\kappa^2 + k^2 c_{sr}^2} +
\frac{\Sigma_g}{\kappa^2 + k^2 c_{sg}^2} \right) > 1,
\end{equation}
where $k = 2\pi/\lambda$ is the wavenumber of the perturbation
considered.  
Jog \& Solomon (1984a) and Romeo (1992) extended this model to include
the effect of the finite thickness of the disk.  Elmegreen (1995) was
able with some effort to derive an effective Toomre parameter that
includes the effects of both stars and gas, but that can only be
analytically computed in the thin disk limit.  To compute it,
independent measures of the velocity dispersion of the stars and of
the gas are, of course, needed. Jog (1996) numerically computed the
effective stability parameter for a wide range of values of stellar
and gas disk parameters. The contribution of the stellar disk may
alone be sufficient to explain the correction factors found by
Kennicutt (1989) and Martin \& Kennicutt (2001).

Magnetic fields offer direct support against collapse through their
magnetic pressure and tension.  However, Chandrasekhar (1954) and
Lynden-Bell (1966) were the first to note that they can also have the
less expected effect of destabilization of a rotating system.  The
magnetic field in this case acts to brake the shear that would
otherwise prevent collapse, redistributing angular momentum and
allowing collapse to occur down field lines.  Elmegreen (1987)
performed a linear analysis of the growth rate of gravitational
instability in a rotating, magnetized disk, which was extended by Fan
\& Lou (1997) to follow the excitation of the different modes.  Kim \&
Ostriker (2001) determined when magnetic field acts to prevent or to
promote collapse.  When shear is strong, as it is in the parts of
galactic disks with flat rotation curves, and the field is moderate or
weak, with plasma $\beta \leq 1$, swing amplification stabilized by
magnetic pressure dominates.  Sufficiently unstable disks, with Toomre
$Q \leq $1.0--1.1 (depending on field strength), collapse due to
nonlinear secondary instabilities despite magnetic stabilization.  On
the other hand, if shear is weak, and fields are stronger ($\beta >
1$), magnetic tension forces act against epicyclic motions, reducing
their stabilizing effect, and producing magneto-Jeans instabilities
along the field lines.  This leads to large regions of gravitational
collapse.  In the outer parts of disks, the collapse rate from swing
amplification is so slow that additional effects such as spiral arm
amplification may be important to drive the formation of observed
regions of star formation.  Kim \& Ostriker (2002) show that the
introduction of spiral arms indeed produce feather-like features
similar to those observed in spiral galaxies, with masses comparable
to the largest star forming regions.  These results suggest that the
presence of magnetic fields can actually enhance the star formation
rate in some parts of galactic
disks.

The temperature and the velocity of the coldest gas in a multi-phase
interstellar medium at any point in the disk may be the determining
factor for gravitational instability, rather than some average
temperature.  Schaye (2003) suggests that the sharp rise in
temperature associated with the lack of molecular gas causes the sharp
drop in the star formation rate at the edges of disk galaxies.  This
reverses the argument of Elmegreen \& Parravano (1994) who suggested
that the lack of gravitational instability prevents cooling. Schaye
(2003) derives the disk surface density required to allow molecule
formation in the presence of the intergalactic ultraviolet background
field and suggests that this is consistent with the observed threshold
column densities.  However, Martin \& Kennicutt (2001) show a wide
variation in the atomic gas fraction at the critical radius (see their
Figure~9a), calling this idea into question.

The balance between gravitation and local shear is argued by Hunter,
Elmegreen, \& Baker (1998) to be a better criterion than the Toomre
(1964) criterion, which balances gravitation against Coriolis
forces. Effectively this substitutes the Oort A constant (Binney \&
Tremaine 1997) for the epicyclic frequency $\kappa$ in Eq.\
(\ref{eqn:Toomre}).  The difference is small (of order 10\%) in
galaxies with flat rotation curves, but can lower the critical density
substantially in galaxies with rising rotation curves, such as dwarf
galaxies.

\subsubsection{Thermal instability}
\label{subsub:thermal}
Thermal instability has been the organizing principle behind the most
influential models of the ISM (Field \etal\ 1969; McKee \& Ostriker
1977; Wolfire \etal\ 1995). Under the assumption of approximate
pressure and thermal equilibrium, thermal instability can explain the
widely varying densities observed in the ISM. It can not explain the
order of magnitude higher pressures observed in molecular clouds,
though, so it was thought that most molecular clouds must be confined
by their own self-gravity. Turbulent pressure fluctuations in a medium
with effective adiabatic index less than unity (that is, one that
cools when compressed, like the ISM) can provide an alternative
explanation for both pressure and density fluctuations. Although
thermal instability exists, it does not necessarily act as the primary
structuring agent, nor, therefore, as the determining factor for the
star formation rate.

Thermal instability occurs when small perturbations from thermal
equilibrium grow.  The dependence on density $\rho$ and temperature
$T$ of the heat-loss function ${\cal L} = \Lambda - \Gamma$, the sum
of
the rate of specific 
energy loss minus gain, 
determines whether instability occurs.  Parker (1953) derived the
isochoric instability condition, while Field (1965) pointed out that
cooling inevitably causes density changes, either due to dynamical
flows if the region is not isobaric, or due to pressure changes if it
is.  He then derived the isobaric instability condition.  The
alternative of dynamical compression in a region large enough to be
unable to maintain isobaric conditions has recently received renewed
attention, as described below.

The isobaric instability condition derived by Field (1965) is
\begin{equation} \label{eq:isobaric}
\left(\frac{\partial {\cal L}}{\partial T}\right)_P =
\left(\frac{\partial {\cal L}}{\partial T}\right)_{\rho} - \frac{\rho_0}{T_0}
\left(\frac{\partial {\cal L}}{\partial \rho}\right)_T < 0,
\end{equation}
where $\rho_0$ and $T_0$ are the equilibrium values.  Optically thin
radiative cooling in the interstellar medium gives a cooling function
that can be expressed as a piecewise power law $\Lambda \propto \rho^2
T^{\beta_i}$, where $\beta_i$ gives the value for a temperature range
$T_{i-1} < T < T_i$, while photoelectric heating is independent of
temperature.  Isobaric instability occurs when $\beta_i < 1$, while
isochoric instability only occurs with $\beta_i < 0$
(e.g. Field 1965).

In interstellar gas cooling with equilibrium ionization, there are two
temperature ranges subject to thermal instability (Pikel'ner 1968,
Field \etal\ 1969).  In the standard picture of the three-phase
interstellar medium governed by thermal instability (McKee \& Ostriker
1977), the higher of these, with temperatures $10^{4.5}\,{\rm{K}}<T<10^7\,$K
(Raymond, Cox, \& Smith 1977), separates hot gas from the warm ionized
medium.  The lower range of $10^{1.7}\,{\rm{K}}<T<10^{3.7}\,$K (Fig.\ 3a of
Wolfire \etal\ 1995) separates the warm neutral medium from the cold
neutral medium.  Cooling of gas out of ionization equilibrium has been
studied in a series of papers by Spaans (1996, Spaans \& Norman 1997,
Spaans \& Van Dishoeck 1997, Spaans \& Carollo 1998) as described by
Spaans \& Silk (2000).  The effective adiabatic index depends quite
strongly on the details of the local chemical, dynamical and radiation
environment, in addition to the pressure and temperature of the gas.
Although regions of thermal instability occur, the pressures and
temperatures may depend strongly on the details of the radiative
transfer in a turbulent medium, the local chemical abundances, and
other factors.

When thermal instability occurs, it can drive strong motions that
dynamically compress the gas nonlinearly.  Thereafter, neither the
isobaric nor the isochoric instability conditions hold, and the
structure of the gas is determined by the combination of dynamics and
thermodynamics (Meerson 1996, Burkert \& Lin 2000, Lynden-Bell \&
Tout 2001, S\'anchez-Salcedo \etal\ 2002, Kritsuk \& Norman 2002). 
V\'azquez-Semadeni, Gazol, \& Scalo (2000) examined the behavior of
thermal instability in the presence of driven turbulence, magnetic
fields, and Coriolis forces and concluded that the structuring effect
of the turbulence overwhelmed that of thermal instability in a
realistic environment.  Gazol \etal\ (2001) and S\'anchez-Salcedo,
V\'azquez-Semadeni, \& Gazol (2002) found that about half of the gas
in such a turbulent environment actually has temperatures
falling in the thermally unstable region, and emphasize that a bimodal
temperature distribution may simply be a reflection of the gas cooling
function, not a signature of a discontinuous phase transition.  Mac
Low \etal\ (2003) examined supernova-driven turbulence and found a
broad distribution of pressures, which were more important than
thermal instability in producing a broad range of densities in the
interstellar gas.

Heiles (2001) confirmed the suggestions of Dickey, Salpeter, \&
Terzian (1978), and Mebold \etal\ (1982) that substantial amounts of
gas lie out of thermal equilibrium.  This has provided observational
support for a picture in which turbulent flows rather than thermal
instability dominates structure formation prior to gravitational
collapse.  Heiles (2001) measured the temperature of gas along lines
of sight through the warm and cold neutral medium by comparing
absorption and emission profiles of the H{\sc i} 21 cm fine structure
line.  He found that nearly half of the warm neutral clouds measured
showed temperatures that are unstable according to the application of
the isobaric instability condition, Eq.\ (\ref{eq:isobaric}), to the
Wolfire \etal\ (1995) equilibrium ionization phase diagram.

Although the heating and cooling of the gas clearly plays an important
role in the star formation process, the presence or absence of an
isobaric instability may be less important than the effective
adiabatic index, or similar measures of the behavior of the gas on
compression, in determining its ultimate ability to form stars.

\subsection{Driving mechanisms}
\label{sub:driving}

Both support against gravity and maintenance of observed motions
appear to depend on continued driving of the turbulence, which has
kinetic energy density $e = (1/2) \rho v_{\rm rms}^2$.  Mac Low
(1999, 2002) estimates that the dissipation rate for isothermal, supersonic
turbulence is 
\begin{equation} \label{eqn:dissip}
\dot{e} \simeq -(1/2)\rho v_{\rm rms}^3/L_{\rm d} 
= -(3 \times 10^{-27} \,\mbox{erg}\,\mbox{cm}^{-3}\,\mbox{s}^{-1}) 
\left(\frac{n}{1\,\mbox{cm}^{-3}}\right)
\left(\frac{v_{\rm rms}}{10\,\mbox{km}\,\mbox{s}^{-1}}\right)^3 
\left(\frac{L_{\rm d}}{100 \,\mbox{pc}}\right)^{-1},
\end{equation}
where $L_{\rm d}$ is the driving scale, which we have somewhat arbitrarily
taken to be 100$\,$pc (though it could well be smaller), and we have
assumed a mean mass per particle $\mu= 2.11\times 10^{-24}$~g.  The
dissipation time for turbulent kinetic energy
\begin{equation} \label{eqn:disstime}
\tau_{\rm d} = e / \dot{e} \simeq L_{\rm d}/v_{\rm rms} = (9.8\, \mbox{Myr})
\left(\frac{L_{\rm d}}{100\,\mbox{pc}}\right)
\left(\frac{v_{\rm rms}}{10\,\mbox{km}\,\mbox{s}^{-1}}\right)^{-1},
\end{equation}
which is just the crossing time for the turbulent flow across the
driving scale (Elmegreen 2000b).  What then is the energy source for
this driving? We here review the energy input rates for a number of
possible mechanisms.

\subsubsection{Magnetorotational instabilities}

One energy source for interstellar turbulence that has long been
considered is shear from galactic rotation (Fleck 1981).  However, the
question of how to couple from the large scales of galactic rotation
to smaller scales remained open.  
Sellwood \& Balbus (1999) suggested 
that the magnetorotational instability (Balbus \& Hawley
1991, 1998) could couple the large
and small scales
efficiently.  The instability generates Maxwell stresses (a positive
correlation between radial $B_R$ and azimuthal $B_{\Phi}$ magnetic
field components) that transfer energy from shear into turbulent
motions at a rate 
\begin{equation}
\label{eqn:stress}
\dot{e} = - T_{R\Phi} (d\Omega / d \ln R) =  T_{R\Phi} \Omega,
\end{equation}
where the last equality holds for a flat rotation curve.
(Sellwood \& Balbus 1999).  Numerical models suggest that the Maxwell
stress tensor $T_{R\Phi} \simeq 0.6 B^2/(8\pi)$ (Hawley, Gammie \&
Balbus 1995).  For the Milky Way, the value of the rotation rate
recommended by the IAU is $\Omega = (220 \mbox{ Myr})^{-1} = 1.4
\times 10^{-16} \mbox{ rad s}^{-1}$, though this may be as much as
15\% below the true value (Olling \& Merrifield 1998, 2000).  The
magnetorotational instability may thus contribute energy at a rate
\begin{equation}
\dot{e} = (3 \times 10^{-29}\,\mbox{erg}\,\mbox{cm}^{-3}\,\mbox{s}^{-1})
\left(\frac{B}{3 \mu\mbox{G}}\right)^2 \left(\frac{\Omega}{(220\,\mbox{Myr})^{-1}}\right). 
\end{equation}
For parameters appropriate to the H{\sc i} disk of a sample small
galaxy, NGC~1058, including $\rho = 10^{-24}\,$g$\,$cm$^{-3}$, Sellwood \&
Balbus (1999) find that the magnetic field required to produce the
observed velocity dispersion of 6~km~s$^{-1}$ is roughly 3 $\mu$G, a
reasonable value for such a galaxy. 
This instability may provide a base value for the velocity dispersion
below which no galaxy will fall.  If that is sufficient to prevent
collapse, little or no star formation will occur, producing something
like a low surface brightness galaxy with large amounts of H{\sc i}
and few stars. This may also apply to the outer disk of our own
Milky Way
and other star-forming galaxies.

\subsubsection{Gravitational instabilities}

Motions coming from gravitational collapse have often been suggested as a
local driving mechanism in molecular clouds, but fail due to the quick decay
of the turbulence (\S~\ref{sub:motions}).  If the turbulence decays in less
than a free-fall time, as suggested by Eq.\ (\ref{eqn:decay}), then it
cannot delay collapse for substantially longer than a free-fall time.

On the galactic scale, spiral structure can drive turbulence in gas
disks.  Roberts (1969) first demonstrated that shocks would form in
gas flowing through spiral arms formed by gravitational instabilities
in the stellar disk (Lin \& Shu 1964, Lin, Yuan, \& Shu 1969).  These
shocks were studied in thin disks by Tubbs (1980) and Soukoup \& Yuan
(1981), who found few vertical motions.  More recently, it has been
realized that in a more realistic thick disk, the spiral shock will
take on some properties of a hydraulic bore, with gas passing through
a sudden vertical jump at the position of the shock (Martos \& Cox
1998, G\'omez \& Cox 2002).  Behind the shock, downward flows of as
much as 20~km~s$^{-1}$ appear (G\'omez \& Cox 2002).  Some portion of
this flow will contribute to interstellar turbulence.  However, the
observed presence of interstellar turbulence in irregular galaxies
without spiral arms, as well as in the outer regions of spiral
galaxies beyond the regions where the arms extend suggest that this
cannot be the only mechanism driving turbulence.  A more quantitative
estimate of the energy density contributed by spiral arm driving
has not yet been done.

The interaction between rotational shear and gravitation can, at least
briefly, drive turbulence in a galactic disk, even in the absence of
spiral arms.  Vollmer \& Beckert (2002) describe the consequences of
assuming that this effect fully supplies the energy to drive the
observed turbulent flow, without demonstrating explicitly that this is
the case.  They base their assumption on 2D, high-resolution
(sub-parsec zones) numerical models described in a series of papers by
Wada \& Norman (1999, 2001), Wada, Spaans, \& Kim (2000), and Wada,
Meurer \& Norman (2002).  However, these numerical models all share
two limitations: they do not include the dominant stellar component,
and gravitational collapse cannot occur beneath the grid scale. The
computed filaments of dense gas are thus artificially supported, and
would actually continue to collapse to form stars, rather than driving
turbulence in dense disks.  S\'anchez-Salcedo (2001) gives a detailed
critique of these models.  In very low density disks, where even the
dense filaments remained Toomre stable, this mechanism might operate,
however.

Wada \etal\ (2002) estimated the energy input from this mechanism
using equation~(\ref{eqn:stress}),
where, in the absence of significant Maxwell stresses, the stress
tensor is given by the Newton stresses resulting from correlations in
the gravitational velocity $u_G$ as $T_{R\Phi} =
\langle \rho u_{GR} u_{G\Phi} \rangle$ (Lynden-Bell \& Kalnajs 1972).
The Newton stresses will only add energy if a positive correlation
between radial and azimuthal gravitational forces exists, however,
which is not demonstrated by Wada \etal\ (2002).  Nevertheless, they
estimate the order of magnitude of the energy input from Newton
stresses as
\begin{eqnarray}
\dot{e} & \simeq & G (\Sigma_g/H)^2 \lambda^2 \Omega \nonumber \\
        & \simeq & (4 \times 10^{-29} \mbox{ erg cm$^{-3}$ s}^{-1})
\left(\frac{\Sigma_g}{10 \mbox{ M$_{\odot}$ pc}^{-2}} \right)^2
\left(\frac{H}{100 \mbox{ pc}} \right)^{-2}
\left(\frac{\lambda}{100 \mbox{ pc}} \right)^{2}
\left(\frac{\Omega}{(220 \mbox{ Myr})^{-1}}\right), 
\end{eqnarray}
where $G$ is the gravitational constant, $\Sigma_g$ the density of
gas, $H$, the scale height of the gas, $\lambda$ the length scale of
turbulent perturbations, and $\Omega$ the angular velocity of the disk.  Values
chosen are appropriate for the Milky Way.  This is two orders of
magnitude below the value required to maintain interstellar
turbulence (Eq.~[\ref{eqn:dissip}]).

\subsubsection{Protostellar outflows}

Protostellar jets and outflows are a popular suspect for the energy
source of the observed turbulence
in molecular clouds.  
We can estimate their average energy input rate
into the overall ISM, 
following McKee (1989), by assuming that some fraction $f_{\rm w}$ of
the mass accreted onto a star during its formation is expelled in a
wind travelling at roughly the escape velocity.  Shu \etal\ (1988)
argue that $f_{\rm w} \approx 0.4$, and that most of the mass is
ejected from close to the stellar surface, where the escape velocity
\begin{equation}
v_{\rm esc} = \left(\frac{2GM}{R}\right)^{1/2} = (200\,{\rm km}\,{\rm
  s}^{-1}) 
\left(\frac{M}{1\,{\rm M}_{\odot}}\right)^{1/2} \left(\frac{R}{10\,{\rm R}_{\odot}}\right)^{-1/2},
\end{equation}
with scaling appropriate for a solar-type protostar with
radius $R = 10\,\mbox{R}_{\odot}$. 
Observations of neutral atomic winds from protostars 
show
outflow velocities of roughly this value (Lizano \etal\ 1988,
Giovanardi \etal\ 2000).

The total energy input from protostellar winds will substantially
exceed the amount that can be transferred to the turbulence because of
radiative cooling at the wind termination shock.  We represent the
fraction of energy lost there by $\eta_{\rm w}$.  A reasonable upper limit
to the energy loss is offered by assuming fully effective radiation
and momentum conservation, so that 
\begin{equation}
\label{eqn:mntm-cons}
\eta_{\rm w} < \frac{v_{\rm rms}}{v_{\rm w}} = 0.05 \left(\frac{v_{\rm
      rms}}{10\,\mbox{km}\,{\rm s}^{-1}}\right) \left(\frac{200\,\mbox{km}\,{\rm s}^{-1}}{v_{\rm w}}\right),
\end{equation}
where $v_{\rm rms}$ is the rms velocity of the turbulence, and we have
assumed that the flow is coupled to the turbulence at typical
velocities for the diffuse ISM.  If we assumed that most of the energy
went into driving dense gas, the efficiency would be lower, as typical
rms velocities for CO outflows are only 1--2~km~s$^{-1}$. The energy
injection rate 
\begin{eqnarray}
\dot{e}& = & \frac12 f_{\rm w} \eta_{\rm w} \frac{\dot{\Sigma}_*}{H} v_{\rm w}^2 \nonumber \\
       & \simeq &  (2 \times 10^{-28} \,\mbox{erg}\,\mbox{cm}^{-3}\,\mbox{s}^{-1})
       \left(\frac{H}{200\,\mbox{pc}}\right)^{-1} 
       \left(\frac{f_{\rm w}}{0.4}\right) \times \nonumber \\
& \times  &    \left(\frac{v_{\rm w}}{200\,\mbox{km}\,\mbox{s}^{-1}}\right) 
       \left(\frac{v_{\rm rms}}{10\, \mbox{km}\,{\rm s}^{-1}}\right) 
       \left(\frac{\dot{\Sigma}_*}{4.5 \times 10^{-9}\, \mbox{M$_{\odot}\,$pc$^{-2}\,$yr$^{-1}$}}\right),
\end{eqnarray}
where $\dot{\Sigma}_*$ is the surface density of star formation, and
$H$ is the scale height of the star-forming disk.  The scaling value
used for $\dot{\Sigma}_*$ is the solar neighborhood value (McKee 1989).

Although protostellar jets and winds are indeed quite energetic, they
deposit most of their energy into low density gas (Henning 1989), as
is shown by the observation of multi-parsec long jets extending
completely out of molecular clouds (Bally \& Devine 1994).
Furthermore, observed motions of molecular gas show increasing power
on scales all the way up to and perhaps beyond the largest scale of
molecular cloud complexes (Ossenkopf \& Mac Low 2002).  It is hard to
see how such large scales could be driven by protostars embedded in
the clouds.

\subsubsection{Massive stars}
In active star-forming galaxies, massive stars 
probably
dominate the driving.  They 
could 
do so through ionizing radiation and stellar winds from O~stars, or
clustered and field supernova explosions, predominantly from B~stars
no longer associated with their parent gas.  The supernovae appear
likely to be most important, as we now show.

\paragraph{Stellar winds}
First, we consider stellar winds.  The total energy input from a
line-driven stellar wind over the main-sequence lifetime of an early
O~star can equal the energy from its supernova explosion, and the
Wolf-Rayet wind can be even more powerful.  However, the mass-loss
rate from stellar winds drops as roughly the sixth power of the star's
luminosity if we take into account that stellar luminosity varies as
the fourth power of stellar mass (Vink, de Koter \& Lamers 2000), while
the powerful Wolf-Rayet winds (Nugis \& Lamers 2000) last only $10^5$
years or so, so only the very most massive stars contribute
substantial energy from stellar winds.  The energy from supernova
explosions, on the other hand, remains nearly constant down to the
least massive star that can explode.  As there are far more lower-mass
stars than massive stars, with a Salpeter IMF giving a power-law in
mass of $\alpha = -2.35$ (Eq.\ \ref{eqn:salpeter}), supernova
explosions inevitably dominate over stellar winds after the first
few million years of the lifetime of an OB association.

\paragraph{Ionizing radiation}
Next, we consider ionizing radiation from OB stars.  The total amount
of energy contained in ionizing radiation is vast.  Abbott (1982)
estimates the integrated luminosity of ionizing radiation in the disk
of the Milky Way to be
\begin{equation}
\dot{e} = 1.5 \times 10^{-24} \mbox{ erg s$^{-1}$ cm}^{-3}.
\end{equation}
However, only a small fraction of this total energy goes to driving
interstellar motions.

Ionizing radiation contributes to interstellar turbulence in two
ways. First, it ionizes the diffuse interstellar gas, heating it to
7,000--10,000~K and adding energy to it.  As this gas cools, it
contracts due to thermal instabilities, driving turbulent flows, as
modeled by Kritsuk \& Norman (2002a,b).  They modeled the flow in a
cooling instability after a sudden increase in heating by a factor of
five, and found that a flow with peak thermal energy of $E_{\rm th}$
gains a peak kinetic energy of roughly $E_{\rm kin} = \eta_c E_{\rm
th}$, with $\eta_c \simeq 0.07$.  Parravano, Hollenbach, \& McKee
(2003) find that the local UV radiation field, and thus the
photoelectric heating rate, increases by a factor of
2--3 due to the formation of a nearby OB association every
100--200~Myr.  However, substantial motions only lasted about 1~Myr
after a heating event in the model by Kritsuk \& Norman (2002b).  We
can estimate the energy input from this mechanism on average by taking
the kinetic energy input from the heating event and dividing by the
typical time $\tau_{\rm OB}$ between heating events.  If we take the
thermal energy to be that of $n=1$~cm$^{-3}$ gas at $10^4$~K (perhaps
a bit higher than typical), we find that
\begin{equation}
\dot{e} = \frac32 n k T \eta_c / \tau_{\rm OB} \simeq
       (5\times 10^{-29} \,\mbox{erg}\,\mbox{cm}^{-3}\,\mbox{s}^{-1})
       \left(\frac{n}{1\,\mbox{cm}^{-3}}\right)
       \left(\frac{T}{10^4\,\mbox{K}}\right)
       \left(\frac{\eta_c}{0.07}\right)
       \left(\frac{\tau_{\rm OB}}{100\,\mbox{Myr}}\right)^{-1}.
\end{equation}
Although comparable to some other proposed energy sources discussed
here, this mechanism appears unlikely to be as important as the
supernova explosions from the same OB stars, as discussed below.

The second way that ionization drives turbulence is through driving
the supersonic expansion of H{\sc ii} regions after photoionization
heating raises their pressures above that of the surrounding neutral
gas.  Matzner (2002) computes the momentum input from the expansion of
an individual H{\sc ii} region into a surrounding molecular cloud, as
a function of the cloud mass and the ionizing luminosity of the
central OB association.  By integrating over the H{\sc ii} region
luminosity function derived by McKee \& Williams (1997), he finds that
the average momentum input from a Galactic region is
\begin{equation}
\langle \delta p \rangle \simeq (260 \,\mbox{km}\,\mbox{s}^{-1})
\left(\frac{N_{\rm H}}{1.5 \times 10^{22}\,\mbox{cm}^{-2}}\right)^{-3/14}
\left(\frac{M_{cl}}{10^6\,\mbox{M}_{\odot}}\right)^{1/14} 
\langle M_* \rangle.
\end{equation}
The column density $N_{\rm H}$ is scaled to the mean value for Galactic
molecular clouds (Solomon \etal\ 1987), which varies little as cloud
mass $M_{cl}$ changes.  The mean stellar mass per cluster in the Galaxy
$\langle M_* \rangle = 440 \,\mbox{M}_{\odot}$ (Matzner 2002).

The number of OB associations contributing substantial amounts of
energy can be drawn from the McKee \& Williams (1997) cluster
luminosity function
\begin{equation}
{\cal N} (> S_{49}) = 6.1 \left(\frac{108}{S_{49}} - 1 \right),
\end{equation}
where ${\cal N}$ is the number of associations with ionizing photon
luminosity exceeding $S_{49} = S/(10^{49}$~s$^{-1})$.  The
luminosity function is rather flat below $S_{49} = 2.4$, the theoretical
luminosity of the highest-mass single star considered
(120~M$_{\odot}$)
so taking its value at $S_{49} = 1$ is about right, giving ${\cal N}(>
1) = 650$ clusters.

To derive an energy input rate per unit volume $\dot{e}$ from the mean
momentum input per cluster $\langle \delta p \rangle$, we need to
estimate the average velocity of momentum input $v_{i}$, the time over
which it occurs $t_{i}$, and the volume $V$ under consideration.
Typically expansion will not occur supersonically with respect to the
interior, so $v_{i} < c_{{\rm s},i}$, where $c_{{\rm s},i} \simeq 10$~km~s$^{-1}$
is the sound speed of the ionized gas.  McKee \& Williams (1997) argue
that clusters go through about five generations of massive
star formation, where each generation lasts $\langle t_* \rangle \approx
3.7$~Myr. The scale height for massive clusters is $H_c \sim 100$~pc
(e.g.\ Bronfman \etal\ 2000), and the radius of the star-forming disk
is roughly $R_{sf} \sim 15$~kpc, so the relevant volume $V = 2 \pi
R_{sf}^2 H_c$.  The energy input rate from H{\sc ii} regions is then
\begin{eqnarray}
\dot{e}& =& \frac{\langle \delta p \rangle {\cal N}(>1) v_{i}}{V
t_{i}} \nonumber \\
       & = & (3 \times 10^{-30} \mbox{ erg s$^{-1}$ cm}^{-3})
\left(\frac{N_{\rm H}}{1.5\times 10^{22} \mbox{ cm}^{-2}}\right)^{-3/14}
\left(\frac{M_{cl}}{10^6 \mbox{ M}_{\odot}}\right)^{1/14}
\left(\frac{\langle M_* \rangle}{440 \mbox{ M}_{\odot}}\right)
\left(\frac{{\cal N}(>1)}{650}\right) \times \nonumber \\
       & \times  &  \left(\frac{v_{i}}{10 \mbox{ km s}^{-1}}\right)
\left(\frac{H_c}{100 \mbox{ pc}}\right)^{-1}
\left(\frac{R_{sf}}{15 \mbox{ kpc}}\right)^{-2}
\left(\frac{t_{i}}{18.5 \mbox{ Myr}}\right)^{-1}, 
\end{eqnarray}
where all the scalings are appropriate for the Milky Way as discussed
above.  Nearly all of the energy in ionizing radiation goes towards
maintaining the ionization and temperature of the diffuse  medium, and hardly
any towards driving turbulence.  Flows of ionized gas may be important
very close to young clusters and may terminate star formation locally
(\S~\ref{sub:termination}), but do not appear to contribute
significantly on a global scale.

\paragraph{Supernovae}
The largest contribution from massive stars to interstellar turbulence
comes from supernova explosions.  To estimate their energy input rate,
we begin by finding the supernova rate in the Galaxy $\sigma_{SN}$.
Cappellaro \etal\ (1999) estimate the total supernova rate in
supernova units to be $0.72 \pm 0.21$ SNu for galaxies of type S0a-b
and $1.21 \pm 0.37$~SNu for galaxies of type Sbc-d, where 1 SNu = 1 SN
(100~yr)$^{-1} (10^{10} L_B/\mbox{L}_{\odot})^{-1}$, and $L_B$ is the
blue luminosity of the galaxy.  Taking the Milky Way as lying between
Sb and Sbc, we estimate $\sigma_{SN} = 1$~SNu.  Using a Galactic
luminosity of $L_B = 2 \times 10^{10} \mbox{ L}_{\odot}$, we find a
supernova rate of (50~yr)$^{-1}$, which agrees well with the estimate
in equation~(A4) of McKee (1989).  If we use the same scale height
$H_c$ and star-forming radius $R_{sf}$ as above, we can compute the
energy input rate from supernova explosions with energy $E_{SN} =
10^{51}\,$erg to be
\begin{eqnarray}
\dot{e} & = &\frac{\sigma_{SN} \eta_{SN} E_{SN}}{\pi R_{sf}^2 H_c} \nonumber \\
       &  = & (3 \times 10^{-26} \mbox{ erg s$^{-1}$ cm}^{-3})
\left(\frac{\eta_{SN}}{0.1} \right)
\left(\frac{\sigma_{SN}}{1 \mbox{ SNu}} \right) 
\left(\frac{H_c}{100 \mbox{ pc}} \right)^{-1}
\left(\frac{R_{sf}}{15 \mbox{ kpc}} \right)^{-2}
\left(\frac{E_{SN}}{10^{51} \mbox{ erg}} \right).
\end{eqnarray}
The efficiency of energy transfer from supernova blast waves to the
interstellar gas $\eta_{SN}$ depends on the strength of radiative
cooling in the initial shock, which will be much stronger in the
absence of a surrounding superbubble (e.g.\ Heiles 1990).  Substantial
amounts of energy can escape in the vertical direction in superbubbles
as well, however.  Norman \& Ferrara (1996) make an analytic estimate
of the effectiveness of driving by SN remnants and superbubbles. The
scaling factor $\eta_{SN} \simeq 0.1$ used here was derived by
Thornton \etal\ (1998) from detailed, 1D, numerical
simulations of SNe expanding in a uniform ISM.  It can alternatively be
drawn from momentum conservation arguments
(eq.~\ref{eqn:mntm-cons}), 
comparing a typical expansion velocity of 100$\,$km$\,$s$^{-1}$ to
typical interstellar turbulence velocity of 10$\,$km$\,$s$^{-1}$. 
Multi-dimensional models of the interactions of multiple SN remnants
(e.g.\ Avillez 2000) are required to better determine the
effective scaling factor.

Supernova driving appears to be powerful enough to maintain the
turbulence even with the dissipation rates estimated in
Eq.~(\ref{eqn:dissip}).  It provides a large-scale
self-regulation mechanism for star formation in disks with sufficient
gas density to collapse despite the velocity dispersion produced by
the magnetorotational instability.  As star formation increases in
such galaxies, the number of OB stars increases, ultimately increasing
the supernova rate and thus the velocity dispersion, which restrains
further star formation.

\subsection{Applications}
\label{sub:applications}

The theory of star formation controlled by supersonic turbulence
offers a unified approach to a wide range of astrophysical objects.
In this section we discuss several illustrative scenarios, moving from
low to high star formation efficiencies.

\subsubsection{Low surface brightness galaxies}

Low surface brightness galaxies have large fractions of their baryonic
mass in gas, whether they have masses typical of massive (Schombert
\etal\ 1992, McGaugh \& de Blok 1997) or dwarf galaxies
(Schombert, McGaugh, \& Eder 2001).  Nevertheless, their star
formation rates lie well below typical values for high surface
brightness galaxies (van der Hulst \etal\ 1993; McGaugh \& de Blok
1997). Their rotation curves have been derived from both H{\sc i}
measurements (van der Hulst \etal\ 1993, de Blok, McGaugh, \& van der
Hulst, 1996), and higher resolution H$\alpha$ measurements (Swaters,
Madore, \& Trewhalla, 2000; McGaugh, Rubin, \& de Blok, 2001, Matthews
\& Gallagher 2002), which may sometimes disagree with the H{\sc
i} in the innermost regions (Swaters \etal\ 2000), but are in
generally good agreement (McGaugh \etal\ 2001).  They have lower gas
and stellar surface densities than high surface brightness galaxies
(van der Hulst \etal\ 1987; de Blok \& McGaugh 1996). The question of
whether their disks have surface densities lying below the Kennicutt
(1989) threshold for star formation has been studied using rotation
curves derived from H{\sc i} measurements for both massive (van der
Hulst \etal\ 1993) and dwarf (van Zee \etal\ 1997) galaxies.  

In the case of massive galaxies, surface densities beneath the
Kennicutt (1989) threshold do indeed appear to explain the lack of
star formation (van der Hulst \etal\ 1993).  The moderate levels of
turbulence required to maintain the observed velocity dispersions may
be produced by magnetorotational instabilities (Sellwood \& Balbus
1999).  Other explanations for the lack of star formation, such as an
inability to form molecular hydrogen (Gerritsen \& de Blok 1999) or to
cool it (Mihos, Spaans, \& McGaugh 1999), were derived from numerical
models that did not include magnetic effects, and thus had no source
of support other than thermal pressure to counteract gravitational
collapse and star formation. If magnetorotational instability is the
dominant support mechanism, then star formation will not be suppressed
in the center, where the rotational shear drops.  This is, in fact, where
star formation is found in low surface brightness galaxies.

In the case of dwarf galaxies (Hunter 1997), the situation appears to
be slightly more complex. On the one hand, Van Zee \etal\ (1997)
demonstrate that the surface density in a sample of low surface
brightness dwarf galaxies falls systematically below the Kennicutt
threshold, with star formation indeed observed in regions that
approach the threshold, and van Zee, Skillman, \& Salzer (1998) show
that blue compact dwarf galaxies have surface densities exceeding the
threshold in their centers, where star formation occurs.  On the other
hand, Hunter, Elmegreen, \& Baker (1998) argue that a criterion based
on local shear correlates better with the observations, especially in
galaxies with rising rotation curves.  Furthermore, another factor
that may contribute to the star formation histories of dwarf
galaxies is that starbursts in the smaller ones (under
$10^8$~M$_{\odot}$) can actually push all the gas well out into the
halo, from where it will take some hundreds of millions of years to
collect back in the center (Mac Low
\& Ferrara 1999).  This last scenario may be consistent with observations
in some galaxies, as summarized by Simpson \& Gottesman (2000).

\subsubsection{Galactic disks}
\label{subsub:disks}

In normal galactic disks, where SNe appear to dominate the driving of
the turbulence, we speculate that most regions will have a star
formation rate just sufficient to produce turbulence that can balance
the local surface density in a self-regulating fashion.  However, as
spiral arms or other dynamical features increase the local density,
this balance would fail, leading to higher local star formation rates.
Because the increase in star formation rate as turbulence is
overwhelmed is not sudden but gradual, the enhanced star formation in
spiral arms and similar structures should not globally approach
starburst rates except when density rises sharply.  Locally, however,
even relatively small regions can reach starburst-like star formation
efficiencies if they exceed the local threshold for turbulent support
and begin to collapse freely.  A classic example of this is the
massive star formation region NGC~3603, which locally resembles a
starburst knot, even though the Milky Way globally does not have a
large star formation rate.  On a smaller scale, even the Trapezium
cluster in Orion seems to have formed with efficiency of $\sil 50$\%
(Hillenbrand \& Hartmann 1998).

Triggering of star formation by compressive shocks from nearby
star-forming regions (Elmegreen \& Lada 1977) is a special case of
global support from turbulence leading to local collapse.  Although
prompt blast waves from winds and early supernovae of OB association
can compress nearby gas and induce collapse, most of the energy from
that association is released at later times as the less massive B
stars explode, driving the larger-scale interstellar turbulence that
provides support against general collapse. The instances of apparent
triggering seen both in linear sequences of OB associations
(e.g. Blaauw 1964) and in shells (e.g. Walborn \& Parker 1992, Efremov
\& Elmegreen 1998b, Kamaya 1998, Barb\'a \etal\ 2003) may represent
this prompt triggering.  It seems unlikely, however, that this prompt
triggering will dominate large-scale star formation as first suggested
by Gerola \& Seiden (1978), and since developed by Neukirch \&
Feitzinger (1988), Korchagin et al. (1995), and Nomura \& Kamaya
(2001).  Compression due to supersonic turbulent flows is suggested to
be the main mechanism leading to stellar birth in gas-rich dwarf
galaxies without spiral density waves, such as Holmberg~II (e.g
Stewart \etal\ 2000).  However, the rate compression-induced star
formation is small compared to the rates expected for global collapse,
which is effectively prevented by the same turbulent flows.

\subsubsection{Globular clusters}

Globular clusters may simply be the upper end of the range of normal
cluster formation.  Whitmore (2000) reviews evidence showing that
young clusters have a power-law distribution reaching up to globular
cluster mass ranges.  The luminosity function for old globular
clusters is log normal, which Fall \& Zhang (2001) attribute to the
evaporation of smaller clusters by two-body relaxation, and the
destruction of the largest clusters by dynamical interactions with the
background galaxy (also see Vesperini 2000, 2001).  Fall \& Zhang
(2001) suggest that the power-law distribution of young clusters is
related to the power-law distribution of molecular cloud masses found
by Harris \& Pudritz (1995).  However, numerical models of
gravitational collapse tend to produce mass distributions that appear
more log-normal, and are not closely related in shape to the
underlying mass distributions of density peaks (Klessen 2001, Klessen
\etal\ 2000).  It remains unknown whether cluster masses are
determined by the same processes as the masses of individual
collapsing objects, but the simulations do not include any physics
that would limit them to one scale and not the other.  Further
investigation of this question will be interesting.

\subsubsection{Galactic nuclei}

In galaxies with low star formation rates, the galactic nucleus is often the
only region with substantial star formation occurring. As rotation curves
approach solid body in the centers of galaxies, magnetorotational
instabilities die away, leaving less turbulent support and perhaps
greater opportunity for star formation. In more massive galaxies, gas is often
funneled towards the center by bars and other disk instabilities, again
increasing the local density sufficiently to overwhelm local turbulence and
drive star formation.

Hunter \etal\ (1998) and Schaye (2003) note that central regions of
galaxies have normal star formation despite having surface densities
that appear to be stable according to the Toomre criterion.  This
could be due to reduced turbulence in these regions decreasing the
surface density required for efficient star formation.  The radial
dependence of the velocity dispersion is difficult to determine,
because H~{\sc i} observations with sufficient velocity resolution to
measure typical turbulent linewidths of 6--12~km~s$^{-1}$ have rather
low spatial resolution, with just a few beams across the galaxy.  Most
calculations of the critical surface density therefore assume a
constant value of the turbulent velocity dispersion, which may well be
incorrect (Wong \& Blitz 2002).

As an alternative, or perhaps additional explanation, Kim \& Ostriker
(2001) point out that the magneto-Jeans instability acts strongly in
the centers of galaxies.  The magnetic tension from strong magnetic
fields can reduce or eliminate the stabilizing effects from Coriolis
forces in these low shear regions, effectively reducing the problem to
a 2D Jeans stability problem along the field lines.

\subsubsection{Primordial dwarfs}

In the complete absence of metals, cooling becomes much more
difficult.  Thermal pressure supports gas that accumulates in dark
matter halos until the local Jeans mass is exceeded.  The first
objects that can collapse are the ones that can cool from H$_2$
formation through gas phase reactions.  Abel, Bryan, \& Norman (2000,
2002) and Bromm, Coppi, \& Larson (1999) have computed models of the
collapse of these first objects.  Abel \etal\ (2000, 2002) used
realistic cosmological initial conditions, and found that inevitably a
single star formed at the highest density peak before substantial
collapse had occurred elsewhere in the galaxy.  Bromm \etal\ (1999)
used a flat-top density perturbation that was able to fragment in many
places simultaneously, due to its artificial symmetry.

Li, Klessen, \& Mac Low (2003) suggest that the lack of fragmentation
seen by Abel \etal\ (2000, 2002) may be due to the relatively stiff
equation of state of metal-free gas.  Li \etal\ (2003) found that
fragmentation of gravitationally collapsing gas is strongly influenced
by the polytropic index $\gamma$ of the gas, with fragmentation
continuously decreasing from $\gamma \sim 0.2$ to $\gamma \sim 1.3$.
The limited cooling available to primordial gas even with significant
molecular fraction may raise its polytropic index sufficiently to
suppress fragmentation.  Abel \etal\ (2000, 2002) argue that the resulting
stars are likely to have masses exceeding 100$\,$M$_{\odot}$, leading
to prompt supernova explosions with accompanying metal pollution and
radiative dissociation of H$_2$.

\subsubsection{Starburst galaxies}

Starburst galaxies convert gas into stars at such enormous rates that
the timescale to exhaust the available material becomes short compared
to the age of the universe (see the review by Sanders \& Mirabel
1996). Starbursts 
typically last for a few tens or hundreds of millions of
years. However, they may occur several times during the life of a
galaxy.  The star formation rates in starburst galaxies can be as high
as 1000$\,$M$_{\odot}\,$yr$^{-1}$ (Kennicutt 1998), some three orders
of magnitude above the current rate of the Milky Way.  Starburst
galaxies are rare in the local universe, but rapidly increase in
frequency at larger lookback times, suggesting that starbursts are
characteristic of early galaxy evolution at high redshifts. The
strongest starbursts occur in galactic nuclei or circumnuclear
regions.

However, in interacting galaxies, strong star formation is also
triggered far away from the nucleus in the overlapping regions, in
spiral arms, or sometimes even in tidal 
tails.
In these interactions a significant number of super-star clusters
form, which may be the progenitors of present-day globular clusters
(Zhang \& Fall 1999, Whitmore \etal\ 1999), or even compact elliptical
galaxies (Fellhauser \& Kroupa 2002).  The Antennae galaxy, the
product of a major merger of the spiral galaxies NGC~4038 and 4039, is
a famous example where star formation is most intense in the overlap
region between the two galaxies (Whitmore \& Schweizer 1995). Merging
events always seem to be associated with the most massive and luminous
starburst galaxies, the ultraluminous IR galaxies identified by
Sanders \& Mirabel (1996).

Gentler minor mergers can also trigger starbursts. Such an event
disturbs but does not disrupt the primary galaxy. It recovers from the
interaction without dramatic changes in its overall morphology. 
This could explain the origin of lower-mass, luminous, blue, compact
galaxies, which often show very little or no sign of interaction
(e.g.\ van Zee, Salzer, \& Skillman 2001). Alternative triggers of the
starburst phenomenon that have have been suggested for these galaxies
include bar instabilities in the galactic disk (Shlosman, Begelman,
\& Frank 1990), or the compressional effects of multiple
supernovae and winds from massive stars (e.g.\ Heckman, Armus, \&
Miley 1990), which then would lead to a very localized burst of star
formation. 

Regardless of the details of the different starburst triggering
mechanisms, they all focus gas into a concentrated region quickly
enough to overwhelm the local turbulence and any additional
turbulence driven by newly formed stars.  Combes (2001) argues
that this can only be accomplished by gravitational torques on the
gas. We suggest that starburst galaxies are just extreme examples of
the continuum of star formation phenomena, with gravity overwhelming
support from turbulent gas motions on kiloparsec scales rather than
the parsec scales of individual OB associations, or the
even smaller scales of low-mass star formation.

\section{CONCLUSIONS}
\label{sec:conclusions}

\subsection{Summary}
\label{sub:summary}

The formation of stars represents the triumph of gravity over a
succession of opponents.  These include thermal pressure, turbulent
flows, magnetic flux, and angular momentum.  
For several decades, magnetic fields were thought to dominate the
resistance against gravity, with star formation occurring
quasistatically as ambipolar diffusion allows collapse of neutral gas
towards the center of magnetically supported cores.  However, a
growing body of observational evidence suggests that when stars do
form they do so quickly and dynamically, with gravitational collapse
occuring at a rate controlled by supersonic turbulence driven at
scales of order a hundred parsecs.  Such turbulence can explain the
superthermal linewidths and self-similar structure observed in
star-forming clouds, while magnetic fields fail to do so.  The
varying balance between turbulence and gravity then provides a natural
explanation for the widely varying star formation rates seen at both
cloud and galactic scales.

Scattered, inefficient star formation is a signpost of turbulent
support, while clustered, efficient star formation occurs in regions
lacking support.  In this picture, gravity has already won in observed
dense protostellar cores: dynamical collapse seems to explain their
observed properties better than the alternatives.  On the other hand,
molecular clouds as a whole may be able to form from turbulent
compression, rather than being dominated by self-gravity.  The mass
distribution of stars then depends at least partly on the density and
velocity structure resulting from the turbulence, perhaps explaining
the apparent local variations of the stellar initial mass function
(IMF) despite its broad universality.

We began by summarizing observations of the structure and properties
of molecular clouds and the mass distribution of stars that form in
them in \S\ \ref{sec:molecular}. Observations of self-similar
structure in molecular clouds ({\S}s \ref{sub:mol-clouds} and
\ref{sub:LSS}) seem to indicate that interstellar turbulence is driven
on scales substantially larger than the clouds themselves (see also
\S\ \ref{sub:driving}).  Molecular clouds appear to actually be
transient objects with lifetimes of several million years that form
and dissolve in the larger scale turbulent flow. Some well-known
descriptions of the clouds like Larson's (1981) size-linewidth
relation may be natural consequences of the turbulent gas flow
observed in projection (\S~\ref{sub:scaling-law}). However, Larson's
mass-radius relation appears to be an observational artifact,
suggesting that most molecular clouds are not in virial equilibrium.

In Section \ref{sec:history} we gave a
historical overview of our understanding of star formation, beginning
with the classical dynamical theory of star formation (\S\
\ref{sub:classical}), which already included turbulent flows, but only
in the microturbulent approximation, treating them as an addition to
the thermal pressure.  We then turned to the development of the standard
theory of star formation (\S\ \ref{sub:standard}) which was motivated
by growing understanding of the importance of the interstellar
magnetic field in the 1960's and 1970's, as discussed in
\S~\ref{sub:classicalprobs}.

The standard theory relies on ion-neutral drift, also known as
ambipolar diffusion, to solve the magnetic flux problem for
protostellar cores, which were thought to be initially
magnetostatically supported. At the same time magnetic tension
acts to brake rotating cores, thus solving the angular
momentum problem as well. The timescale for ambipolar diffusion to
remove enough magnetic flux from the cores for gravitational collapse
to set in can exceed the free-fall time by as much as an order of
magnitude, suggesting that magnetic support could also explain low
observed star formation rates.  Finally, magnetic fields were also
invoked to explain observed supersonic motions as Alfv\'en waves.

Recently, however, both observational and theoretical results have
begun to cast doubt on the standard theory. In \S\
\ref{sub:standardprobs} we summarized theoretical limitations
of the singular isothermal sphere model that forms the basis for many
of the practical applications of the standard theory.  We then discussed
several observational findings that put the fundamental assumptions of
that theory into question. The observed magnetic field strengths in
molecular cloud cores appear too weak to support against gravitational
collapse. At the same time, the infall motions measured around star
forming cores extend too broadly, while the central density profiles
of cores are flatter than expected for isothermal spheres.
Furthermore, the chemically derived ages of cloud cores are
comparable to the free-fall time instead of the much longer ambipolar
diffusion timescale. Observations of young stellar objects also appear
discordant.  Accretion rates appear to decrease rather than remaining
constant, far more embedded objects have been detected in cloud cores
than predicted, and the spread of stellar ages in young clusters does
not approach the ambipolar diffusion time.

New theoretical and numerical studies of turbulence that point beyond
the standard theory while looking back to the classical dynamical
theory for inspiration have now emerged (\S\ \ref{sec:paradigm}).
Numerical models show that supersonic turbulence decays rapidly, in
roughly a crossing time of the region under consideration, regardless
of magnetic field strength.  Under molecular cloud conditions, such
turbulence decays in less than a free-fall time.  This implies that
the turbulence in star-forming clouds needs to be continuously driven
in order to maintain the observed motions.  Driven turbulence has long
been thought capable of supporting gas against gravitational collapse.
A numerical test demonstrated that turbulence indeed can offer global
support, but at the same time can lead to local collapse on small
scales. In strongly compressible turbulence, gravitational collapse
occurs in the density enhancements produced by shocks.  The rate of
local collapse depends strongly on the strength and driving scale of
the turbulence. This gives a natural explanation for widely varying
star formation rates.  Magnetic fields not strong enough to provide
static support make a quantitative but not a qualitative
difference. They are capable of reducing the collapse rate somewhat,
but not of preventing collapse altogether.  They may still act to
transfer angular momentum so long as they are coupled to the gas,
however.

We outlined the shape of the new theory in \S\ \ref{sub:new}.  Rather
than relying on quasistatic evolution of magnetostatically supported
objects, it suggests that supersonic turbulence controls star
formation.  Inefficient, isolated star formation is a hallmark of
turbulent support, while efficient, clustered star formation occurs in
its absence.  When stars form, they do so dynamically, collapsing on
the local free-fall time.  The initial conditions of clusters appear
largely determined by the properties of the turbulent gas, as is the
rate of mass accretion onto these objects.  The balance between
turbulent support and local density then determines the star formation
rate. Turbulent support is provided by some combination of supernovae and
galactic rotation, along with possible contributions from other
processes.  Local density is determined by galactic dynamics including
galaxy interactions, along with the balance between heating and cooling in a
region.  The initial mass function is at least partly determined by
the initial distribution of density resulting from turbulent flows,
although a contribution from stellar feedback and interactions with
nearby stars cannot yet be ruled out. 

We explored the implications of the control of star formation by
supersonic turbulence at the scale of individual stars and stellar
clusters in \S\ \ref{sec:local}.  We examined how turbulent
fragmentation determines the star forming properties of molecular
clouds (\S\ \ref{sub:multiple}), and then turned to discuss
protostellar cores (\S\ \ref{sub:cores}), binary stars (\S\
\ref{sub:binary}), and stellar clusters (\S~\ref{sub:clusters}) in
particular.  Strongly time-varying protostellar mass growth rates may
result as a natural consequence of competitive accretion in nascent
embedded clusters (\S\ \ref{sub:accretion}).  Turbulent models predict
protostellar mass distributions (\S~\ref{sub:imf}) that appear roughly
consistent with the observed stellar mass spectrum (\S\
\ref{sub:IMF-observed}), although more work needs to be done to arrive
at a full understanding of the origin of stellar masses.

The same balance between turbulence and gravity that seems to
determine the efficiency of star formation in molecular clouds also
works at galactic scales, as we discussed in \S~\ref{sec:galactic}.
The transient nature of molecular clouds suggests that they form and
are dispersed in either of two ways.  One possibility is that they
form during large-scale gravitational collapse, and are dispersed
quickly thereafter by radiation and supernovae from the resulting
violent internal star formation.  The other possibility is that
large-scale turbulent flows in galactic disks compress and cool
gas. These same flows will continue to drive the turbulent motions
observed within the clouds.  Some combination of turbulent flow, free
expansion at the sound speed of the cloud, and dissociating radiation
from internal star formation will then be responsible for their
destruction on a timescale of 5--10~Myr (\S\ \ref{sub:clouds}).

Having considered the formation of molecular clouds from the
interstellar gas, we then discuss the role of differential rotation
and thermal instability competing and cooperating with turbulence to
determine the overall star formation efficiency in
\S~\ref{sub:efficient}.  We examined the physical mechanisms that
could drive the interstellar turbulence, focusing on the energy
available from each mechanism in \S\ \ref{sub:driving}.  In
star-forming regions of disks, supernovae appear to overwhelm all
other possibilities.  In outer disks and low surface brightness
galaxies, on the other hand, the situation is not so clear:
magnetorotational or gravitational instabilities look most likely to
drive the observed flows but further work is required on these
regions.  Finally, in \S\ \ref{sub:applications}, we gave examples of
how this picture may apply to different types of objects, including
low surface brightness, normal, and starburst galaxies, as well as
galactic nuclei and globular clusters.  We argue that efficient star
formation occurs at all scales when gravity overwhelms turbulence,
with the result ranging from a single low-mass star at the very
smallest scale to a starburst at the very largest scale.

\subsection{Future research problems}
\label{sub:future}

Although the outline of a new theory of star formation has emerged, it
is by no means complete.  The ultimate goal of a predictive,
quantitative theory of the star formation rate and stellar initial mass
function remains elusive.  It may be that the problem is intrinsically
so complex, like terrestrial climate, that no single solution exists,
but only a series of temporary, quasi-steady states.  Certainly our
understanding of the details of the star formation process can be
improved, though.  Eventually, coupled models capturing different
scales will be necessary to follow the interaction of the turbulent
cascade with the thermodynamics, chemistry, and opacity of the gas at
different densities.  We can identify several major questions that
summarize the outstanding problems. As we merely want to summarize
these open issues in star formation, we refrain from giving an
in-depth discussion and the associated references, which may largely
be found in the body of the review.

{\em How can we describe turbulence driven by astrophysical
processes?}  There is really no single driving scale, because of the
non-uniformity of explosions, and perhaps of other drivers. 
However, a good description of the structure around the driving scales
remains essential, as the largest perturbations lie at the largest
scales in any turbulent flow.  This remains to be found.  The length
of the self-similar turbulent cascade also depends on the scales on
which the driving acts. The self-similarity of the turbulent cascade
is further perturbed by the drastic changes in the equation of state
that occur as increasing densities lead first to stronger radiative
cooling, and then to the reduction of heating by the exclusion of
first ionizing radiation, and then cosmic rays.  Finally, at small
scales, diffusion and dissipation mechanisms determine the structure.
Although ambipolar diffusion probably limits the production of
small-scale magnetic field structures, there is increasing theoretical
support for additional density and velocity structure at scales below
the ambipolar diffusion cutoff, whose interaction with self-gravity
needs to be investigated.

{\em What determines the masses of individual stars?}  One factor must
be the size of the initial reservoirs of collapsing gas, determined by
turbulent fragmentation in the complex flow just described.
Subsequent accretion from the turbulent gas, perhaps in competition
with other stars, or even by collisions between either protostellar
cores or stars, could also be important, but must still be shown to
occur, especially in a magnetized medium.  The properties of
protostellar objects depend on the time history of the
accretion. Feedback from the newly formed star itself, or from its
neighbors, in the form of radiation pressure, ionizing radiation, or
stellar winds and jets, may yet prove to be another bounding term on
stellar mass.

{\em At what scales does the conservation of angular momentum and
magnetic flux fail?}  That they must fail is clear from the vast
discrepancy between galactic and stellar values.  Protostellar jets
almost certainly form when magnetic fields redistribute angular
momentum away from accreting gas. This demonstrates that the
conservation of flux and angular momentum must be coupled at least at
small scales.  However, the observational hint that molecular cloud
cores may be lacking substantial flux from the galactic value suggests
that flux may already be lost at rather large scales and low
densities.
Conversely, the prevalence of binary stars suggests that magnetic
braking cannot be completely efficient at draining angular momentum
from collapsing protostars, and indicates that ambipolar diffusion or
some other process limits the effectiveness of braking.  This has not
yet been modeled.

{\em What determines the initial conditions of stellar groups and
clusters?} The spatial distribution, initial velocity dispersion, and
binary distribution of stars of different masses in a stellar cluster
or association are all determined at least partly by the properties of
the turbulent flow from which the stars formed.  A quantitative
analytic model for the retardation of collapse by hydrodynamical or
MHD turbulence remains needed.  It further remains unknown how much
the final properties of a stellar group or cluster depend on the
initial state of the turbulence, and how much they depend on the
properties of gravitationally collapsing gas.  The influence of
magnetic fields on these properties also remains almost unexplored,
although the ability of the field to redistribute angular momentum
suggests that they must play at least some role.

{\em What controls the distribution and metallicity of gas in
star-forming galaxies?}  At the largest scale, gas follows the
potential of a galaxy just as do all its other constituents.  The
dissipative nature of gas can allow it to quickly shed angular
momentum in disturbed potentials and fall to the centers of galaxies,
triggering starbursts.  Even in normal galaxies, gravitational
instability may determine the location of the largest concentrations
of gas available for star formation.  How important is turbulence in
determining the location and properties of molecular clouds formed
from that gas?  Are the molecular clouds destroyed again by the same
turbulent flow that created them, or do they decouple from the flow,
only to be destroyed by star formation within them?  How slowly do
turbulent flows mix chemical inhomogeneities, and can the scatter of
metallicities apparent in stars of apparently equal age be explained
by the process?

{\em Where and how fast do stars form in galaxies?}  The existence of
the empirical Schmidt Law relating gas column density to star
formation rate, with a threshold at low column density, still needs to
be definitively explained.  Can the threshold be caused by a universal
minimum level of turbulence, or by a minimum column density below
which it is difficult for gas to cool?  In either case, examination of
low-metallicity and dwarf galaxies may well provide examples of
objects sufficiently different from massive disk galaxies in both
cooling and rotation to demonstrate one or the other of these
possibilities.

{\em What determines the star formation efficiency of galaxies?}  The
relative importance of turbulence, rotation, gravitational
instability, and thermal instability remains unresolved.  At this
scale, turbulence can only play an instrumental role, transmitting the
influence of whatever drives it to the interstellar gas.  One
possibility is that galaxies are essentially self-regulated, with
supernovae from recent star formation determining the level of
turbulence, and thus the ongoing star formation rate.  Another
possibility is that a thermal or rotational bottleneck to star
formation exists, and that galaxies actually form stars just as fast
as they are able, more or less regardless of the strength of the
turbulence in most reasonable regimes.   Finding observational and
theoretical means to distinguish these scenarios represents the great
challenge of understanding the large-scale behavior of star formation
in galaxies.

\section*{ACKNOWLEDGMENTS}

We have benefited from long-term collaborations, discussions, and
exchange of ideas and results with a large number of people. We
particularly mention (in alphabetical order) M.\ A.\ de Avillez, J.\
Ballesteros-Paredes, P.\ Bodenheimer, A.\ Burkert, B.\ G.\ Elmegreen,
C.\ Gammie, L.\ Hartmann, F.\ Heitsch, P.\ Kroupa, D.\ N.\ C.\ Lin,
C.\ F.\ McKee, \AA.\ Nordlund, V.\ Ossenkopf, E.\ C. Ostriker, P.\
Padoan, F. H. Shu, M.\ D.  Smith, J.\ M.\ Stone, E.\
V\'azquez-Semadeni, H.\ Zinnecker, and E.\ G.\ Zweibel.  This review
also benefited from two detailed anonymous reviews and extended
comments from E.\ Falgarone, F.\ Heitsch, M.\ K.\ R.\ Joung, A.\
Lazarian, M.\ D.\ Smith, and E.\ V\'azquez-Semadeni.  Finally, we must
thank our editors V.\ Trimble and J.\ Krolik for their patience and
encouragement. M-MML was supported by CAREER grant no.\ AST 99-85392
from the US National Science Foundation, and by the US National
Aeronautics and Space Administration (NASA) Astrophysics Theory
Program under grant no.\ NAG5-10103.  RSK was supported by the Emmy
Noether Program of the Deutsche Forschungsgemeinschaft (grant no.\
KL1358/1) and by the NASA Astrophysics Theory Program through the
Center for Star Formation Studies at NASA's Ames Research Center, UC
Berkeley, and UC Santa Cruz.  Preparation of this work made extensive
use of the NASA Astrophysical Data System Abstract Service.



{\narrowtext
\begin{table}
\caption{\label{tab:larson-penston}
Properties of the Larson-Penston solution of isothermal
  collapse.} 
\begin{tabular}{@{\extracolsep{\fill}}p{2.3cm}p{3cm}p{3cm}}
 & before core formation  & after core formation\\
 & $(t<0)$                  & $(t>0)$               \\
\tableline
density profile   & $\rho \propto (r^2+r_0^2)^{-1}$    & $\rho
\propto r^{-3/2}, r\rightarrow 0$\\ 
                  & ($r_0
 \rightarrow 0$ as $t\rightarrow 0_{-}$)  & $\rho
\propto r^{-2},\,\,\,\, r\rightarrow \infty$\\ 
& {\em \mbox{flattened~isothermal} sphere} & \\
velocity profile  & $v\propto r/t$ as $t \rightarrow 0_{-}$  &
$v \propto r^{-1/2}$, \,\,\,\,$r\rightarrow 0$\\
& $v\approx -3.3\,c_{\rm s}$, $r\rightarrow \infty$ & $v\approx -3.3\,c_{\rm s}$, $r\rightarrow \infty$\\
accretion rate & &  $\dot{M}= 47\,c^3_{\rm s}/G$ 
\end{tabular}
\end{table}
}

{\narrowtext
\begin{table}
\caption{\label{tab:shu}
Properties of the Shu solution of isothermal
  collapse.} 
\begin{tabular}{@{\extracolsep{\fill}}p{2.3cm}p{3cm}p{3cm}}
& before core formation  & after core formation\\
 & $(t<0)$                  & $(t>0)$               \\
\tableline
density profile   & $\rho \propto r^{-2}$, $\forall$  $r$    & $\rho
\propto r^{-3/2}$, $r\le c_{\rm s}t$\\ 
& {\em\mbox{singular~isothermal} sphere} &  $\rho
\propto r^{-2}$, \,\,\,\,$r> c_{\rm s}t$\\
velocity profile  & $v\equiv 0$, $\forall$  $r$ &
$v \propto r^{-1/2}$, $r\le c_{\rm s}t$\\
& & $v\equiv 0$, \,\,\,\,\,\,\,\,\,\,\,$r>c_{\rm s}t$\\
accretion rate & &  $\dot{M}
=0.975\,c^3_{\rm s}/G$ 
\end{tabular}
\end{table}
}

{\widetext
\begin{table}
{\caption{\label{tab:MC-prop}
 Physical properties of interstellar clouds}}
\begin{tabular}[t]{p{3.5cm}p{3.5cm}p{3.0cm}p{3.0cm}p{3.0cm}}
& GIANT MOLECULAR CLOUD COMPLEX & MOLECULAR CLOUD & STAR-FORMING CLUMP &
PROTOSTELLAR CORE$^a$ \\
\tableline
Size (pc)                           & $10 - 60$  &$2 - 20$   & $0.1-2$    &$\sil 0.1$\\
Density ($n({\rm H}_2)/{\rm cm}^3$) & $100-500$  &$10^2-10^4$& $10^3-10^5$&$>10^5$ \\
Mass (M$_{\odot}$)                  & $10^4-10^6$&$10^2-10^4$& $10 - 10^3$&$0.1-10$ \\
Line width (km$\,$s$^{-1}$)         & $5-15$     &$1 - 10$   & $0.3-3$    &$0.1-0.7$\\
Temperature (K)                     & $7-15$     &$10-30$    & $10-30$    &$7-15$ \\
Examples                            & W51, W3, M17, Orion\-Monoceros,
Taurus-Auriga-Perseus complex & L1641, L1630, W33, W3A, B227, L1495,
L1529&& see \S~\ref{sub:cores}\\
\end{tabular}
{\footnotesize \vspace{0.2cm} $^a$~Protostellar cores in the "prestellar" phase,
  i.e.\ before the formation of the protostar in its interior.
}
\end{table}
}




\end{document}